\documentclass[notitlepage, onecolumn]{revtex4-1}
\usepackage{mathrsfs}  
\usepackage{amsfonts}
\usepackage{amsmath}
\usepackage{amssymb}
\usepackage{mathtools}
\usepackage{epsfig}
\usepackage{color}
\usepackage{array}
\DeclareMathAlphabet{\mathbfsf}{\encodingdefault}{\sfdefault}{bx}{n}
\newcommand{\mri}{\mathrm{i}}

\newcommand{\inner}[2]{\left\langle #1,\,#2\right\rangle}
\newcommand{\norm}[1]{\left\| #1\right\|}
\newcommand{\arccot}{\mathrm{arccot}}

\newcommand{\continuous}{\mathcal{C}^0\left(\left[0,\pi\right]\right)}

\newcommand{\odd}{^{(o)}}
\newcommand{\even}{^{(e)}}
\newcommand{\sym}{^{(s)}}
\newcommand{\anti}{^{(a)}}

\newcommand{\elemcss}{\mathcal{S}^{cs}}
\newcommand{\elemssc}{\mathcal{S}^{sc}}
\newcommand{\elemscs}{\mathcal{C}^{ss}}
\newcommand{\elemccc}{\mathcal{C}^{cc}}
\begin{document}
\title{Hybrid Chebyshev function bases for sparse spectral methods in parity-mixed PDEs on an infinite domain}
\author{\surname{Benjamin} Miquel}
\email[Corresponding author:]{benjamin.miquel@colorado.edu}
\affiliation{Department of Applied Mathematics, University of Colorado
, Boulder, CO 80309, USA}
\author{\surname{Keith} Julien}
\affiliation{Department of Applied Mathematics, University of Colorado
, Boulder, CO 80309, USA}

\date{28 July 2017}
\begin{abstract}
We present a numerical spectral method to solve systems of differential equations on an infinite interval $y\in (-\infty, \infty)$ in presence of linear differential operators of the form $Q(y) \left(\partial/\partial_y\right)^b$ (where $Q(y)$ is a rational fraction and $b$ a positive integer). Even when these operators are not parity-preserving, we demonstrate how a mixed expansion in interleaved Chebyshev rational functions $TB_n(y)$ and $SB_n(y)$ preserves the sparsity of their discretization. This paves the way for fast $O(N\ln N)$ and spectrally accurate 
mixed implicit-explicit time-marching of sets of linear and nonlinear equations in unbounded geometries.
\end{abstract}
\maketitle

\section{Introduction}
Different numerical strategies (see~\cite{boyd} for a review) are available to solve differential equations on an infinite interval $y\in (-\infty, \infty)$. The first subcategory of methods consists in artificially introducing a bounded interval, the width of which is large compared to the extension of the exact solution. Over this finite interval, the classic weaponry of finite differences methods or spectral methods~\cite{boyd, fornberg, orszag} 
may be deployed. Other approaches rely on a representation of the solution using a basis of functions of infinite support. Usual candidates contain the Hermite functions, the Whittaker sinc function, or rational Chebyshev functions. In this paper, the focus is on the latter rational Chebyshev functions since fast Fourier transform algorithms can be advantageously utilized to compute the direct and inverse transforms from physical to spectral space. 

Fast transform algorithms render amenable mixed implicit-explicit (IMEX)  time-marching schemes
for coupled systems of differential equations 
\begin{equation}
\mathscr{M}\partial_t \boldsymbol{\psi} = \mathscr{L} \boldsymbol{\psi} +  \mathscr{N}(\boldsymbol{\psi})
\label{gov_general_form_a}
\end{equation}
with  linear operators $\mathscr{M},  \mathscr{L}$ and where a popular choice is to treat pseudo-spectrally the nonlinear terms $\mathscr{N}$. Often, the system of differential equations are accompanied with a set of $M_b$ linear boundary conditions, labelled by the integer $m\in[1,M_b]$:
\begin{equation}
\mathscr{B}_m \boldsymbol{\psi}(\boldsymbol{r}) = b_m(\boldsymbol{r})\quad \mathrm{for} \: \boldsymbol{r} \in \partial \Omega\,.
\label{bc_general_form_a}
\end{equation}
The stability and accuracy of implicit time-stepping schemes render them very desirable over explicit schemes, but comes at the cost of one or several linear algebraic solves  of the form $\mathbf{A x}=\mathbf{b}$ for each time-step. As the resolution $N$ increases, such solves become prohibitively costly and necessitate $O(N^3)$ operations if the chosen discretization results in a dense representation of $\mathscr{L}$. An implicit treatment of the linear terms, as in the popular IMEX schemes, hence advocates for a spectral decomposition that yields a sparse representation of both the operators $\mathscr{L}$ and $\mathscr{M}$. Additionally, solving the corresponding generalized eigenproblem $\lambda \mathscr{M} \boldsymbol{\psi} = \mathscr{L} \boldsymbol{\psi}$  and (\ref{bc_general_form_a}) in order to analyze the linear dynamics also benefit from sparse, and ideally well-conditioned, discretized operators.
  
In the case of a scalar differential equation, Boyd~\cite{boydJCP87} has shown that the rational Chebyshev functions potentially provide a banded representation for operators $\mathscr{L}$ composed of multiplications by monomials and differential operators $y^a\left( \partial/\partial y \right)^b$. However, sparsity relies on the parity of $\mathscr{L}$, and is usually ruined if $\mathscr{L}$ is composed of operators of different parities. 

In this paper, we introduce hybrid Chebyshev expansions based upon the definition of two families of function $TS_n(y)$ and $ST_n(y)$. These are obtained by interweaving Chebyshev rational functions $TB_n(y)$ and $SB_n$. Upon leveraging these hybrid expansions, we expose a method that yields sparse representations for rational differential operators, even in cases where parity mixing would destroy the sparsity with regular Chebyshev expansions. Following the spirit of this approach for a single scalar equation, a recipe for obtaining a sparse discretized system for sets of coupled differential equations is also exposed.
 
For the sake of completeness and clarity, the authors have elected to use a pedagogical tone. As such the first sections will be familiar to seasoned users of spectral methods. The complexity of the examples developed is gradually ramped up throughout the paper. The role of these examples is hence to illustrate the mechanics of our expansions in different situations, so that to provide the reader with models that he can adapt to his own problem.

The paper is organized as follows. A brief summary of spectral methods and discretization is presented in section~\ref{sec_specmet}. The rational Chebyshev functions and hybrid expansions are presented in section~\ref{sec_ratcheb}. Their convergence properties are established and their utility for producing sparse discretized system is illustrated on a fundamental textbook example: the quantum harmonic oscillator. 
 Section~\ref{sec_QAHO} is dedicated to the resolution of scalar equations with mixed-parity operator, through the example of the anharmonic oscillator. The method is extended to sets of coupled equations in Section~\ref{sec_eqwaves}, through the geophysical fluid dynamics example of the equatorial $\beta-$plane equations. Section~\ref{sec_2d} is dedicated to the generalisation to multidimensional domains. The applicability of the method to time-stepping schemes for non linear equations is described in section~\ref{sec_KH} and illustrated on the example of the Kelvin-Helmholtz instability in shear flows. Finally, section~\ref{sec_conclusion} is composed of concluding remarks. Throughout, details of numerical approach are relegated to the appendices.

\section{From equations to discrete representations} 
\label{sec_specmet}

Three classes of spatial discretization methods are generally available for governing equations (\ref{gov_general_form_a}) and (\ref{bc_general_form_a}). The venerable finite differences method, first introduced in 1910 by Richardson~\cite{richardsonPRSA10} evaluates derivatives on a mesh and can be extended to the finite volume method on an irregular mesh~\cite{leveque}. Finite spectral element methods~\cite{pateraJCP84} decompose $\Omega$ in sub-domains where the unknowns are approximated by a small sample of basis functions. Finally, spectral methods, of interest here, express the unknowns  as truncated expansion of $N$ orthogonal basis functions $\big\{ \phi_k(\boldsymbol{r})\big\}_{1 \le k \le N}$ on the whole domain $\Omega$~\cite{boyd,fornberg,orszag}:
\begin{equation}
\label{eqn:spdec}
\boldsymbol{\psi}_i(\boldsymbol{r}) = \sum_{k=1}^{N} \widetilde{\mathsf{X}}_{i,k} \phi_k(\boldsymbol{r}).
\end{equation} 
Upon substitution of the truncated expansion above into the governing equations  (\ref{gov_general_form_a}), projecting each individual equation onto the $\big\{\phi_j\big\}$ basis yields the following algebraic system: 
\begin{equation}
\mathbfsf{M} \partial_t \widetilde{\mathsf{X}} = \mathbfsf{L} \widetilde{\mathsf{X}} + \mathsf{N} [\widetilde{\mathsf{X}} ]\, .
\label{algebraic}\end{equation}
In the context of spectral methods boundary conditions can be enforced following three routes. 
\emph{Galerkin} methods use basis functions $\phi_k(\boldsymbol{r})$, possibly tailored by basis recombination, that intrinsically obey the boundary conditions on $\partial\Omega$, so that no additional step is required for the whole expansion to comply with boundary conditions. For basis functions that violate the boundary conditions, two approaches are available. The \emph{Collocation} method explicitly solves the governing equations on the physical space, i.e. on the Gauss-Lobatto or Gauss-Chebyshev grids, and enforces the boundary condition on the boundary locations $\partial\Omega$. The \emph{tau method} evaluates the boundary conditions in coefficient space and incorporates the resulting tau lines in the matrices $\mathbfsf{M}$ and $\mathbfsf{L}$ at the expense of higher order projections of the governing equations. A more detailed discussion is dedicated to boundary conditions in section~\ref{sec_2d}. 

The linear limit of the algebraic system~(\ref{algebraic}) is obtained by setting $ \mathsf{N} [\widetilde{\mathsf{X}} ]=0$. In the case of autonomous sets of equations, a generalized eigenvalue problem where solutions are sought in the form of normal modes $\widetilde{\mathsf{X}}(t) = \widetilde{\mathsf{X}}_a \exp(\lambda t)$ with growth rate $\lambda$ can be solved to characterize the linear dynamics:
\begin{equation}
\label{eigenproblem}
\lambda\, \mathbfsf{M}\, \widetilde{\mathsf{X}}_a = \mathbfsf{L}\, \widetilde{\mathsf{X}}_a\, .
\end{equation}
When the full spectrum is sought after, QR/QZ methods are available in linear algebra libraries (e.g. LAPACK). Regardless of the sparsity of the matrices, such methods are very costly with an $O(N^3)$ operations count. Moreover, it is often the case
that the growthrate and frequency of the most unstable mode only is of primary interest. If this is so, using sparse iterative methods such as Arnoldi's method will prove much cheaper and faster, typically of complexity $O(N)$ for sparse matrices. Iterative methods are therefore appealing for continuation in parameters. Granted, these methods suffer a certain lack of robustness: they usually require to be initialized with a good guess and commonly miss eigenvalues. However, a possible approach (previously described in~\cite{boyd}) for a practical problem such as computing marginal stability curves or surfaces, or computing the optimal growthrate in a parameter space, is to first sweep through the parameter space along a coarse mesh and to compute the full spectra at a modest resolution using a QR/QZ algorithm. Once a first intuition is gained, a much finer sweeping can be carried out using iterative methods and a higher resolution (potentially ruling out under resolved modes, for instance).

An alternate approach suitable to the study of instabilities consists in time-stepping the linearized equation: one may then observe the growth of the most unstable mode. A large variety of time-stepping schemes are now available. Having applications in fluid mechanics in mind, the focus in this paper is on developing, for systems on an infinite line, spectral methods that are suitable for IMEX time-steppers. IMEX schemes evaluate the linear term in  (\ref{algebraic}) implicitly at the end of the time-step, as opposed to  evaluation at the start of the time-step for explicit time-stepping schemes. This result in a greater numerical stability and accuracy in comparison with their explicit counterparts, particularly desirable when stiff operators are present. As a pedagogical example, the simplest implicit time-stepper is the first order accurate Backward Euler scheme, which obtains the value of the unknown $\widetilde{\mathsf{X}}^{(n+1)}$ after a time-step $\tau$ by solving:
\begin{equation}
\left(\mathbfsf{M} - \tau\mathbfsf{L}\right) \widetilde{\mathsf{X}}^{(n+1)} =
 \mathbfsf{M} \widetilde{\mathsf{X}}^{(n)}\, , \label{linear_solve}
\end{equation} 
where the value at the beginning of the time-step $\widetilde{\mathsf{X}}^{(n)}$ is known. 
Higher-order L-stable implicit schemes~\cite{ascherANM97,groomsJCP11}, and more recently memory-efficient schemes~\cite{cavaglieriJCP15} have been developed. The performance bottleneck of such schemes is the complexity of the linear solves of the form given in equation~\ref{linear_solve}. Such linear algebraic systems are costly to solve when the matrix is dense: the complexity of the solve algorithm is of order $O(N^3)$. In strong contrast, for a discretization that yields a banded operator $\left(\mathbfsf{M} - \tau\mathbfsf{L}\right)$, this complexity is drastically reduced to $O(N)$. From this observation, a successful implementation of an implicit algorithm is conditional on the existence of sparse matrix representations $\mathbfsf{M}$ and $\mathbfsf{L}$  for operators $\mathscr{M}$ and $\mathscr{L}$.

Beyond the linear dynamics, one may investigate the fully nonlinear dynamics, for instance in view of understanding the saturation of an instability. To avoid a costly direct computation of the non linear term in spectral space by means of a convolution product, the pseudo-spectral approach is utilized where potential derivatives are first evaluated in spectral space and products are then computed in physical space. The complexity of this procedure is dominated by the complexity of the transforms required between the physical and spectral space. When implemented as matrix multiplications, these transforms usually necessitate $O(N^2)$ operations. This is largely the reason why spectral methods did not gain popularity until a fast Fourier transform (FFT) algorithm was published in 1965 by Cooley and Tukey~\cite{cooleyMC65}, revisiting an idea of Gauss~\cite{gauss}. The FFT algorithm implements transforms in $O(N\ln N)$ operations only.

The great strength of spectral methods is their accuracy: the truncated expansion employed in spectral methods converges exponentially towards the true solution, when analytic, as opposed to algebraically with finite differences. Further, spectral methods have negligible spurious numerical dissipation, a well known feature of finite differences methods. A conservative numerical method is highly desirable for weakly dissipative equations, such as high Reynolds number fluid dynamics problems. The accuracy of spectral method comes at a price. First, they suffer from a poor flexibility concerning the geometry of the domain. Regular domains (e.g. rectangles, parallelepipeds, spheres, cylinders, \emph{etc.}) are good candidates for a spectral treatment, but spectral methods are ill-suited for irregular geometries. Second, we already mentioned the necessity of transforms: this point strongly advocates for Chebyshev or Fourier basis, which both utilize fast Fourier transform algorithms. Third, the discretized representation of the operators are often dense. 

As a conclusion of these observations, the emphasis of this paper is on choices of rational Chebyshev function expansions for functions on the infinite line. The central idea here is that sensibly chosen hybrid expansions obtained by interleaving different bases
preserve the sparsity of discretized operators and exploit existing FFT algorithms.

\section{The rational Chebyshev functions}\label{sec_ratcheb}
In this paper, we introduce the basis of rational Chebyshev functions as remapped cosine and sine functions on the infinite line. Not only does this approach lead to an insightful definition of the collocation points on the infinite line, but it also provides an understanding of which 
operators will lend themselves to a sparse discrete representation when approximated under the basis of rational Chebyshev functions. We thus begin by 
focussing on functions on the interval $[0,\pi]$ prior to the remapping. 
 
 \subsection{Trigonometric basis functions on a finite interval}
We consider $\continuous$, 
the Hilbert space of real valued continuous functions on the interval $[0,\pi]$, that are also piecewise of class $\mathscr{C}^1$ (i.e. with a continuous first derivative) and that vanishes at the end of the interval $f(0) = f(\pi) = 0$. We define the inner product:
\begin{equation}
\inner{f}{g} = \frac{1}{\pi} \int_0^\pi f(\theta)g(\theta) \mathrm{d}\theta\, ,
\end{equation}
and the corresponding 2-norm $\norm{f} = \sqrt{\inner{f}{f}}$. 
The families of functions $\left\{\cos(n\theta)\right\}_{n\in\mathbb{N}}$, indexed by nonnegative integers,  and $\left\{\sin(n\theta)\right\}_{n\in\mathbb{N}^*}$, indexed by positive integers $\mathbb{N}^*$, each form an orthogonal and countable basis of $\continuous$. Hence a function $f\in\continuous$ can be expressed as a cosine series:
\begin{equation}\label{purecos_series}
f(\theta) = \sum_{n\ge 0} \widetilde{f}_n\cos(n\theta)\, ,
\end{equation}
with
\begin{subequations}\label{cos_coefs}
\begin{align}
\mathrm{for}\: n=0:\quad \widetilde{f}_0 &= \frac{1}{\pi}\int_0^\pi f(\theta)\mathrm{d}\theta\,,\\
\mathrm{for}\: n>0:\quad \widetilde{f}_n &=  \frac{\inner{\cos(n\theta)}{f(\theta)}}{\norm{\cos(n\theta)}^2} = \frac{2}{\pi}\int_0^\pi\cos(n\theta) f(\theta)\mathrm{d}\theta\, ,
\end{align}
\end{subequations}
 or alternatively as a sine series:
\begin{equation}\label{puresine_series}
f(\theta) = \sum_{n\ge 1} \widehat{f}_n \sin(n\theta)\, ,
\end{equation}
with 
\begin{align}
\label{sine_coefs}
\widehat{f}_n &=  \frac{\inner{\sin(n\theta)}{f(\theta)}}{\norm{\sin(n\theta)}^2} = \frac{2}{\pi}\int_0^\pi\sin(n\theta) f(\theta)\mathrm{d}\theta\, .
\end{align}
 For functions in $\continuous$, both expansions converge uniformly to the function $f$. This does not imply that the expansions are equivalent: a discussion of the convergence properties of different representations, based on the analyticity of the possible extensions of $f(\theta)$ to the interval $[-\pi,\pi]$ (e.g., see ~\cite{boyd, fornberg, orszag}), can be found in section~\ref{sec:choosing} below.
\subsection{Hybrid trigonometric bases}
\label{sec:mixedsc}
\begin{figure}
\begin{center}
\includegraphics[height = 0.28 \textwidth]{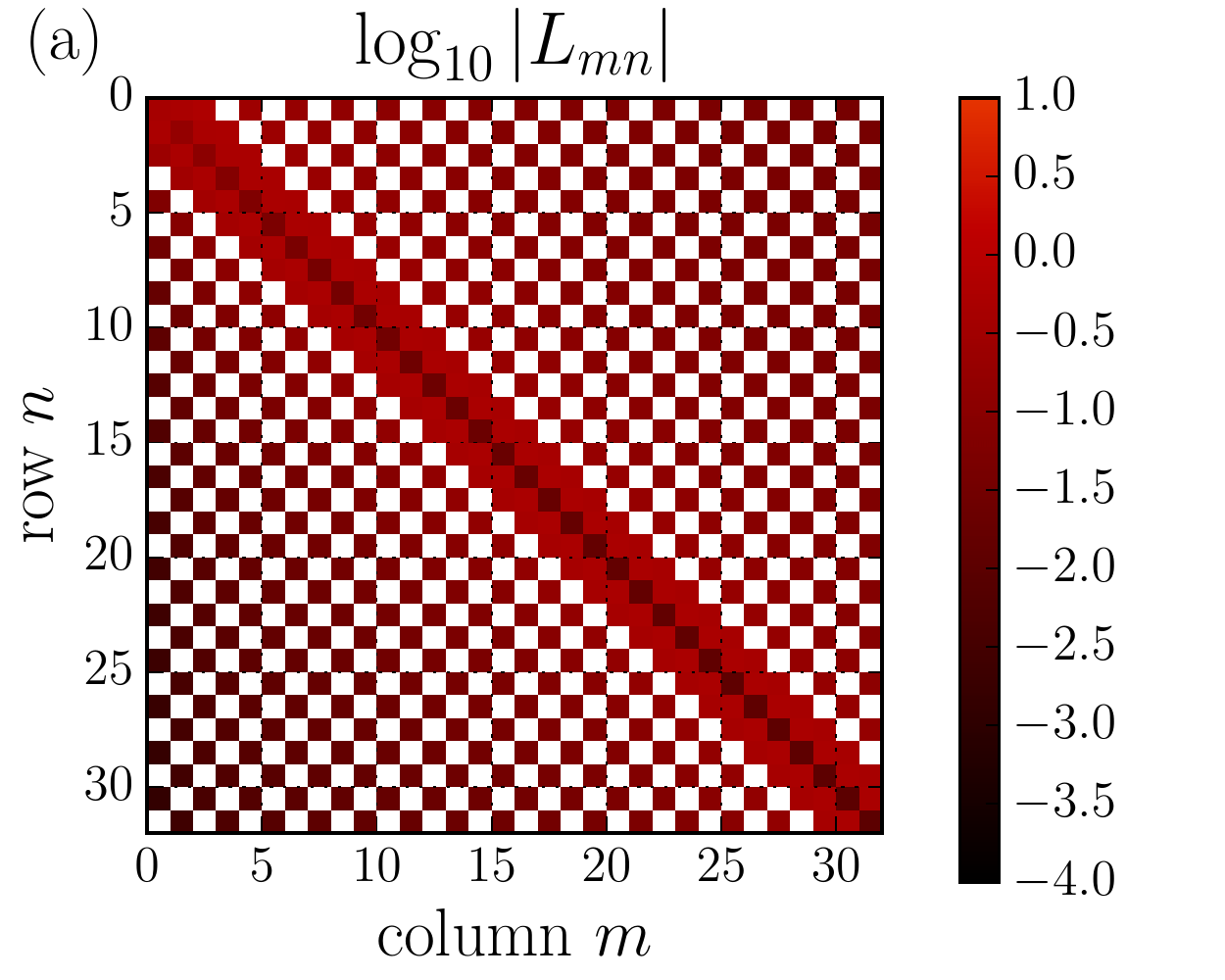}
\includegraphics[height = 0.28 \textwidth]{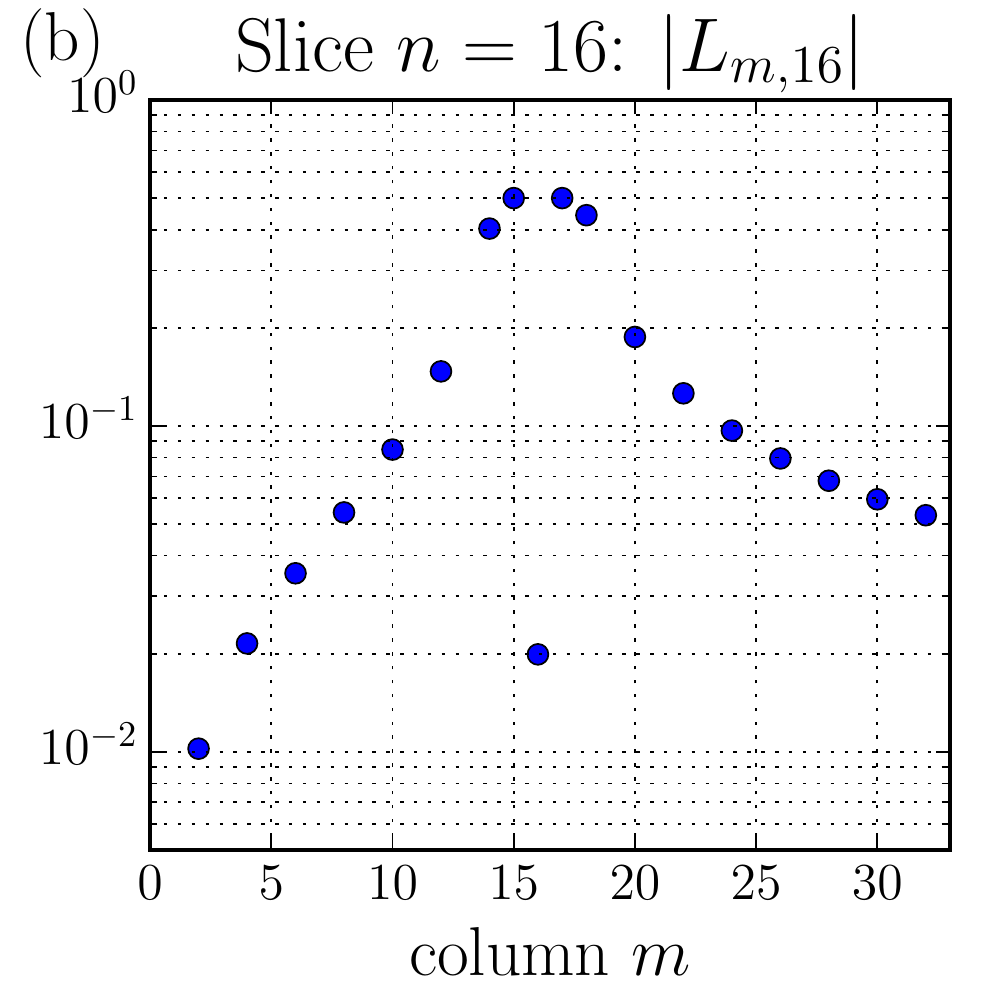}
\includegraphics[height = 0.28 \textwidth]{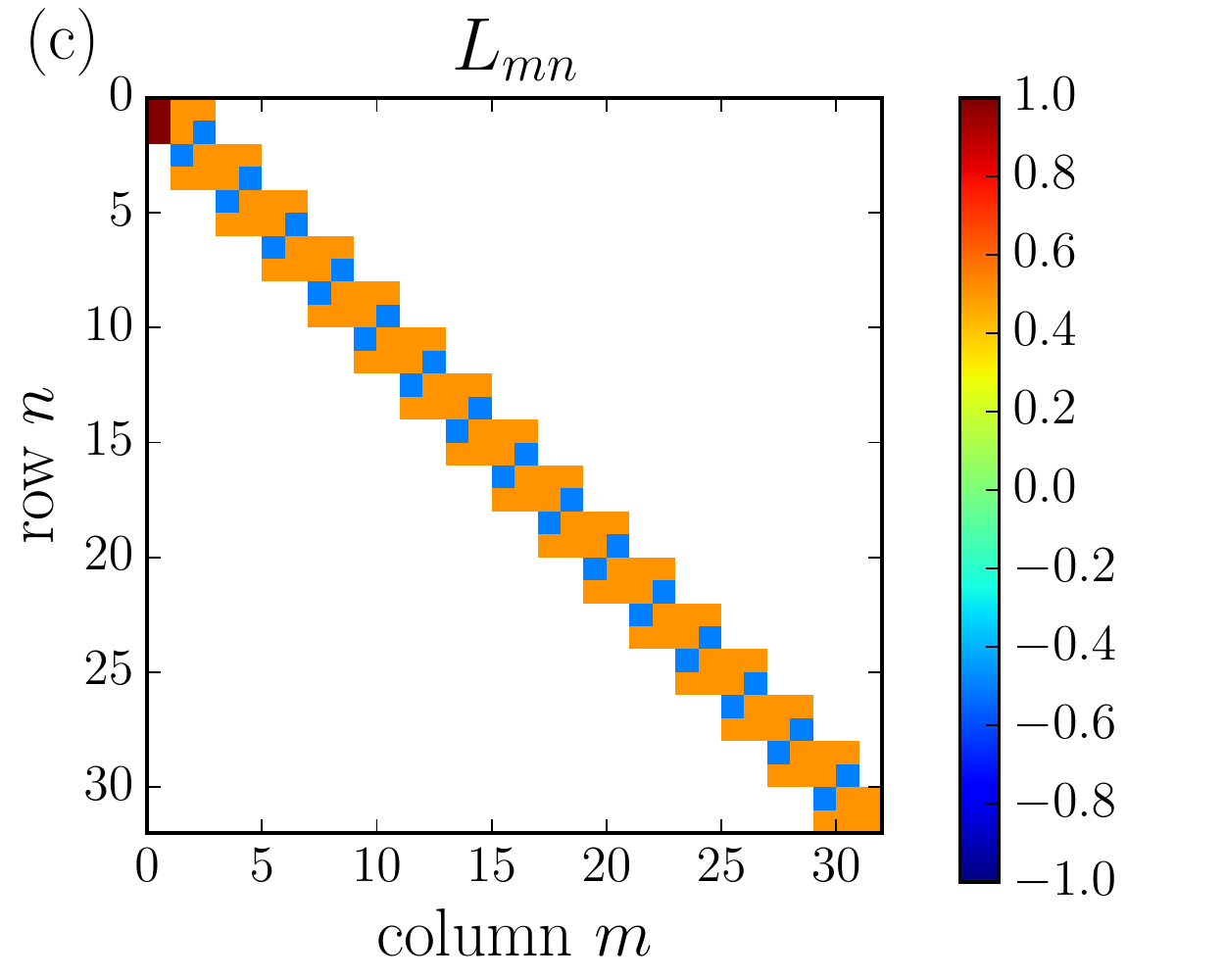}
\caption{\label{fig_dense_matrix}
 Discretizations of $\mathscr{L} = \cos\theta + \sin\theta$ on a pure trigonometric series (figures a, b), or a mixed trigonometric series (figure c). Sparsity is illustrated by displaying the zero elements in white. Non zero elements are color-coded. (a) Discretization on a sine basis $L_{mn}= \frac{\inner{\sin(m\theta)}{\mathscr{L}\sin(n\theta)}}{\norm{\sin(m\theta)}}$ (see equation~\ref{eq_dense_matrix}). (b) Slice of matrix $L_{mn}$ along row 16 (zero elements are ignored by the logarithmic scale), displaying a moderately rapid decay of the matrix elements away from the diagonal. (c) Matrix elements obtained with an expansion on $\breve{\beta}_n$ and a projection on $\mathring{\beta}_n$, namely $L_{mn}=\frac{\inner{\mathring{\beta}_m(\theta)}{\mathscr{L}\breve \beta_n(\theta)}}{\norm{\mathring{\beta}_m(\theta)}}$.
 }
\end{center}
\end{figure}
Pure cosine or sine bases (presented above in equations (\ref{purecos_series},\ref{puresine_series})) suffer the curse of dense discretization for symmetry breaking operators. For instance, consider $\mathscr{L}=\cos\theta + \sin\theta$ as example of an operator with no particular symmetry around $\pi/2$. To avoid confusion with the symmetry around $\theta=0$ that stems in our discussion about convergence (section~\ref{sec:choosing} and appendix~\ref{app:convergence}), we emphasize that the symmetry that pertains to the present discussion is the one around $\pi/2$, the center of the segment $[0,\pi]$. Regardless of the selection of the pure basis (\ref{purecos_series}) or (\ref{puresine_series}) for expansion and projection, we would invariably obtain dense representations:
\begin{equation}
\frac{\inner{\cos(m\theta)}{\mathscr{L}\cos(n\theta)}}{\norm{\cos(m\theta)}},\:
\frac{\inner{\sin(m\theta)}{\mathscr{L}\cos(n\theta)}}{\norm{\sin(m\theta)}},\:
\frac{\inner{\cos(m\theta)}{\mathscr{L}\sin(n\theta)}}{\norm{\cos(m\theta)}},\:
\mathrm{or}\:
\frac{\inner{\sin(m\theta)}{\mathscr{L}\sin(n\theta)}}{\norm{\sin(m\theta)}},
\end{equation}
all of which are band-unlimited. For example, the latter element of this list, represented on figure~\ref{fig_dense_matrix}, yields:
\begin{equation}
\frac{\inner{\sin(m\theta)}{\mathscr{L}\sin(n\theta)}}{\norm{\sin(m\theta)}} = \frac{1}{2}\left( \delta^m_{n+1} + \delta^m_{n-1}\right) 
+ \frac{m\left( 1+ \left(-1\right)^{m+n} \right)}{\pi(m-n+1)(m+n-1)}
+ \frac{m\left( 1+ \left(-1\right)^{m+n} \right)}{\pi(m-n-1)(m+n+1)}
\, .
\label{eq_dense_matrix}
\end{equation}

To circumvent the apparition of dense matrices, 
 the strategy is to introduce mixed trigonometric bases~\cite{vasilJCP08}, described below. The core idea to this paper then consists in generating the corresponding mixed Chebyshev rational function bases when working on an unbounded interval.

The continuous functions $f\in \continuous$ may be uniquely decomposed into symmetric and antisymmetric components about the center of the segment $[0,\pi]$, namely $\theta=\pi/2$:
\begin{subequations}
\begin{gather}
f(\theta) = f\sym(\theta) + f\anti (\theta)\,,
\intertext{where}
f\sym(\theta) = f\sym(\pi -\theta)\quad \mathrm{and}\quad f\anti(\theta) = -f\anti(\pi -\theta)\, .
\end{gather}
\end{subequations}
In contrast to the ``pure'' expansion in sine and cosine presented above in equations (\ref{purecos_series},\ref{puresine_series}), we define the two ``hybrid'' (or ``mixed'') cosine-sine expansions:
\begin{subequations}\label{mixed_trig_series}
	\begin{align}
	f(\theta) = f\sym (\theta) + f\anti(\theta) &\approx \sum_{p=0}^{N}\widetilde f_{2p} \cos(2p\theta) + \sum_{p=0}^{N}\widehat f_{2p}\sin(2p\theta)\label{cossin_series}\,, \intertext{or:}
	f(\theta) = f\sym (\theta) + f\anti(\theta) 	& \approx \sum_{p=0}^{N}\widehat f_{2p+1}\sin\left(\left[2p+1\right]\theta\right) +  \sum_{p=0}^{N}\widetilde f_{2p+1} \cos\left(\left[2p+1\right] \theta\right)\label{sincos_series}\,.
	\end{align}
\end{subequations}
As deduced from these mixed representations, each component $f\sym$ and $f\anti$ has two possible basis representations. For instance, $\left\{ \cos\left(2p\theta\right)\right\}_{p\in\mathbb{N}}$ and $\left\{ \sin\left(\left[2p+1\right]\theta\right)\right\}_{p\in\mathbb{N}}$ each form an independent basis for symmetric functions about $\pi/2$, whereas, $\left\{\cos\left([2p+1]\theta\right)\right\}_{p\in\mathbb{N}}$ and $\left\{ \sin\left(2p\theta\right)\right\}_{p\in\mathbb{N}^*}$ form an independent basis for antisymmetric functions about $\pi/2$. 
Since this paper leverages the virtue of mixed expansions (\ref{cossin_series},\ref{sincos_series}) and makes a heavy use of them in producing sparse discretizations, each of these composite expansions is seen hereafter as a unique expansion by introducing the following compact notations: 
\begin{subequations}
\begin{align}
f(\theta) &\approx \sum_{n=1}^{N} \breve{f}_n \breve{\beta}_n(\theta) \quad \mathrm{with}\quad \Big\{ \breve{f}_n\Big\}_{n\ge 1} = \Big\{ \widetilde{f}_0, \widehat{f}_2, \widetilde{f}_2, \widehat{f}_4,\widetilde{f}_4,\cdots\Big\}\,,
\intertext{or:}
     f(\theta)      &\approx \sum_{n=1}^{N} \mathring{f}_n \mathring{\beta}_n(\theta)\quad \mathrm{with}\quad \Big\{ \mathring{f}_n\Big\}_{n\ge 1} = \Big\{ \widetilde{f}_1, \widehat{f}_1, \widetilde{f}_3, \widehat{f}_3, \widetilde{f}_5, \widehat{f}_5,\cdots\Big\}\,.
\end{align}
\end{subequations}
The two distinct hybrid bases obtained, $\{\breve{\beta}_n\}$ and $\{\mathring{\beta}_n\}$, can be thought of as interleaved collections of sines and cosines:
\begin{subequations}
\begin{gather}
\breve{\beta}_n(\theta) = \left\{\begin{aligned}
&\cos (2p\theta)    &\quad \mathrm{for}\: n=2p+1\,,\\
&\sin (2p\theta)& \mathrm{for}\: n=2p \,,
\end{aligned} \label{def:beta_breve_condensed}\right. \\ 
\mathring{\beta}_n(\theta) = \left\{\begin{aligned}
&\cos ([2p-1]\theta)    &\quad \mathrm{for}\: n=2p-1 \,,\\
&\sin ([2p-1]\theta)&\quad \mathrm{for}\: n=2p\,,
\end{aligned}\right.\label{def:beta_mathring_condensed}
\end{gather}\label{def:betas}
\end{subequations}
or more explicitly:
\begin{subequations}
\label{def:beta_explicit}
\begin{gather}
\Big\{\breve{\beta}_n(\theta)\Big\}_{n\ge 1} = \Big\{1 , \sin 2\theta, \cos 2\theta, \sin 4\theta, \cos 4\theta,\cdots\Big\} \, ,\label{def:beta_breve_explicit}\\ 
\Big\{\mathring{\beta}_n(\theta)\Big\}_{n\ge 1} = \Big\{\cos\theta, \sin\theta, \cos 3\theta, \sin 3\theta, \cos 5\theta, \sin 5\theta, \cdots\Big\}\, .
\end{gather}
\end{subequations}
Both bases are orthogonal sets, as proven in appendix~\ref{app:ortho}.
The benefits of using such functions become evident by revisiting the introductory example of this paragraph, $\mathscr{L}=\cos \theta + \sin\theta$. From the trigonometric identities for products, 
we deduce easily that the following matrix representations for $\mathscr{L}$ (one of which is represented on figure~\ref{fig_dense_matrix}) are both sparse:
\begin{equation}
\frac{\inner{\mathring{\beta}_m(\theta)}{\mathscr{L}\breve \beta_n(\theta)}}{\norm{\mathring{\beta}_m(\theta)}},
\quad
\frac{\inner{\breve{\beta}_m(\theta)}{\mathscr{L}\mathring \beta_n(\theta)}}{\norm{\breve{\beta}_m(\theta)}}\,.
\end{equation}
Therefore, a sparse discretization for $\mathscr{L}$ is obtained by considering an expansion in $\{\breve\beta_n\}$ and a projection on the $\{\mathring{\beta}_n\}$ basis (or vice versa). This observation can be generalized in the following proposition.

\subsubsection*{Proposition 1 --}
Let $\mathscr{L}$ be a linear operator of the form
\begin{equation}
\mathscr{L} = \sum_{i=1}^{N_{\mathscr{L}}} \left(\alpha_i\cos(n_i \theta) + \beta_i \sin(m_i \theta)\right) \left(\frac{\mathrm{d}}{\mathrm{d}\theta}\right)^{\gamma_i}
\end{equation}
where $N_\mathscr{L}\in \mathbb{N}$, and for all integers $1<i<N$: $\alpha_i, \beta_i\in\mathbb{R}$, $\gamma_i\in\mathbb{N}$, and $(n_i,m_i)\in\mathbb{N}^2$ such that all integers $n_i$ and $m_i$ are of the same parity (even or odd). If all these integers are: 
\begin{itemize}
\item odd, then $\frac{\inner{\mathring{\beta}_m(\theta)}{\mathscr{L}\breve \beta_n(\theta)}}{\norm{\mathring{\beta}_m(\theta)}}$ and $\frac{\inner{\breve{\beta}_m(\theta)}{\mathscr{L}\mathring \beta_n(\theta)}}{\norm{\breve{\beta}_m(\theta)}}$ are band-limited matrices,
\item even, then $\frac{\inner{\breve{\beta}_m(\theta)}{\mathscr{L}\breve \beta_n(\theta)}}{\norm{\breve{\beta}_m(\theta)}}$ and $\frac{\inner{\mathring{\beta}_m(\theta)}{\mathscr{L}\mathring \beta_n(\theta)}}{\norm{\mathring{\beta}_m(\theta)}}$ are band-limited matrices.
\end{itemize}
\paragraph*{Proof--}
These matrices are easily built as linear combinations of matrix product of the banded elementary matrices $\mathcal{S}$, $\mathcal{C}$, and $\mathcal{D}$ given in appendix~\ref{app:chebfunc}.

The requirement that the $n_i$ and $m_i$ be of given parity will be naturally satisfied by operators originating from a remapping of rational operators on the infinite line, as discussed in the following sections.

For completeness, we close up this section by mentioning alternative expansions: analogous to the $\{\breve{\beta}_n\}_{n\ge 1}$ basis is the ,$\{\exp\left(\mathrm{i}[2p+1]\theta\right)\}_{p\in\mathbb{Z}}$ basis. Similarly, the $\{\exp\left(\mathrm{i}[2p+1]\theta\right)\}_{p\in\mathbb{Z}}$ basis is analogous to the $\{\mathring{\beta}_n\}_{n\ge 1}$ basis. The changes of basis are straightforward, so that convergence properties (discussed below in section~\ref{sec:choosing}) will be unaffected and choosing between the trigonometric and the complex exponential bases is purely a matter of taste. In the rest of this paper, we discuss and utilize the hybrid trigonometric expansions $\{\breve\beta\}$ and $\{\mathring\beta\}$ only.


\subsection{Rational Chebyshev basis functions on the infinite line}
Given the mixed representations on the $[0,\pi]$ interval, following Cain \emph{et al.}~\cite{cainJCP84} we now map this interval onto the infinite line using:
\begin{subequations}\label{def_mapping}
\begin{gather}
y = L \cot \theta\, , \\
\theta = \arccot \left(y/L\right)\, ,
\end{gather}
\end{subequations}
where $L>0$ is the mapping parameter. Derivatives on the two intervals are connected according to
\begin{equation}
\partial_y = -\frac{\sin^2 \theta}{L^2} \partial_{\theta}.
\label{chain_rule}
\end{equation}
By use of composite functions, a function on the infinite line $\psi(y)$ can be bijectively mapped on a function $f(\theta)$ on the segment $\theta\in[0,\pi]$:  
\begin{equation}
\psi(y) = \psi(y(\theta)) = f(\theta)\, .
\end{equation}
Hence, we define the Chebyshev functions $TB_n(y)$ and $SB_n(y)$ as the mapped cosine and sine functions stretched on the infinite line:
\begin{subequations}
\begin{gather}
TB_n(y) = \cos\left(n\, \mathrm{arccot} \left(y/L\right)\right)\, ,\\ SB_n(y) = \sin\left(n\, \mathrm{arccot} \left(y/L\right)\right)\, .
\end{gather}
\end{subequations}
From their definition and our discussion above, the Chebyshev functions inherit the properties of the trigonometric functions from which they are derived: they form two independent orthogonal bases on the infinite line and a truncated sum can approximate an arbitrary continuous bounded function:
\begin{subequations}\label{purecheb_series}
\begin{gather}
\psi(y) \approx \sum_{n=0}^{N}\widetilde{\psi}_n TB_n \left(y\right) \label{TB_series}\,, \\
\psi(y) \approx \sum_{n=1}^{N}\widehat{\psi}_n SB_n\left(y\right) \label{SB_series}\,.
\end{gather}
\end{subequations}
These expansions are the infinite line counterparts of the cosine and sine expansion given in equations (\ref{purecos_series},\ref{puresine_series}). The $TB_n$ (respectively, $SB_n$) coefficients  $\widetilde{\psi}_n$  (respectively, $\widehat{\psi}_n$)  are equal to the cosine (respectively, sine) coefficients $\widetilde{f}_n$ (respectively, $\widehat{f}_n$). However, they can be computed directly on the infinite line, without mapping back to the interval $[0,\pi]$, by using
\begin{equation}
\int_0^{\pi}f(\theta)\mathrm{d}\theta = \frac{1}{L}\int_{-\infty}^{\infty}\frac{\psi(y)}{1+y^2}\mathrm{d}y
\end{equation}
to generalize equations (\ref{cos_coefs},\ref{sine_coefs}) into:
\begin{subequations}
\begin{align}
\mathrm{for}\: n=0:\quad \widetilde{\psi}_0 &= \frac{1}{\pi L}\int_{-\infty}^\infty \frac{\psi(y)}{1+y^2}\mathrm{d}y\, ,\\
\mathrm{for}\: n>0:\quad \widetilde{\psi}_n &=  \frac{2}{\pi L}\int_{-\infty}^\infty \frac{TB_n(y)\,\psi(y)}{1+y^2}\mathrm{d}y\, ,\\
\widehat{\psi}_n &=  \frac{2}{\pi L}\int_{-\infty}^\infty \frac{SB_n(y)\,\psi(y)}{1+y^2}\mathrm{d}y\, .\end{align}
\end{subequations}
\subsection{Hybrid Chebyshev bases}
\label{hybrid_chebyshev_bases}
By means of the mapping (\ref{def_mapping}), our partition of symmetric and asymmetric functions around $\theta = \pi/2$ on the interval $[0,\pi]$ results in a partition of functions $\psi(y)$ into odd and even components $\psi\even(y)$ and $\psi\odd(y)$ around $y=0$. We emphasize here that in the following, symmetry properties are implicitly relative to $\theta=\pi/2$ for functions on the $[0,\pi]$ segment, and relative to $y=0$ on the infinite line $(-\infty, \infty)$. The odd and even components of $\psi(y)$ can be expanded independently with either a $TB_n$ or $SB_n$ basis. Hence, the analogous extension of the hybrid sine-cosine expansions (eq. \ref{mixed_trig_series}) are the hybrid $SB$-$TB$ expansions:
\begin{subequations}\label{mixed_cheb_series}
\begin{align}
\psi(y) = \psi\even (y) + \psi\odd(y) &\approx \sum_{p=0}^{N/2-1}\widetilde \psi_{2p}\, TB_{2p-2} \left(y\right)  + \sum_{p=1}^{N/2}\widehat \psi_{2p}\, SB_{2p} \left(y\right)\, ,\label{TBSB_series}\, \\
& \approx \sum_{p=0}^{N/2}\widehat \psi_{2p+1}\, SB_{2p+1}\left(y\right) +  \sum_{p=0}^{N/2}\widetilde \psi_{2p+1}\,
 TB_{2p+1}\left(y\right)\label{SBTB_series}\,.
\end{align}
\end{subequations}
Following section~\ref{sec:mixedsc}, we introduce the following notations for the two bases $ST(y)$ and $TS(y)$:
\begin{subequations}
\begin{gather}
TS_n(y) = \left\{\begin{aligned}
	&TB_{2p}(y)&\quad \mathrm{for}\: n=2p+1\,, \\
&SB_{2p} (y)&\quad \mathrm{for}\: n=2p\,,
\end{aligned}\right. \\ 
ST_n(y) = \left\{\begin{aligned}
	&TB_{2p-1} (y)&\quad \mathrm{for}\: n=2p-1\,, \\
&SB_{2p-1} (y)&\quad \mathrm{for}\: n=2p\,.
\end{aligned}\right.
\end{gather}
\end{subequations}
More explicitly,
\begin{subequations}\label{def:TSST}
\begin{gather}
\Big\{TS_n(y)\Big\}_{n\ge 1} = \Big\{1 , SB_2(y), TB_2(y), SB_4(y), TB_4(y),\cdots\Big\} \, ,\\ 
\Big\{ST_n(\theta)\Big\}_{n\ge 1} = \Big\{TB_1(y), SB_1(y), TB_3(y), SB_3(y), TB_5(y), SB_5(y),\cdots\Big\}\, 
\end{gather}
\end{subequations}
such  that the expansions (\ref{cossin_series},\ref{sincos_series}) become respectively:
\begin{subequations}
\begin{align}
\psi(y) &\approx \sum_{n=1}^{N} \breve{\psi}_n TS_n(y) \quad \mathrm{with}\quad \Big\{ \breve{\psi}_n\Big\}_{n\ge 1} = \Big\{ \widetilde{\psi}_0, \widehat{\psi}_2, \widetilde{\psi}_2, \widehat{\psi}_4,\widetilde{\psi}_4,\cdots\Big\}\,,\\
          &\approx \sum_{n=1}^{N} \mathring{\psi}_n ST_n(y)\quad \mathrm{with}\quad \Big\{ \mathring{\psi}_n\Big\}_{n\ge 1} = \Big\{ \widetilde{\psi}_1, \widehat{\psi}_1, \widetilde{\psi}_3, \widehat{\psi}_3, \widetilde{\psi}_5,  \widehat{\psi}_5, \cdots\Big\}\,.
\end{align}
\end{subequations}
In the remainder of the paper, we discuss how the sparsity of discretized systems is affected by the choice of this expansion. Particularly, we  highlight cases when a mixed expansion is necessary to conserve sparsity, by extending our Proposition 1 that pertains to $\theta\in[0,\pi]$ onto the infinite line $y\in(-\infty, \infty)$ by means of the mapping $y=L\cot\theta$.
\subsection{Choosing between $\{TB_n\}$, $\{SB_n\}$, $\{TS_n\}$, $\{ST_n\}$ expansions and alternatives}
\label{sec:choosing}At this stage, it is necessary to discuss the global relevance of hybrid $\{TS_n\}$ and $\{ST_n\}$ expansions for a given function $\psi(y)$. Granted the fact that they prove useful in providing sparse discretization of parity-flipping operators, the question arises of under which conditions they inherit the convergence properties of the regular $\{TB_n\}$ and $\{SB_n\}$ expansions? The answer to this question is related to the analyticity of the continuations of the remapped function $f(\theta)= \psi(L \cot \theta)$ to the $[-\pi,\pi]$ interval. A detailled discussion is given in appendix~\ref{app:convergence}, and boils down to the following conclusions.\\
Exponentially decaying functions that are dominated by $\exp(-b|y|^a)$ with $a,b>0$ have subgeometrically converging coefficients for each of the four expansions. That is to say, hybrid expansions $\{TS_n\}$ and $\{ST_n\}$ will converge as fast as pure $\{TB_n\}$ and $\{SB_n\}$ expansions. For this reason, hybrid expansions should be given preference over the pure expansions in cases where they yield a sparse discretization, as discussed above and exemplified below. Therefore the question of using hybrid Chebyshev expansions instead of alternative methods such as expansions based on Hermite functions or Cardinal functions is really equivalent to the choice between regular Chebyshev expansions versus Hermite or Cardinal functions. Hence, we refer the reader to the abundant existing literature (a possible starting point is~\cite{boyd}). The bottom line is that in most cases these three options are equivalent and none really outperforms the others from the perspective of an optimal representation. In practice however, as stated in our introduction, it is our opinion that Chebyshev rational functions have an edge whenever a sparse representation is available. Indeed, Hermite functions prove to be inpractical in the context of non linear dynamics due to the absence of implementation of fast transforms in common linear algebra libraries--despite the existence of such algorithms~\cite{orszag86,boydJCP92}. Using the Cardinal functions collocation method does not require any transform but results
 in dense discretizations, which yield a much slower time-stepping or use of iterative eigensolvers.

\subsection{The relevance of mapping before discretizing}
As mentioned above, one could very well introduce an expansion of the forms given in equations (\ref{purecheb_series}) or (\ref{mixed_cheb_series}) and discretize directly a system of the form:
\begin{equation}
\mathscr{L}_y \psi(y) = b(y)\, .
\end{equation} However, more often than not, the path of least resistance to obtain sparse matrices consists in first mapping the equation to the $[0,\pi]$ segment before proceeding to the discretization. The reason behind this is that our intuition with the action of operators on the trigonometric functions is much more developed than our insight with the Chebyshev functions. As we exemplify in the next paragraph with the case of the quantum harmonic oscillator, tweaking the equations with a view to obtaining a sparse system is much easier with trigonometric functions, whereas dealing directly with Chebyshev functions remains obscure.  
\subsection{A fundamental example: the Quantum Harmonic Oscillator}
\begin{figure}
	\begin{center}
		\includegraphics[height= 0.32\textwidth]{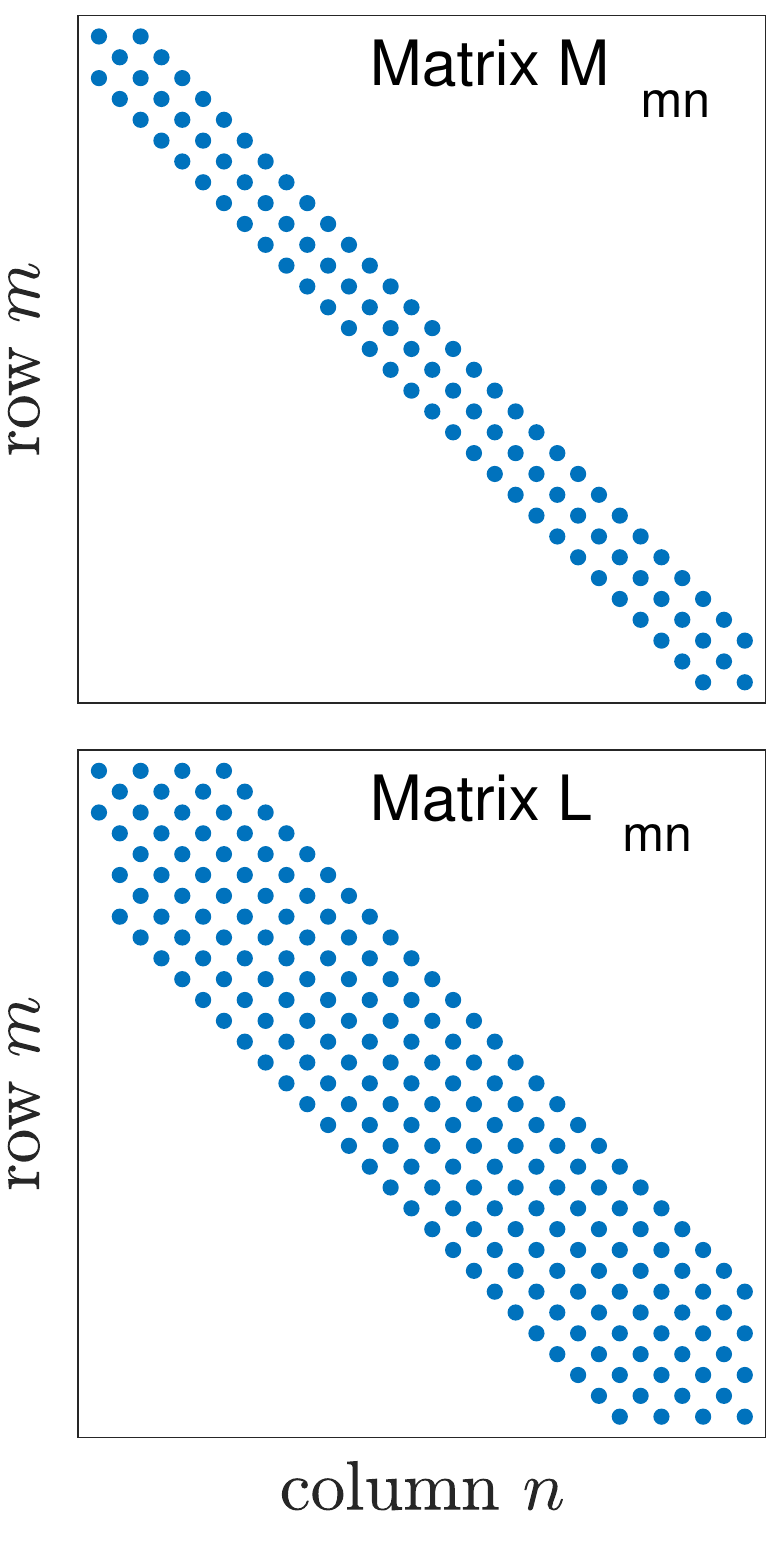}
		\includegraphics[height = 0.32\textwidth]{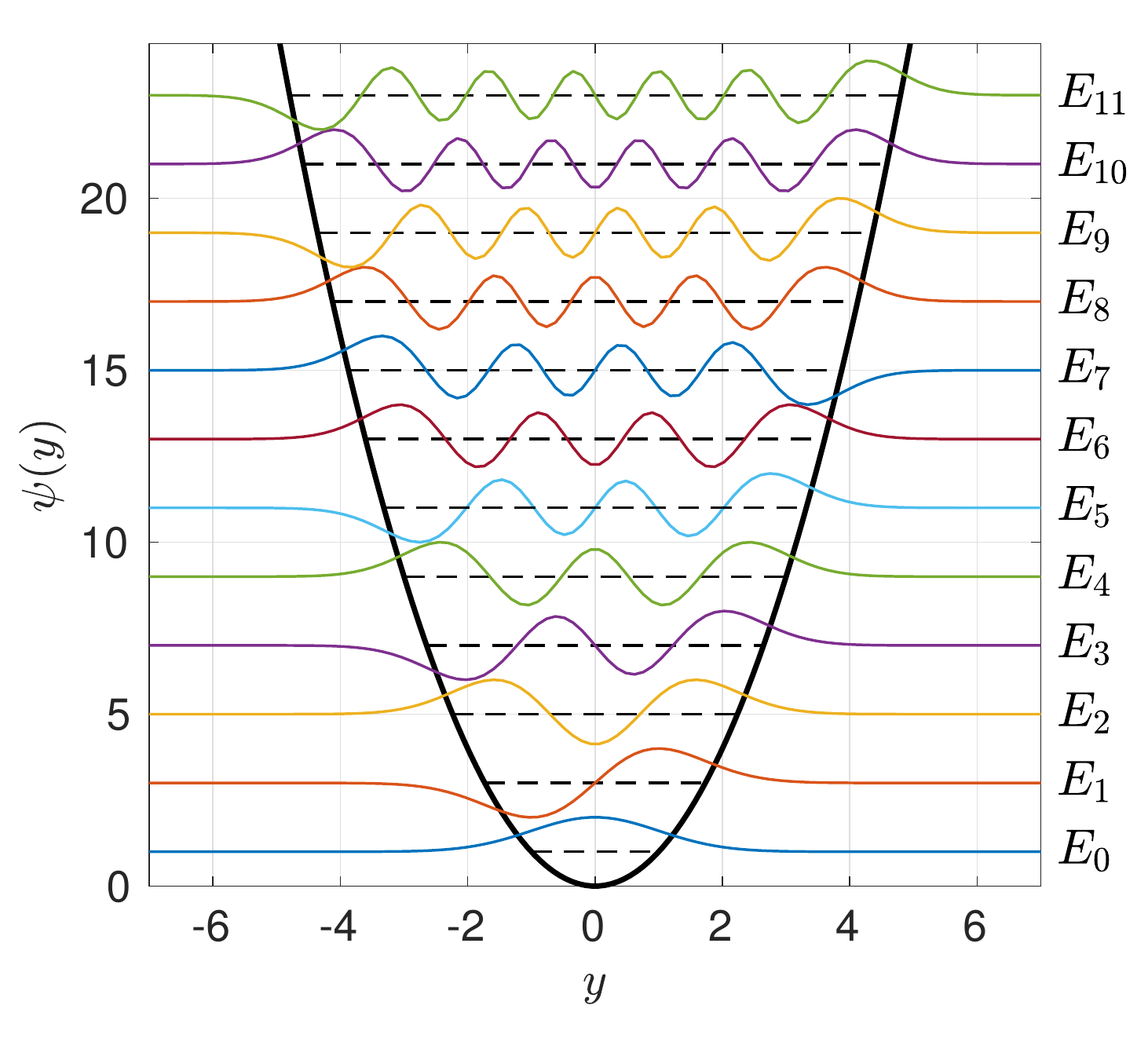}
		\includegraphics[height = 0.32\textwidth]{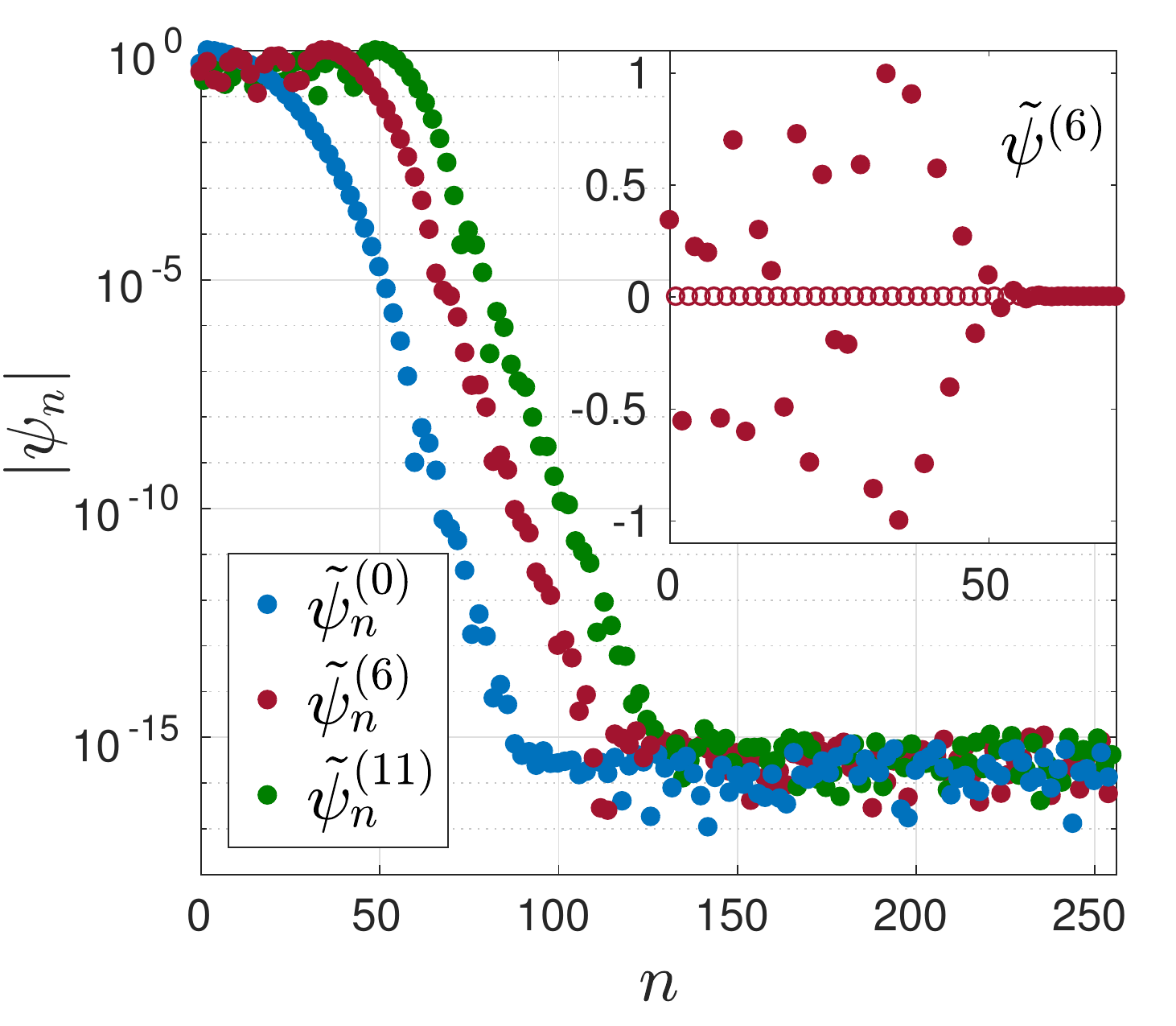}
		\caption{\label{QHO_eigenmodes} Illustration of the Chebyshev spectral decomposition of  the quantum harmonic oscillator using basis functions $TB_n(y)$. Leftmost panel: structure of the $\mathbfsf{M}$ (upper plot) and $\mathbfsf{L}$ (lower plot) matrices for 32 basis functions. Middle panel: the first 12 eigenmodes modes $\psi^{n}(y)$ of the parabolic well (continuous black parabola) obtained with 256 basis functions (solid color lines), shifted vertically by their energy level $E_n$ (dashed horizontal lines). Right panel: Spectra $| \widetilde{\psi}_n|$ for $n=0,\,6,\,11$. Observe the exponential convergence of the expansion until machine precision is reached. Insert: Spectral coefficients of mode $\psi^{(6)}$. Solid circles are non zero coefficients; open circles represent coefficients which are \emph{rigorously} zero (\emph{beyond} mere machine precision). The even symmetry of $\psi^{(6)}$ is preserved by our spectral method, as odd-degree coefficients $\widetilde{\psi}_{2p+1}$ are exactly zero.  }
	\end{center}
\end{figure}
\begin{figure}
	\begin{center}
		\includegraphics[width = 0.32\textwidth]{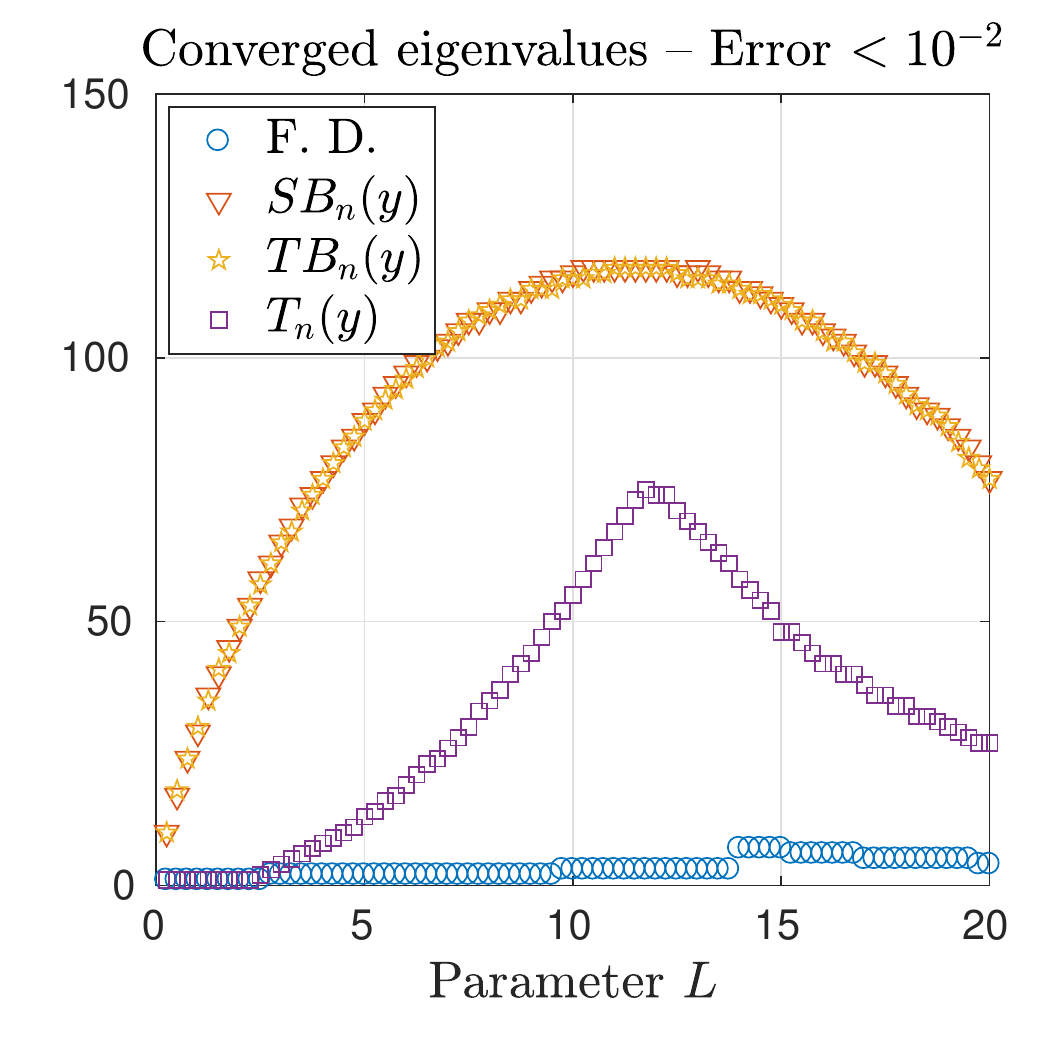}
		\includegraphics[width = 0.32\textwidth]{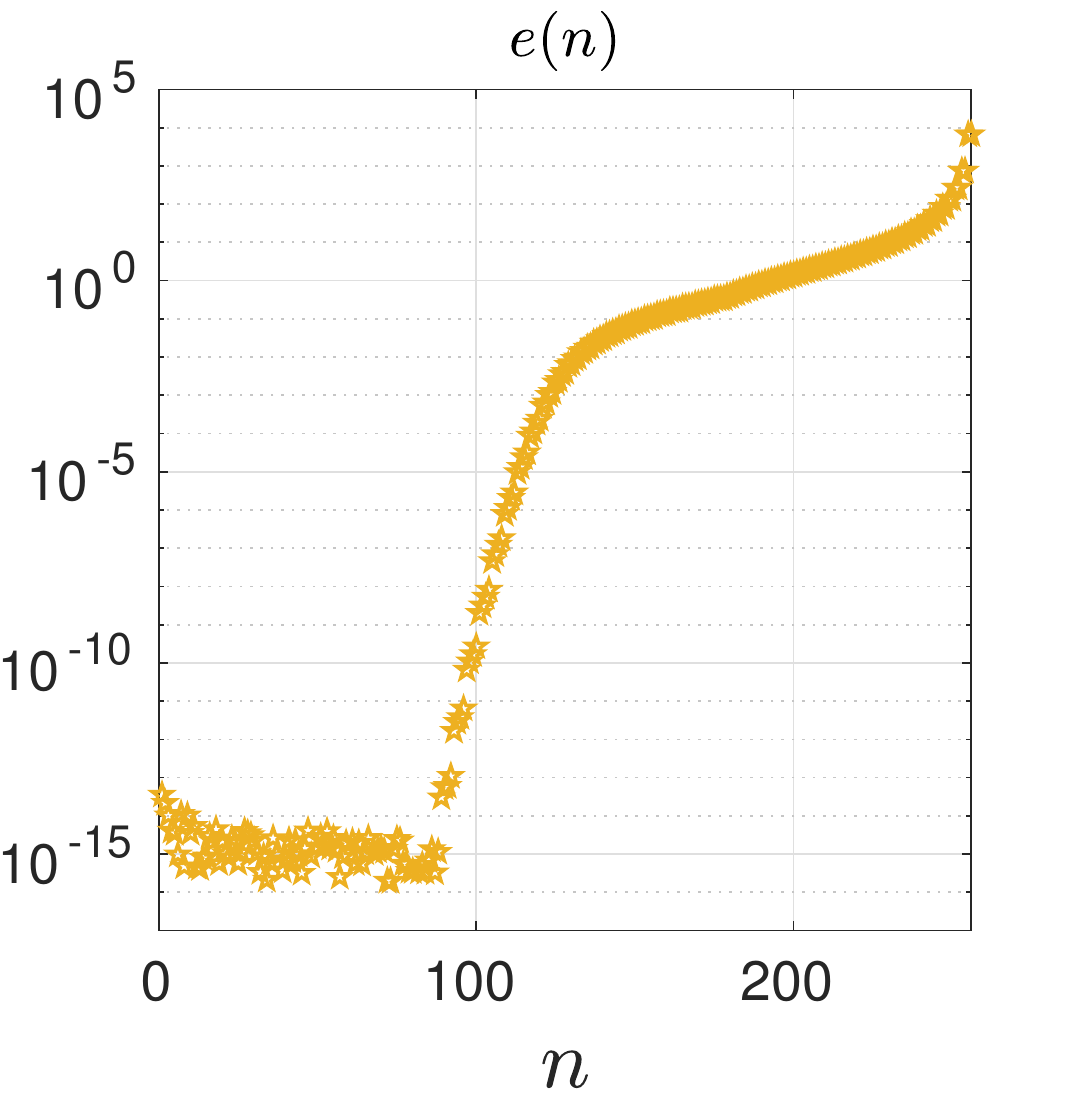}
		\includegraphics[width = 0.32\textwidth]{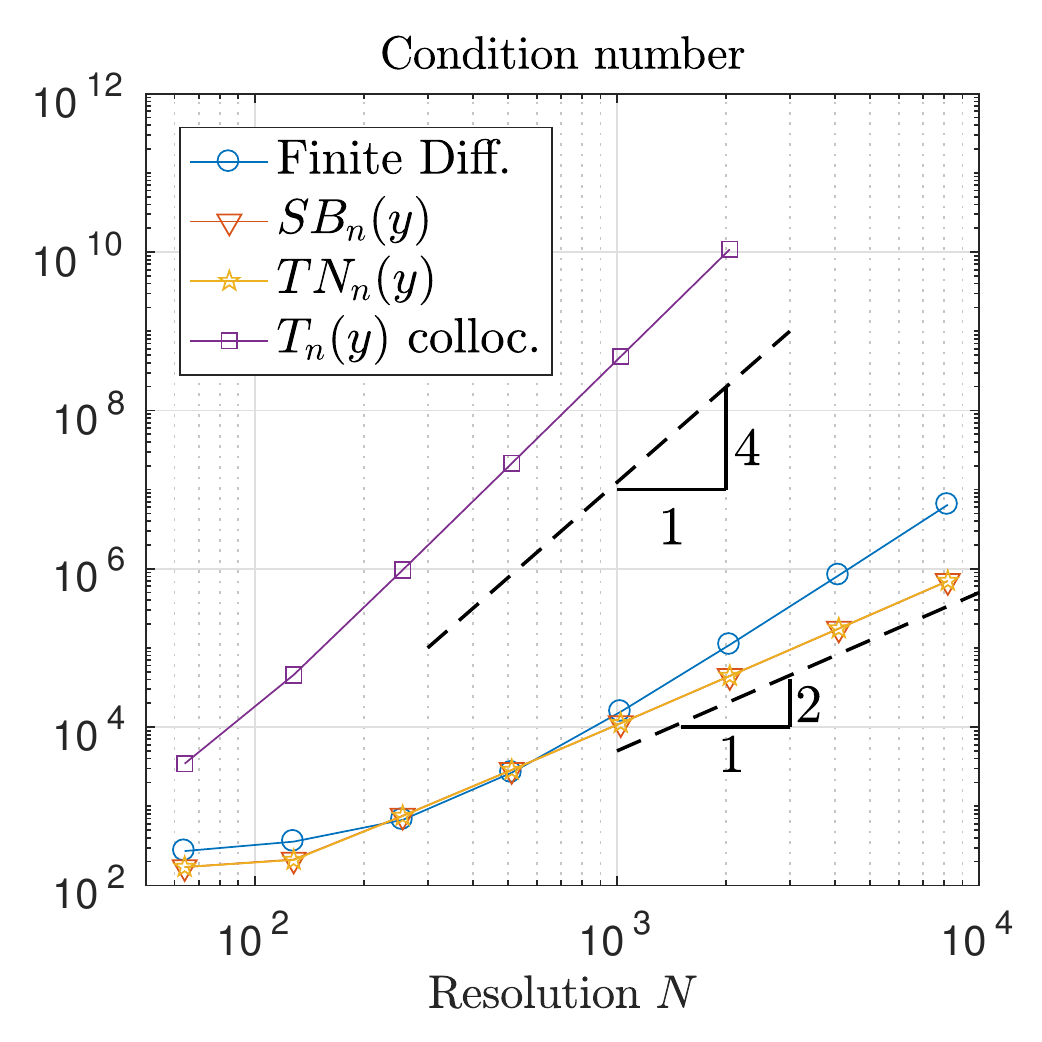}
		\caption{\label{QHO_basic}Comparison of the different methods used to solved the QHO eigenproblem with $N=256$ basis functions or grid points: spectral methods on the infinite line using Chebyshev functions $TB_n(y)$ (pentagrams) and $SB_n(y)$ (triangles) with the mapping parameter $L$; or methods on a finite segment $[-L,L]$: Finite differences (circles) or Chebyshev polynomial $T_n(y)$ collocation (squares). Left panel: number $N_{<0.01}$ of  eigenvalues correct within a relative error of $0.01$, as a function of $L$. In the remaining two panels, we set independently for each method $L=L_{\mathrm{opt}}$, the optimal parameter that maximises $N_{<0.01}$ in the left panel. Center panel: relative error $e(n)=\frac{E_n-(2n+1)}{2n+1}$ of the numerical eigenvalues $E_n$ as a function of $n$. Right panel: Condition number of matrices $\mathbfsf{L}$.}
	\end{center}
\end{figure}
We illustrate the mechanics of the Chebyshev decomposition on the classic example of the Quantum Harmonic Oscillator (QHO). The QHO is 
a fundamental example of a particle propagating in a potential $V(y)$. The wave function $\psi$ and the eigenenergies $E$ of the particle are the solutions to the Schr\"odinger equation:
\begin{equation}
E \psi(y) = \left(V(y) -\partial_{yy}\right) \psi(y)\, .
\end{equation}
 The potential $V(y)=y^2$ is  well-known to possess a point spectrum of exact analytic solutions, indexed by the integer $n \ge 0$:
\begin{equation}
E_n = 2n +1,\quad \psi^{(n)}(y) = \mathrm{H}_n(y)\exp \left(-y^2 /2\right),\, \end{equation}
where $\mathrm{H}_n$ is the Hermite polynomial of degree $n$. As described above, we first solve this equation by mapping it to the segment $[0,\pi]$ using ~(\ref{def_mapping}). Doing so yields the corresponding eigenproblem for $f(\theta) = \psi(y)$:
\begin{equation}
E f(\theta) = -\frac{1}{L^2}\left(\sin^4 \theta \partial_{\theta \theta} + 2\sin^3 \theta \cos \theta \partial_\theta \right) f(\theta) + \frac{L^2 \cos^2 \theta}{\sin^2\theta} f(\theta)\, .
\end{equation}
In this form, the last term with a $\sin^2\theta$ denominator leads to a dense discretization regardless of which expansion is adopted for $f(\theta)$. Therefore, we regularize the equation by the multiplication $\sin^2\theta$ throughout. Doing so readily yields the generalized eigenproblem of the form 
\begin{subequations}
\begin{gather}E\mathscr{M}f(\theta) = \mathscr{L}f(\theta)\,, \intertext{where}
\mathscr{M} = \sin^2 \theta \,, \\
\mathscr{L} = -\frac{1}{L^2}\left(\sin^6 \theta \partial_{\theta \theta} + 2\sin^5 \theta \cos \theta \partial_\theta \right) + L^2 \cos^2 \theta \, .
\end{gather}
\end{subequations}
In this simple example, the discretization is a straightforward step: anticipating no particular difficulty, we adopt as a first guess a pure cosine expansion of $N$ terms for $f(\theta)$. By projecting  the equations onto the $N$ basis functions $\left\{ \cos(m\theta) \right\}_{0\le m \le (N-1)}$, we obtain a $N\times N$ system of the form:
\begin{equation}
E\,\mathbfsf{M} \boldsymbol{\widetilde{f}} = \mathbfsf{L} \boldsymbol{\widetilde{f}}\quad \mathrm{with}\quad \boldsymbol{\widetilde{f}} = \left(\widetilde{f}_0,\,\widetilde{f}_1,\, \cdots,\, \widetilde{f}_{N-1}\right)^{\mathrm{T}} \,, 
\end{equation} 
where the matrix coefficients $\mathsf{M}_{mn}$ and $\mathsf{L}_{mn}$ are given by (for $1\le m,n \le N$):
\begin{subequations}
\begin{gather}
\mathsf{M}_{mn} = \frac{\inner{\cos([m-1]\theta)}{\mathscr{M}\cos([n-1]\theta)}}{\norm{\cos([m-1]\,\theta)}^2}\, , \\ 
\mathsf{L}_{mn} = \frac{\inner{\cos([m-1]\theta)}{\mathscr{L}\cos([n-1]\theta)}}{\norm{\cos([m-1]\,\theta)}^2}\, .
\end{gather}
\end{subequations}
The exact expression of matrices $\mathbfsf{M}$ and $\mathbfsf{L}$ is omitted, but was obtained by using the procedure given in appendix~\ref{app:chebfunc}. Their banded form can be observed on figure~\ref{QHO_eigenmodes}. This sparse structure could have been easily anticipated, as the discretized operators $\mathbfsf{M}$ and $\mathbfsf{L}$ are composed exclusively by sparse representations on the cosines basis: i.e., multiplication by $\cos \theta$, $\cos^2(\theta)$, $\sin^2(\theta)$, second order derivation $\partial_{\theta\theta}$, and $\sin\theta \partial_\theta$. Note that an alternative choice consisting in expressing $f(\theta)$ as a sine expansion and projecting on the $N$ functions $\left\{\sin(m\theta)\right\}_{1\le m \le N}$ would yield a similarly sparse linear system $E\, \mathbfsf{M}\boldsymbol{\widehat f} =  \mathbfsf{L}\boldsymbol{\widehat f}$ where:
\begin{subequations}
\begin{gather}
\mathsf{M}_{mn} = \frac{\inner{\sin(m\theta)}{\mathscr{M}\sin(n\theta)}}{\norm{\sin(m\,\theta)}^2}\, , \\ 
\mathsf{L}_{mn} = \frac{\inner{\sin(m\theta)}{\mathscr{L}\sin(n\theta)}}{\norm{\sin(m\,\theta)}^2}\, .\end{gather}
\end{subequations}In figure~\ref{QHO_basic}, we compare different methods used to solved the QHO equations, all of which contain a free length parameter $L$: the problem is solved on the infinite line using either a $TB_n(y)$ expansion (cosines), or a $SB_n(y)$ expansion. For comparison, we also solve our problem on a large segment $y\in[-L,L]$ by means of either a classic Chebyshev collocation method or finite differences. A Dirichlet boundary condition $\psi(L)= \psi(-L)=0$ is imposed for the latter methods. In all cases, the number of grid points (finite differences and collocation method) or modes (spectral methods) is set to $N=256$. The parameter $L$ is increased between $0.25$ and $20$ by increments of $0.25$. We display as a function of $L$ the number $N_{<0.01}$ of eigenvalues correct within a relative error of $0.01$ obtained by each method. For each method, an optimal value $L_{opt}$ is found which corresponds to a optimal ratio $Q_{opt}=N_{<0.01}/256$ of  accurate eigenvalues. The results are summarized in Table~\ref{accuracy_table}. From a sheer accuracy criteria, the $SB_n$ and $TB_n$ functions on the infinite line are sensibly preferable to Chebyshev polynomials. All three spectral methods are substantially more accurate than the method of finite differences. Finally, note on figure~\ref{QHO_basic} that the poor conditioning of differentiation matrices on the Gauss-Lobatto grid (the collocation grid for $T_n$ polynomials) is visible, as the condition numbers increases as $N^4$. Discretizations based on rational Chebyshev functions do not suffer this phenomenon and their condition number scales quadratically with $N$.

We emphasize that in the present example as well as in the following examples, the computation of the full spectrum using a standard QR/QZ algorithm and its comparison to the analytic spectrum had been intended to provide a purely pedagogical illustration of the accuracy of the method. For real life applications, the sparsity of discretizations using Chebyshev functions should be leveraged through the use of iterative eigensolvers or sparse solvers.
\begin{table}[h]
\begin{tabular}{c || c |c |c}
	\hline
	\hline
Method & $L_{opt}$ & $N_{<0.01}$ & $Q_\mathrm{opt}$ (\%)  \\
\hline
$SB_n(y)$ & $11$  &  $132$  & $51.6\%$ \\
$TB_n(y)$ & $11$  &  $132$  & $51.6\%$ \\
	\emph{Collocation} $T_n(y)$ & $11.75$  &  $75$  & $29.3\%$ \\
Finite Differences & $14$  &  $7$  & $2.7\%$ \\
		\hline
\hline	
\end{tabular} 
\caption{\label{accuracy_table} Comparison of accuracy for the eigenmodes obtained by a spectral method on the infinite line ($SB_n$ or $TB_n$), a Chebyshev collocation method, or first order finite differences with 256 modes or grid points. $L_{opt}$ is the optimal mapping parameters for methods on the infinite line or the optimal size of the domain for the two other methods. $N_{<0.01}$ denotes the number of eigenvalues accurate within a relative error of $10^{-2}$; $Q_\mathrm{opt}= N_{<0.01}/256$ is the corresponding proportion of accurate eigenvalues.}
\end{table}
\begin{figure}
	\begin{center}
		\includegraphics[height = 0.32\textwidth]{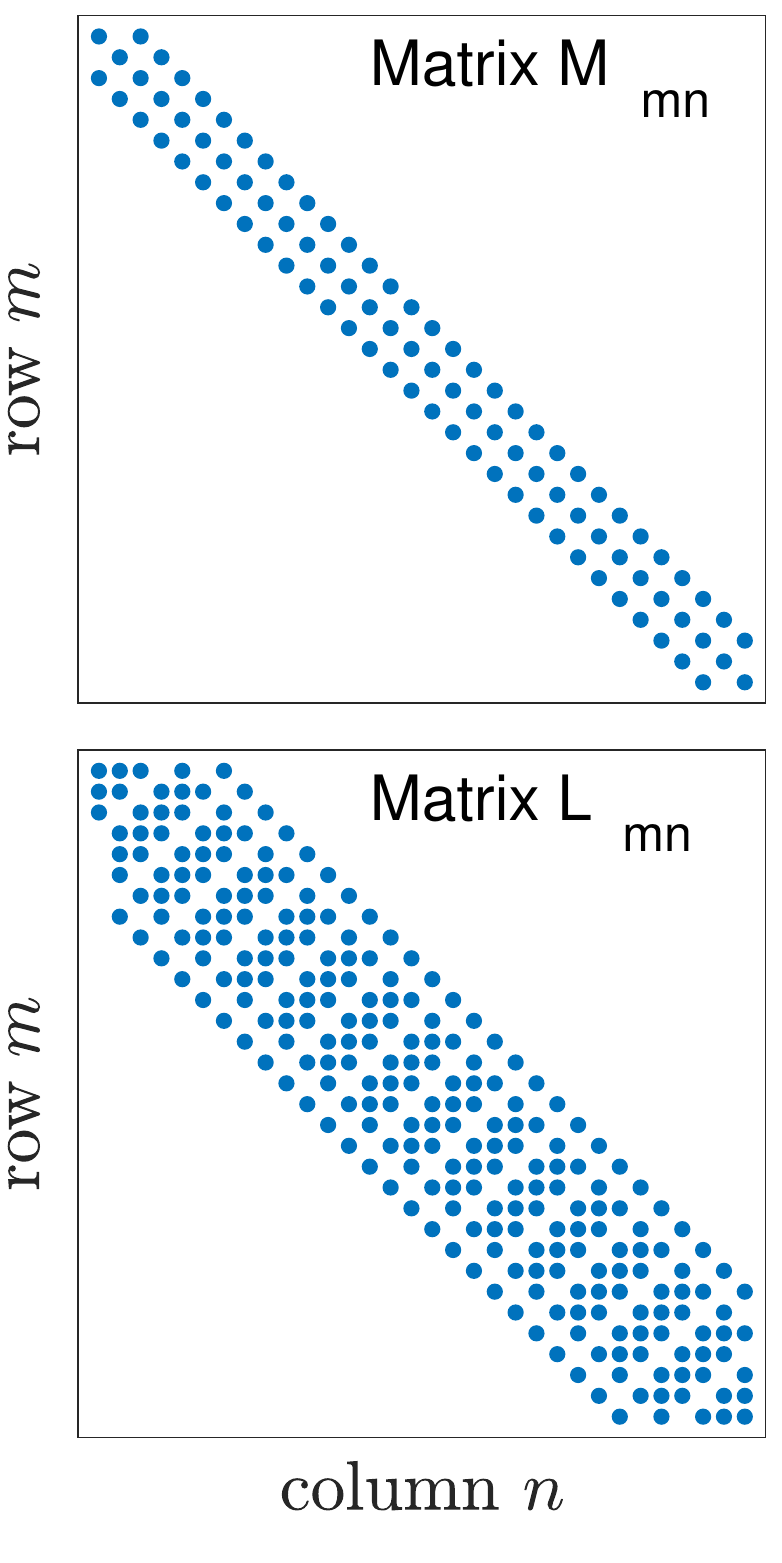}
		\includegraphics[height = 0.32\textwidth]{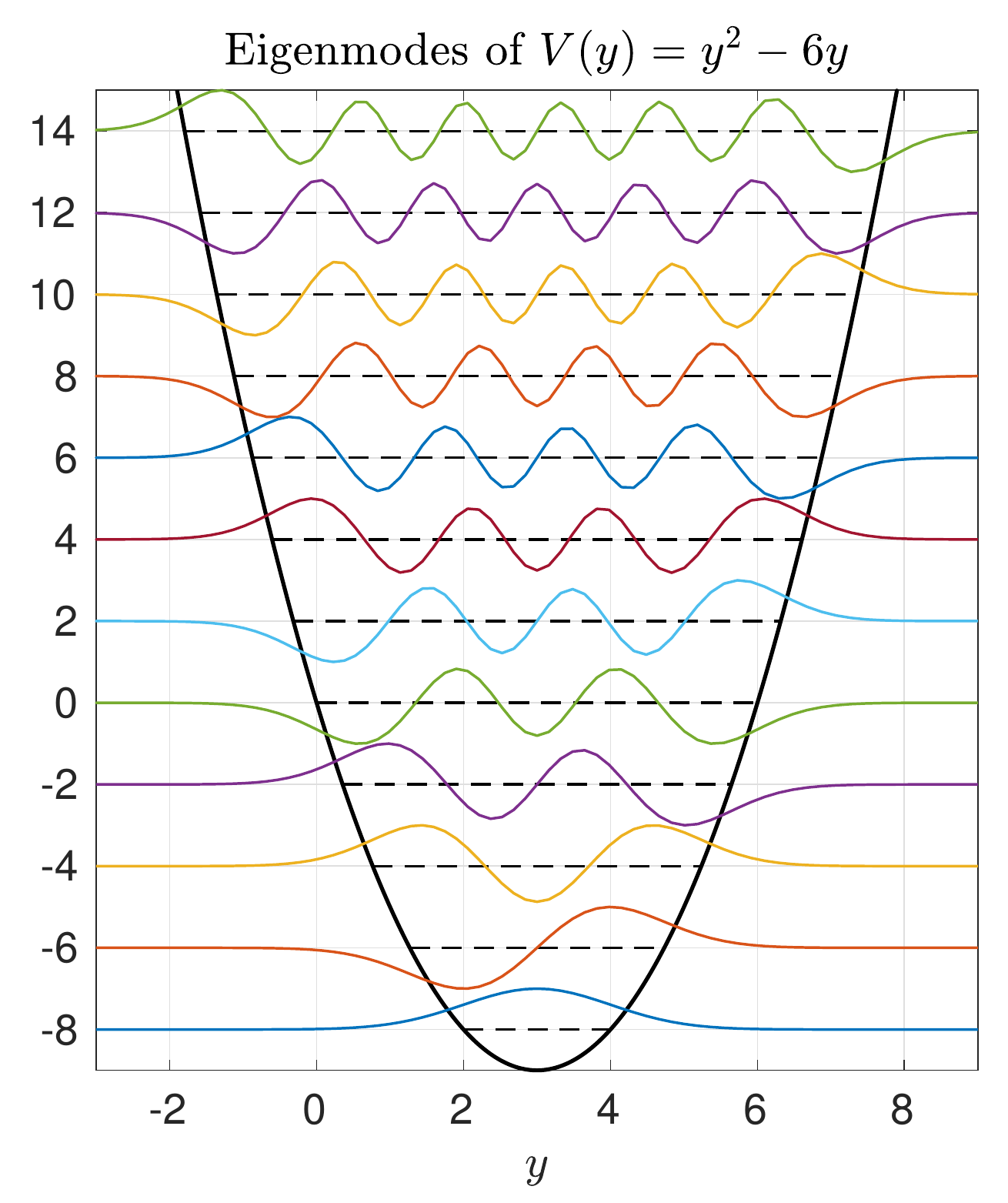}
		\includegraphics[height = 0.32\textwidth]{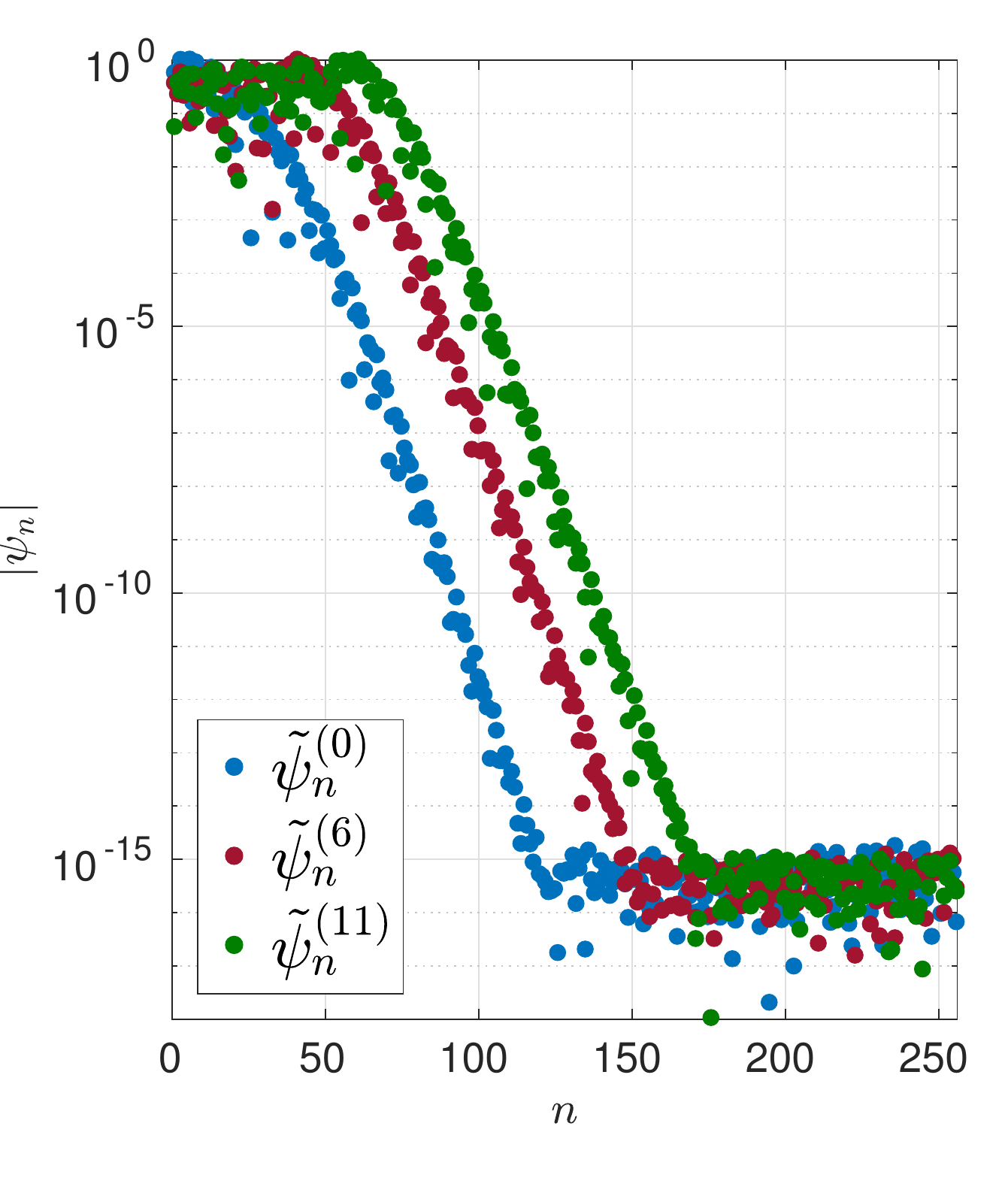}
		\includegraphics[height = 0.32\textwidth]{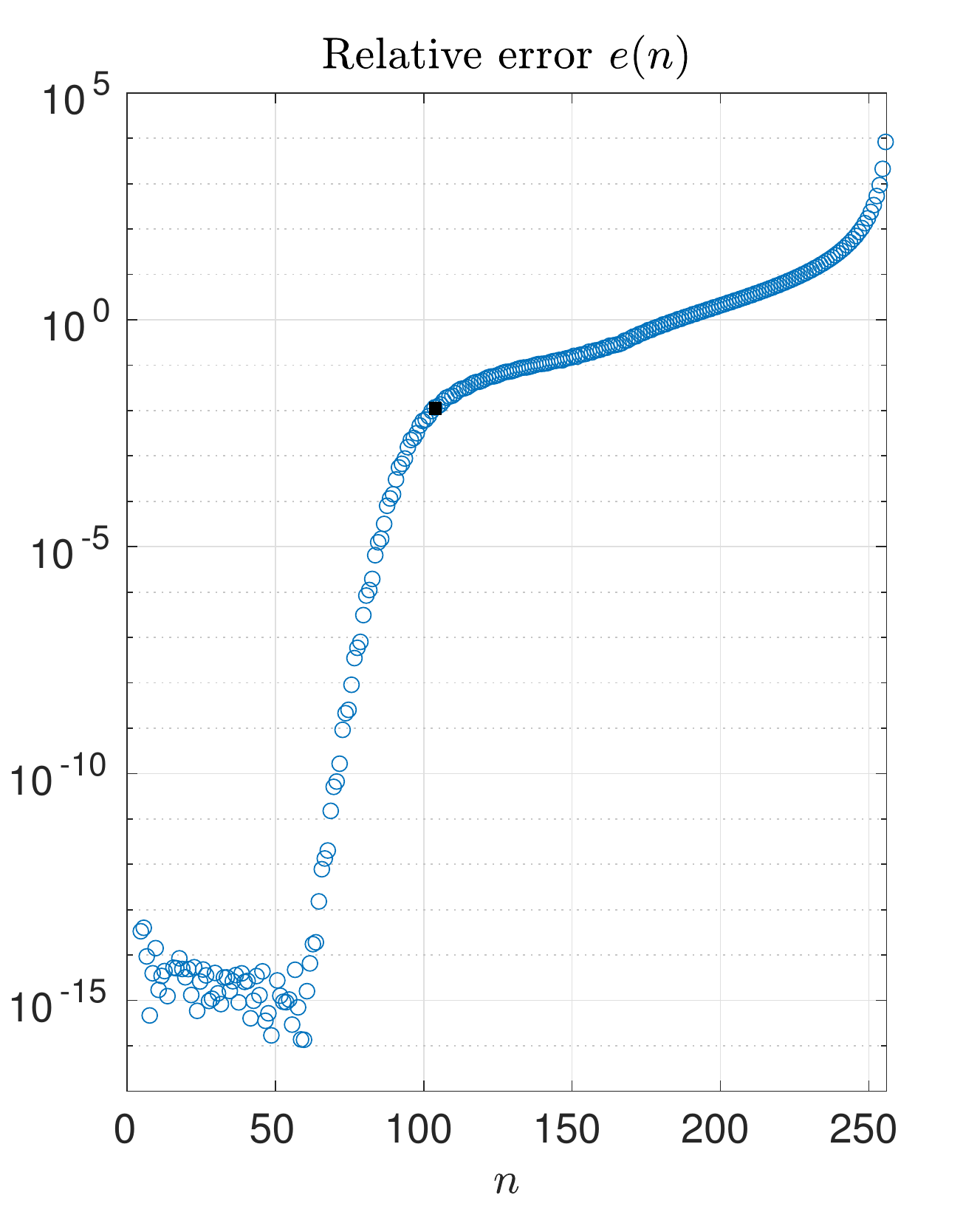}
		\caption{
			\label{sparsity_QHO_shifted} From left to right: Sparse matrices $\mathbfsf{M}$ and $\mathbfsf{L}$ for the translated harmonic oscillator (equation \ref{tho_y}) for $N=32$; Wave functions for $N=256$, translated vertically by their energy (dashed horizontal lines); Spectra of modes $0$, $6$ and $11$ illustrating exponential convergence; Relative error of the eigenenergies.
		}
	\end{center}
\end{figure}

\section{Parity splitting operators in a scalar equation}\label{sec_QAHO}
\subsection{Particle in a translated harmonic well: Chebyshev rational functions}\label{sec_THO}
We have illustrated how approximations using Chebyshev basis functions on the infinite line yields an efficient and accurate resolution of the QHO equation. However, the operators $y^2$ and $\partial_{yy}$ are both even and hence parity-preserving. In this section, we first illustrate how the sparsity of the approximation method is ruined in presence of both even and odd parity operators.  A resolution is then provided to restore sparsity for a large class of polynomials and rational differential operators. 

Our introductory example is the case of a particle in a asymmetric parabolic potential $V_a(y) = y^2 - 6y$. The potential $V_a$ is a mere translation of the harmonic potential accompanied by a renormalization of the ground state energy: applying the transformation $y= y'+3$ and $E= E'-9$, this problem can be mapped back to the QHO treated in the previous section. Hence we refer to this case as the ``translated quantum harmonic oscillator'' (TQHO) and we confront our approach against the known analytical solutions of this equation. 
Following a similar approach as above, we rewrite the Schr\"odinger equation for $\psi(y)$
	\begin{equation}
	E\psi(y) = \left(y^2 - 6y  -\partial_{yy} \right) \psi(y)\, , \label{tho_y}
	\end{equation}
	in  $\theta$-space, and we regularize by multiplication by $\sin^2 \theta$, yielding:
	\begin{equation}
	E\sin^2 \theta f(\theta) = -\frac{1}{L^2}\left(\sin^6 \theta \partial_{\theta \theta} + 2\sin^5 \theta \cos \theta \partial_\theta \right) f(\theta) + \left(L^2 \cos^2 \theta- 6L\sin\theta \cos \theta  \right) f(\theta) \label{tho_theta} \,.
	\end{equation}
	We identify on this simple example the trouble that accompanies parity flipping operators. Equation~(\ref{tho_y}) contains two classes of operators: the identity operator on the left-hand-side and the operator $\left( y^2 - \partial_{yy}\right)$ preserve the parity of the function they act on. By contrast, the operator $-6y$ reverses the parity. As a consequence, were $\psi$ to be written as a ``pure'' $TB$ or $SB$ expansion, at least one term will be band-unlimited and the coupling matrix would be dense.  	
Suppose we assume a pure cosine or sine expansion (\ref{purecos_series},\ref{puresine_series}) for $f$, as we did for the QHO. A discrete set of equations of the form $E\, \mathbfsf{M}\boldsymbol{\widetilde{f}} =\mathbfsf{L}\boldsymbol{\widetilde{f}}$ could be obtained by projection onto the cosine or sine basis. However, by contrast with the QHO and regardless of which one of the four options we elect, this system would never yield a sparse representation. 
Indeed, the two possible representations of $\mathscr{L}$:
\begin{equation}
\frac{\inner{ \cos(m\theta)}{\mathscr{L} \cos(n\theta)}}{\norm{\cos(m\theta)}^2}\,,
\frac{\inner{ \sin(m\theta)}{\mathscr{L} \sin(n\theta)}}{\norm{\sin(m\theta)}^2}\,, 
\end{equation} are all dense matrices due the existence the parity-break term that results in a band-unlimited representation.

We note, however, that the differential operators present in equation~\ref{tho_theta} obey the requirement of our Proposition 1. Therefore, the sparsity can be restored if the odd and even components of $\psi$ are expanded by a different basis.
A solution consists in assuming an hybrid expansion in sines and cosines of the form (\ref{mixed_trig_series}) for the symmetric and antisymmetric components $f\sym$ and $f\anti$. For instance, if we choose the basis $\{\breve\beta_n\}$ (i.e. a cosine and sine expansion (see equation~\ref{cossin_series}) for $f\sym$ and $f\anti$, respectively), then we can compute the matrices representing the action of the operators that compose $\mathscr{L}$ on this basis:
\begin{subequations}
\begin{gather}
 \mathbfsf{A} =\frac{\inner{ \breve{\beta}_m}{\sin^6\theta\,\partial_{\theta\theta} \breve{\beta}_n}}{\norm{\breve{\beta}_m}^2},\qquad 
  \mathbfsf{B} =\frac{\inner{ \breve{\beta}_m}{\cos\theta\sin^5\theta\,\partial_{\theta} \breve{\beta}_n}}{\norm{\breve{\beta}_m}^2}\,, \tag{43a,b} \label{AB} \\ 
   \mathbfsf{C} =\frac{\inner{ \breve{\beta}_m}{\cos^2\theta\breve{\beta}_n}}{\norm{\breve{\beta}_m}^2},\qquad  
   \mathbfsf{D} =\frac{\inner{ \breve{\beta}_m}{\sin^2\theta\breve{\beta}_n}}{\norm{\breve{\beta}_m}^2},\qquad  
    \mathbfsf{E} =\frac{\inner{ \breve{\beta}_m}{\cos\theta\sin\theta\breve{\beta}_n}}{\norm{\breve{\beta}_m}^2}\,, \tag{43c,d,e} \label{CDE}
\end{gather}
\end{subequations}
all of which are sparse. With these building blocks in hand, the expression of which is given in appendix~\ref{app:chebfunc}, the discretized system:
\begin{equation}
E\, \mathbfsf{M}\,\boldsymbol{\breve{f}} = \mathbfsf{L}\boldsymbol{\breve{f}}
\end{equation} is readily obtained and the sparse structure of matrices $\mathbfsf{L}$ and $\mathbfsf{M}$ is illustrated in figure~\ref{sparsity_QHO_shifted}:
\begin{subequations}
\begin{gather}
\mathbfsf{M} = \mathbfsf{D}\,, \\
\mathbfsf{L} = -\frac{1}{L^2} \left( \mathbfsf{A} + 2\mathbfsf{B}\right) + L^2 \mathbfsf{C} - 6L \mathbfsf{E}   \, .
\end{gather}
\end{subequations}
As can also be observed in figure~\ref{sparsity_QHO_shifted} when compared to figure~\ref{QHO_eigenmodes}, neither the exponential convergence of the series nor the excellent accuracy of the method are adversely affected by our using of a mixed basis. This illustrates the discussion of convergence given in section~\ref{sec:choosing}. For exponentially decaying solutions, expansions based upon the $\Big\{\mathring{\beta}_n\Big\}$ basis or the $\Big\{\breve{\beta}_n\Big\}$ basis yield a similarly sparse representations and converge equally rapidly. 

 \begin{figure}
 	\begin{center}
 	\includegraphics[height = 0.32\textwidth]{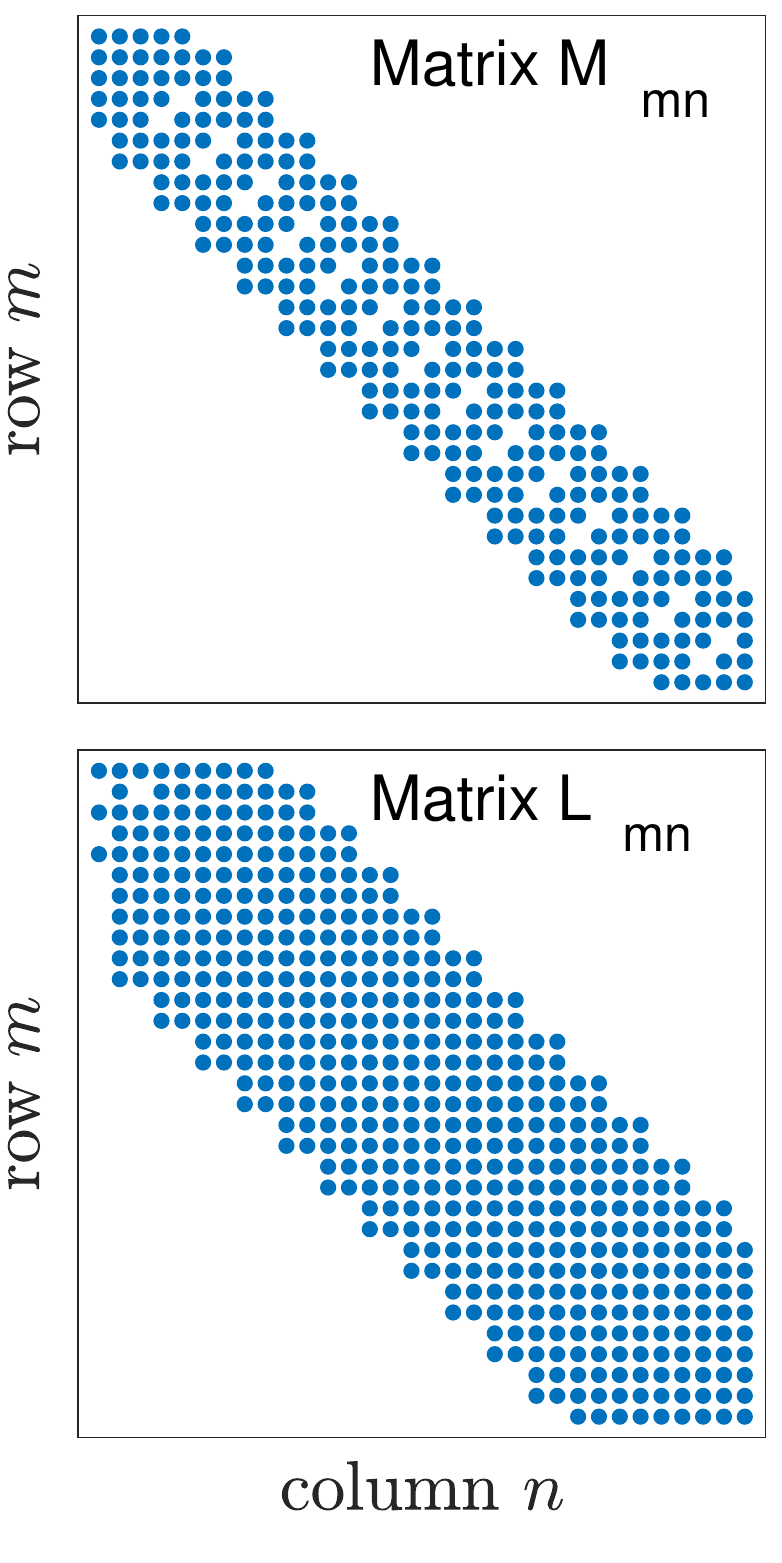}
 	\includegraphics[height = 0.35\textwidth]{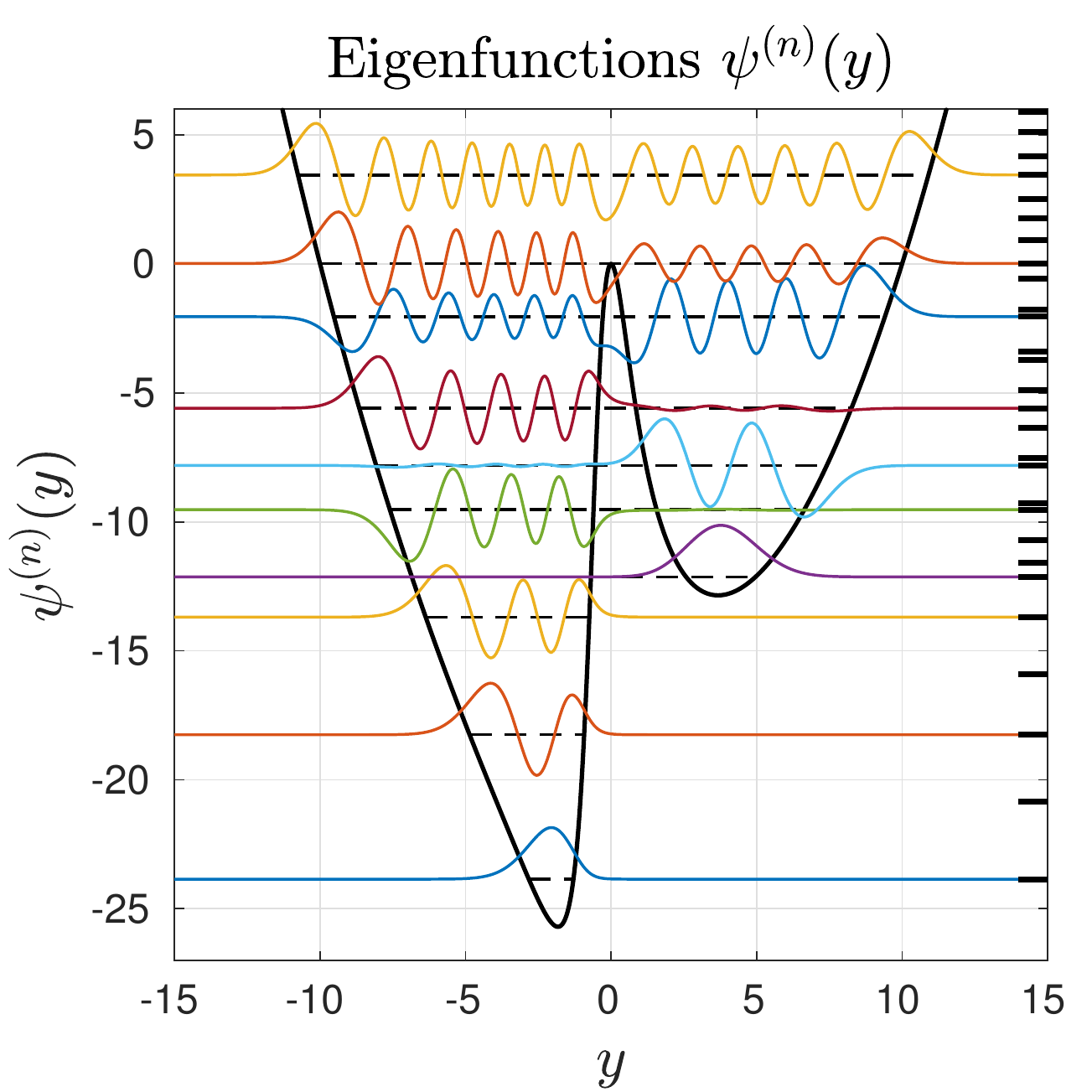}
 	\includegraphics[height = 0.35\textwidth]{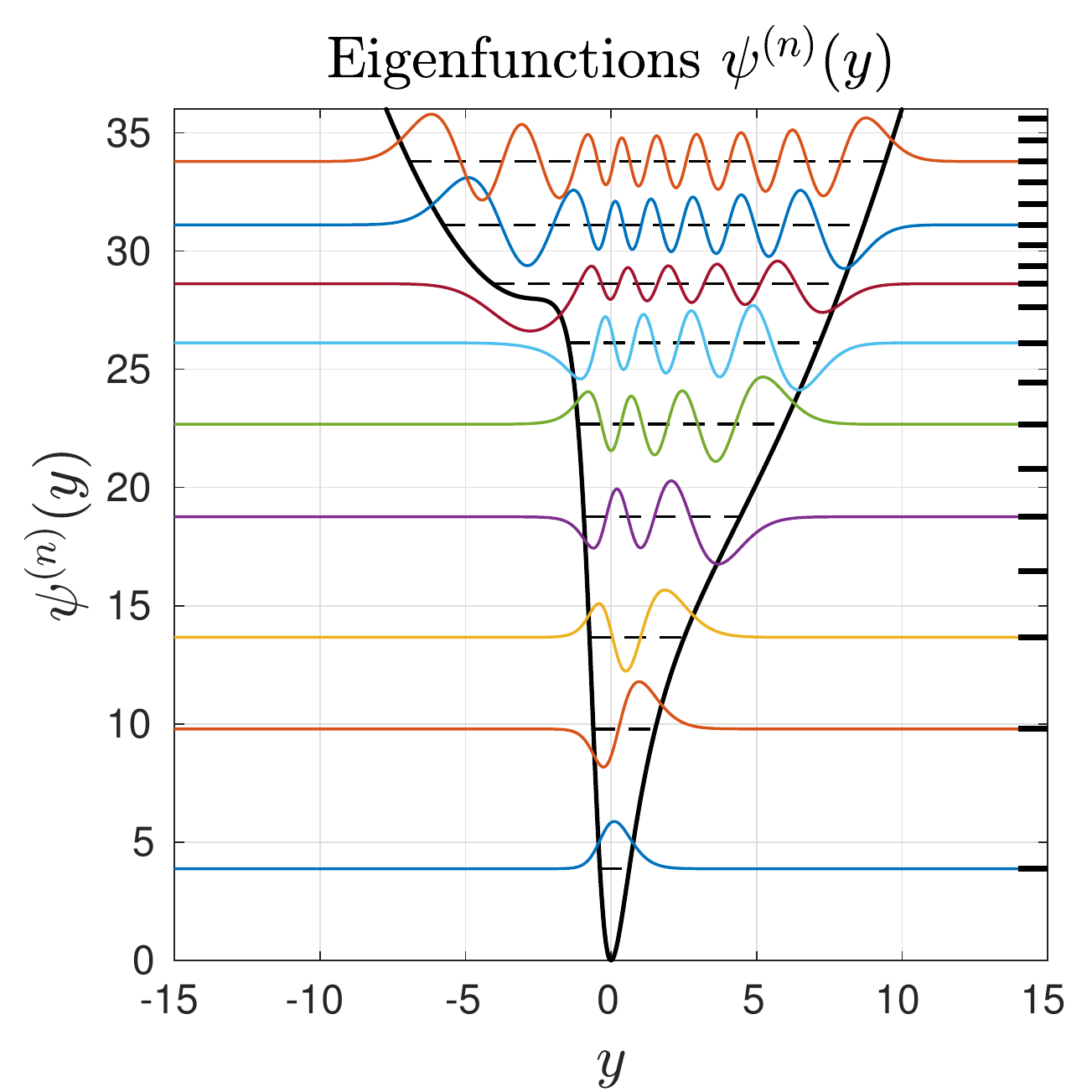}
 	\caption{\label{sparsity_QAHO_shifted} A quantum particle in an asymmetric potential well (\ref{arbitrary_potential}). Left: Discretization of the Schr\"odinger equation. Eigenfunctions of potential (\ref{arbitrary_potential}) with $V_0=20$, $a_4=0.01$ and $a_2=1$ (center) or $a_2=-1$ (right). Only a subset of the eigenfunctions (shifted vertically by the corresponding eigenenergy) are shown for legibility. All the eigenenergies are marked by thick ticks on the right of each figure.}
 	\end{center}
 \end{figure}

\subsection{The generalized case} \label{general_case}Having motivated, illustrated and benchmarked our method on a simple example, we now summarize and 
formalize the procedure to tackle a broader class of eigenproblems:
\begin{equation}
E \psi(y) = \mathscr{H}\psi(y)\,,\label{EP}\end{equation}
where the operator $\mathscr{H}$ is a sum of rational fractions and derivatives of maximal order $N_\mathscr{H}$:
\begin{equation}
\mathscr{H} = \sum_{i=0}^{N_{\mathscr{H}}} \frac{P_i(y)}{Q_i(y)}\left(\frac{\mathrm{d}}{\mathrm{d}y}\right)^{\! i}\,, \label{hamiltonian}
\end{equation}
and $P_i$ and $Q_i$ are polynomials. Note that even more general equations such as $E \mathscr{H}_1\psi = \mathscr{H}_2\psi$ (with $\mathscr{H}_1$ and $\mathscr{H}_2$ of the same form as given above in~\ref{hamiltonian}) can be considered. However, for the sake of brevity and clarity, we present the method in the case of equation (\ref{EP}) only. 

The key idea is that eigenproblems of this form will necessarily fall back into the domain of validity of Proposition 1 after remapping~(\ref{def_mapping}). In contrast, we point out the following limitation to our technique: transcendental operators will not lend themselves to a sparse discretization. The method to obtain sparse representations for rational differential operators proceeds as follows.

As a first step, multiply the equation by $C(y)$, the least common multiple polynomial of the $Q_i(y)$ polynomials:
\begin{equation}
E\, C(y) \psi = \sum_{i=0}^{N_{\mathscr{H}}} P'_i(y) \left(\frac{\mathrm{d}}{\mathrm{d}y}\right)^{\! i}\,,  
\end{equation}
where the polynomials $P'_i(y)=C(y)P_i(y)/Q_i(y)$. As a second step, map the equation to  $\theta$-space, using the mapping (\ref{def_mapping}) and the chain rule (\ref{chain_rule}). Determine the highest power $h_p$ of $\sin\theta$ present as a denominator: multiply the equation by $(\sin\theta)^{h_p}$ to cancel out all denominators, yielding to an equation of the form:
\begin{equation}
E \mathscr{M} f(\theta) = \mathscr{L}f(\theta)\, . \label{general_form}
\end{equation}
Finally, obtain a sparse discretization of this system using a mixed expansion (\ref{mixed_trig_series}).

\subsection{Particle in an arbitrary asymmetric well}
\label{sec_arbitrary_well}
The general method described above is quickly illustrated on an example of greater complexity than the translated parabolic well considered in section~\ref{sec_THO}. We consider a potential of the form:
\begin{equation}
V(y) = V_0 \frac{a_4 y^4 +a_2 y^2 }{y^2 + y + 1}\, , \label{arbitrary_potential}
\end{equation}which corresponds to the Hamiltonian:
\begin{equation}
\mathscr{H} = -\partial_{yy} + V_0\frac{a_4 y^4 +a_2 y^2 }{y^2 + y + 1}\,,
\end{equation}
of the form given in equation (\ref{hamiltonian}). We first eliminate the rational fractions:
\begin{equation}
E \left(y^2 + y + 1 \right) \psi(y) = - \left(y^2 + y + 1 \right) \partial_{yy}\psi(y) +V_0\left( a_4 y^4 +a_2 y^2 \right) \psi(y)\,. \end{equation}
Mapping this equation to $\theta$-space and eliminating the $\sin\theta$ denominators by multiplying by $\sin^4\theta$ gives, in this case: 
\begin{multline}E\sin^2\theta  \left(L^2\cos^2 \theta + L\sin\theta\cos\theta + \sin^2\theta \right)f(\theta) =
V_0\left( a_4 L^4\cos^4\theta +a_2 L^2\cos^2\theta\sin^2\theta \right) f(\theta) \\  -\sin^4\theta\left(L^2\cos^2 \theta + L\sin\theta\cos\theta + \sin^2\theta \right)
 \left(\frac{\sin^2 \theta}{L^2} \partial_{\theta\theta}  + \frac{2\cos\theta\sin\theta}{L^2}\partial_\theta\right) f(\theta)\, . \label{eq_theta_asym_well} \end{multline} 
This equation readily yields a sparse discretization when a mixed expansion is used for $f$. The sparse matrices involved are displayed in figure~\ref{sparsity_QAHO_shifted}. The eigenfunctions of two different wells are computed to illustrate the variety of potentials amenable with our method.
%
%
%
%
%

\section{Coupled systems: the equatorial $\beta-$plane}\label{sec_eqwaves}
\subsection{The equatorial $\beta-$plane}
We now motivate and illustrate the application of the method for coupled systems of equations, namely, the linearized equatorial $\beta-$plane model \cite{gillBOOK} relevant for tropical atmospheric or oceanic dynamics. On the surface of a sphere covered with a shallow layer of fluid, the azimuthal velocity $u$,  meridional velocity $v$ and height of the fluid layer  $h$ obey the shallow water governing equations around the equator:
\begin{subequations}
\begin{gather}
\partial_t u = -a u + yv - \partial_x h\, , \\
\partial_t v = -yu -av -\partial_y h \, ,\\
\partial_t h = -\partial_x u - \partial_y v\, ,
\end{gather}
\end{subequations}
where $x$ is the azimuthal coordinate, $y$ is the meridional coordinate, and $a$ is the Rayleigh friction coefficient. 

As with the harmonic oscillator, we first present the analytical results against which the numerical method will be validated. In absence of friction, i.e., $a=0$, the solutions are waves and these equations can be solved analytically by considering normal modes $\left( U\left(x,y\right),V\left(x,y\right),H\left(x,y\right)\right)  = \left(u(y) ,v(y) ,h(y) \right)\, \exp \left( \mri\left[ kx-\omega t\right]\right)$. The variables $u(y)$ and $h(y)$ can both be expressed as functions of $v(y)$:
\begin{subequations}
\begin{align}
\left(\omega^2 - k^2\right) h &= \mri \left(  ky  -\omega \partial_y \right)v\,,\\
\left(\omega^2 - k^2\right) u &= \mri \left( \omega y - k\partial_y \right)v\, .
\end{align} \label{uhvsv}
\end{subequations}
The variable $v$ itself is determined as the solution of a Weber equation, similarly to the quantum harmonic oscillator:
\begin{equation}
\left( y^2 - \partial_{yy}\right)v = \Omega\, v\, \quad \mathrm{where}\quad \Omega = \left(\omega^2 -k^2 -\frac{k}{\omega}\right)\, . \label{OMEGA}
\end{equation}
As above, the solutions are parabolic cylinder functions and $\Omega$ is quantized:  \begin{subequations}
\begin{gather}
\Omega_n = 2n+1\,,\quad n\in \mathbb{N}\,, \label{rossby_freq}\\
v_n(y) = v_0\, H_n(y) \mathrm{e}^{-y^2/2}\,. \label{equato_v}
\end{gather} \label{generalsol}
\end{subequations}
Each value $\Omega_n$ corresponds to three different modes of frequency  $\omega_n^{(1)}$, $\omega_n^{(2)}$, and $\omega_n^{(3)}$ found as the roots of  equation~(\ref{OMEGA}). The wave functions of each mode $u^{(1)}_n(y)$, $h^{(1)}_n(y)$, $u^{(2)}_n(y)$, $h^{(2)}_n(y)$, $u^{(3)}_n(y)$, and $h^{(3)}_n(y)$, are obtained by using formulas~(\ref{uhvsv}). For $n\ne0,-1$ these modes represent  (westward and eastward propagating) inertio-gravity waves and a (westward propagating) Rossby wave. Two exceptions appear to this picture: for $n=0$, one of the three roots, $\omega = -k$, must be discarded. Indeed, this root corresponds to an exponentially growing solution $v(y)  = \exp \left( y^2 /2\right)$. The remaining root represents the mixed Rossby-gravity (Yanai) wave.  Further, the Kelvin waves are not captured by expression~(\ref{generalsol}). Sometimes dubbed the ``$n = -1$ mode'', these waves correspond to $v$ identically zero, yielding:
\begin{subequations}
\begin{gather}
\label{kelvin_freq}
\omega_{-1} = k \,, \\v_{-1}(y)= 0\, , \\
u_{-1}(y) = h_{-1}(h) \propto \exp\left(-y^2/2\right)\,. 
\end{gather}
\end{subequations}
\begin{figure}
\begin{center}
\includegraphics[height= 0.30 \textwidth]{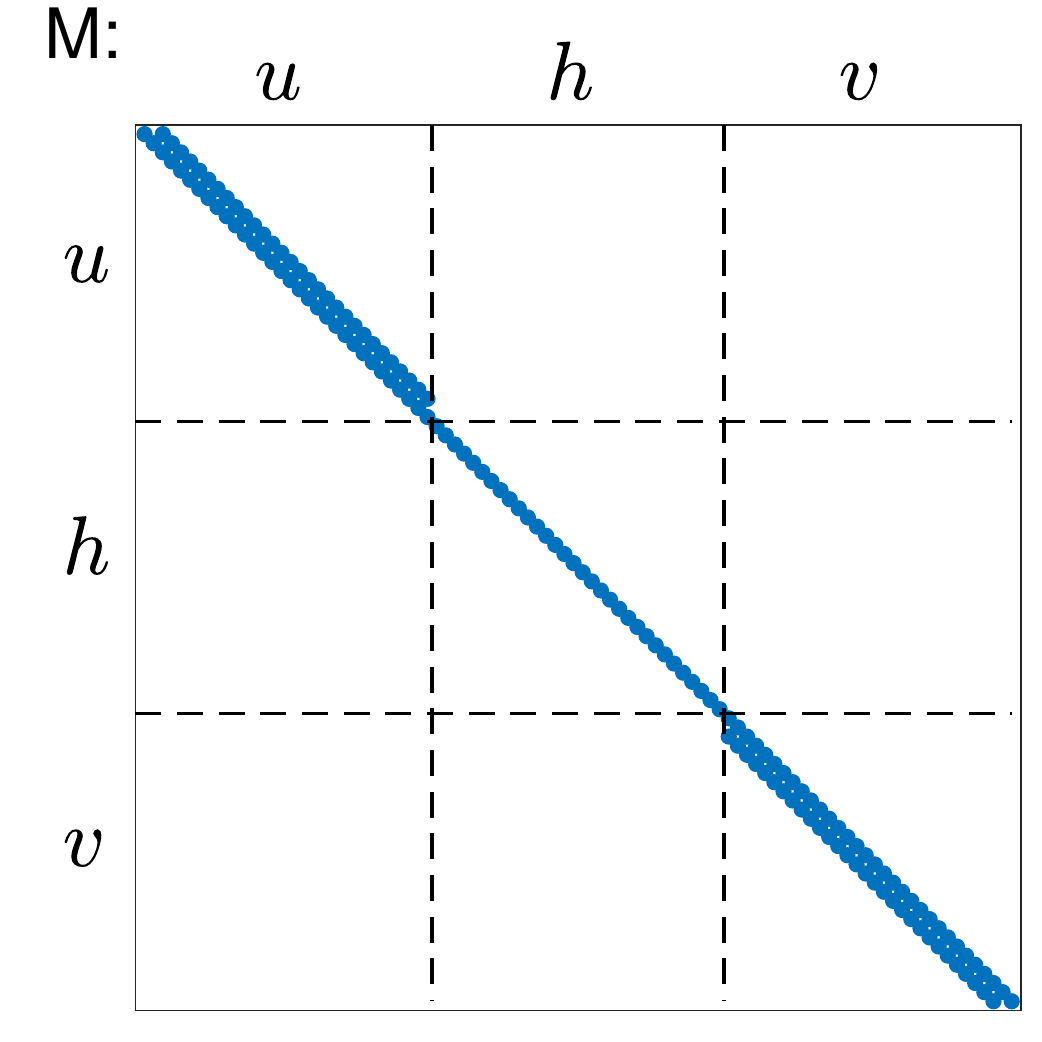}
\includegraphics[height= 0.30 \textwidth]{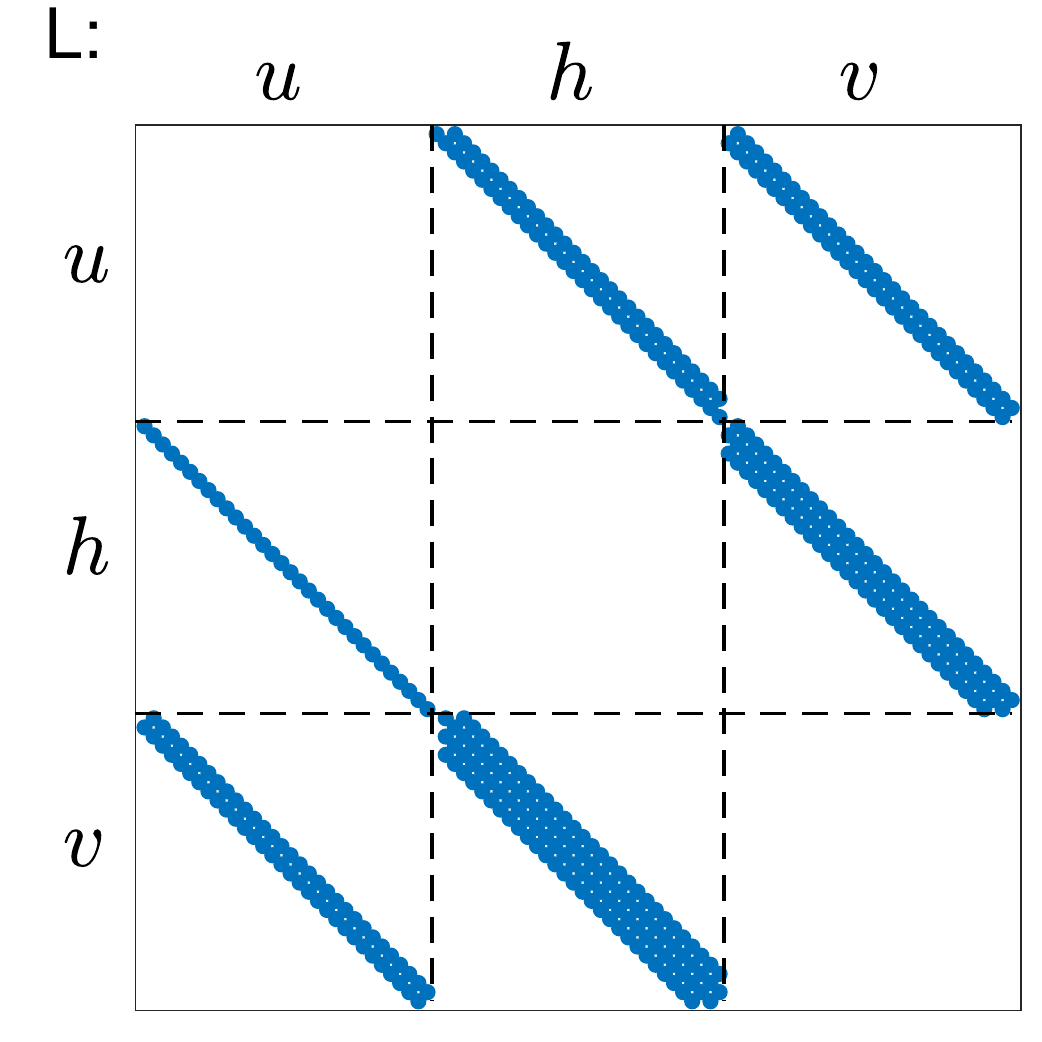}
\caption{ \label{equato_spy} Discretization of system of equations~(\ref{equato_theta}) using expansions~(\ref{equato_dagger}) and ignoring friction, i.e. $a=0$.}
\end{center}
\end{figure}
\begin{figure}[h!]
\begin{center}
\includegraphics[height=0.45\textwidth]{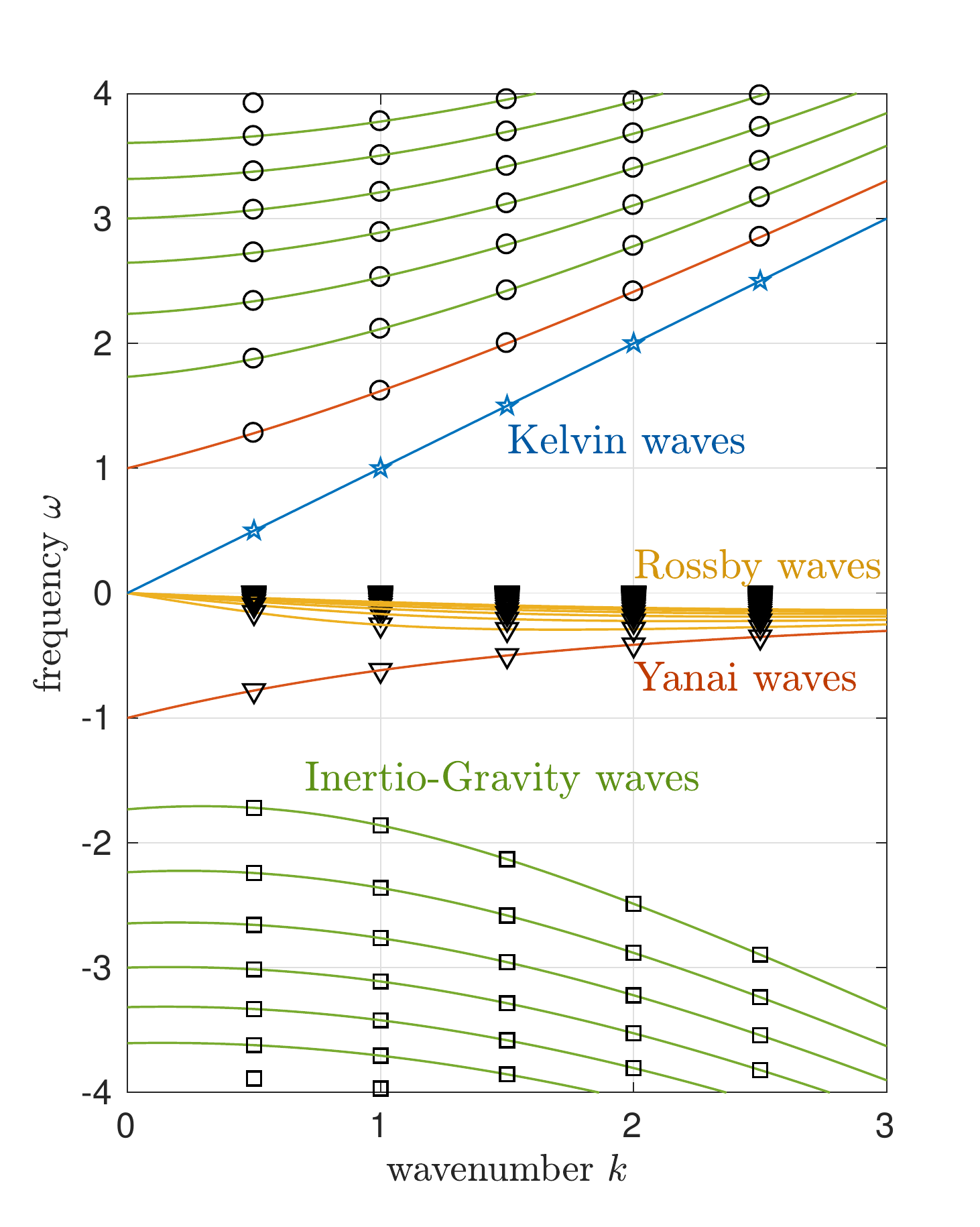}
\includegraphics[height=0.45\textwidth]{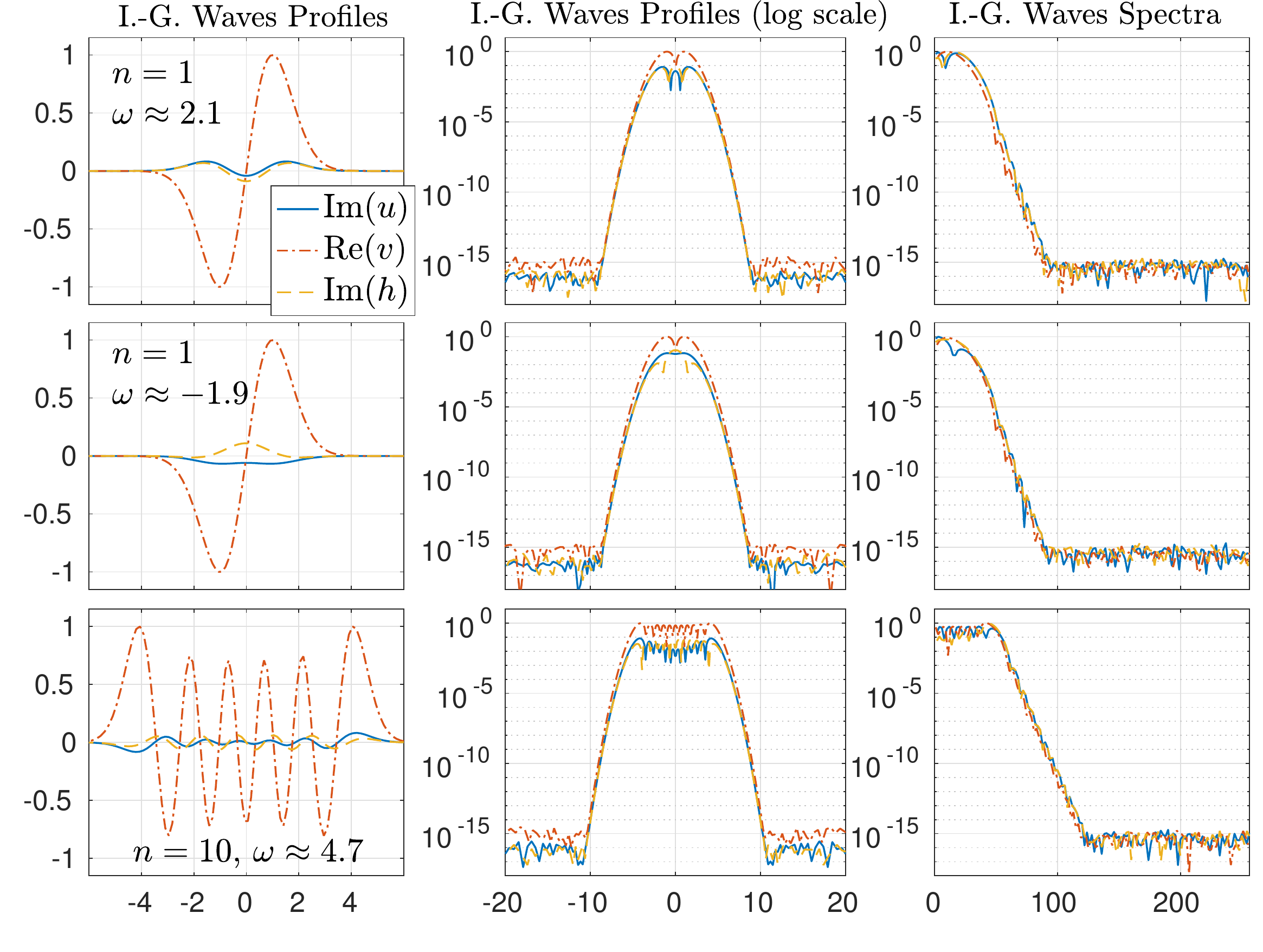}
\caption{\label{equatorial_waves_eigenproblem}Eigenmodes of the frictionless beta-plane. Left panel: the plain lines indicate the different branches of the dispersion relation; the blue line correspond to Kelvin waves ($n = -1$ mode), the orange lines to the Yanai wave ($n = 0$ mode), and  the  yellow and black lines the inertio-gravity and Rossby waves  respectively ($\Omega_n = 2n+1$).
The symbols correspond to the frequency determined numerically, for $k=1/2,\,1,\,3/2,\,2,$ and $5/2$. Right panels concern Inertio-Gravity waves: the first column displays meridional profiles of eigenmodes $u$, $v$ and $h$ as a function of $y$; the second column uses a logarithmic scale to illustrate the decay of the modes to zero within machine precision on the collocation grid; the third column illustrates the exponential convergence of the expansion. First two rows: two different modes for $n=1$ and $\Omega_1 = 3$: note the common value for $v$ (given by (\ref{equato_v})) but the different $u$ and $h$ profiles. The third row corresponds to $n=10$ and $\Omega_{10}=21$.}
\end{center}
\end{figure}\begin{figure}[h!]
\begin{center}
\includegraphics[width=\textwidth]{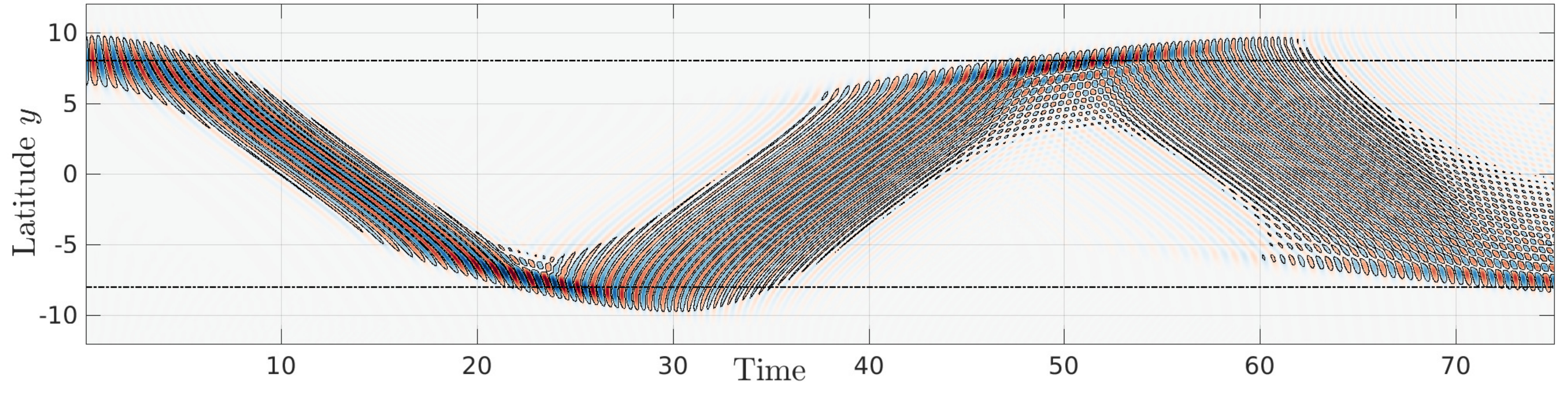}
\includegraphics[width=\textwidth]{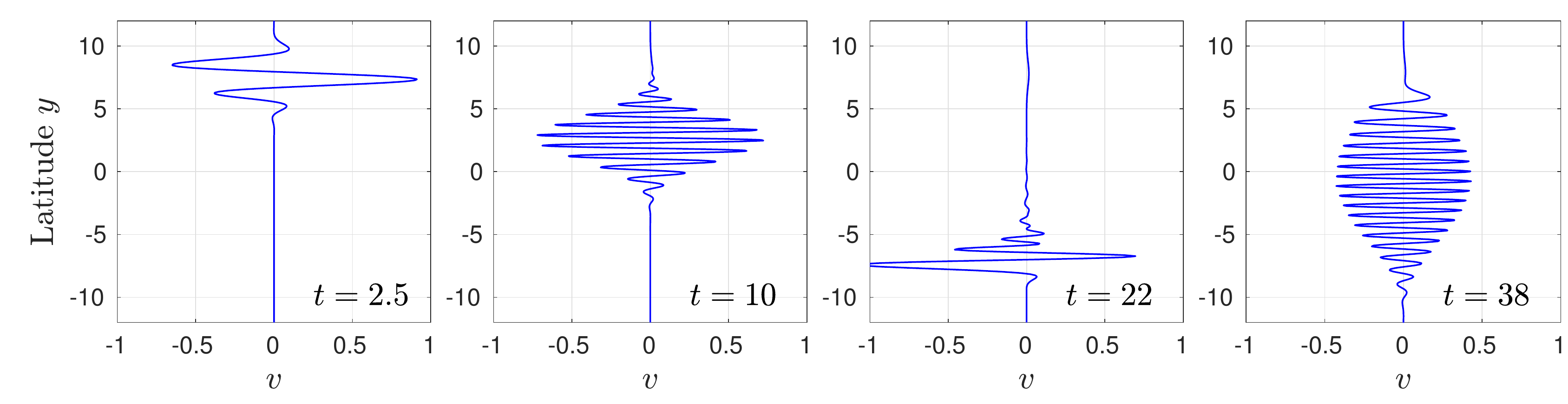}
\caption{\label{equatorial_waves_dynamics}Upper panel: Space-time diagram of a latitudinally trapped wave packet, with the initial condition: $u=0$, $h=0$, $v=\cos(x) e^{-\frac{\left(y-8\right)^2}{2} }$. We display $v(y,t)$ at $x=0$ as a function of $y$ and $t$. Lower panels: corresponding velocity profiles $v(y)$ at $x=0$ for different times. }
\end{center}
\end{figure}
\subsection{Numerical treatment}
When Rayleigh friction is present, the complex growth rate $\lambda$ of a normal mode of azimuthal wave number $k$ obeys the eigenproblem:
\begin{equation}
\lambda \begin{pmatrix}
u  \\ h \\ v
\end{pmatrix} = \mathscr{L} \begin{pmatrix}
u \\ h \\ v
\end{pmatrix}\quad \mathrm{where}\quad \mathscr{L} =\left(
\begin{array}{|cc|c@{}c|} 
\cline{1-2}
-a & -\mri k & & \multicolumn{1}{c}{y} \\
-\mri k & \multicolumn{1}{c|}{-a} & & \multicolumn{1}{c}{-\partial_y} \\
\cline{1-2}\cline{4-4}
\multicolumn{1}{c}{-y} & \multicolumn{1}{c}{-\partial_y} & & \multicolumn{1}{|c|}{-a} \\
\cline{4-4}
\end{array}
\right) \, . \label{Lblocks}
\end{equation}
Observe that variables form two groups: $u$ and $h$ form the first group and $v$ forms the second. The reason behind this segregation becomes clear when observing the blocks which form $\mathscr{L}$. In equation~(\ref{Lblocks}), the diagonal blocks (marked with rectangles) are composed of parity-preserving operators, whereas the off-diagonal blocks contain exclusively parity-flipping operators. Hence, in the spirit of the mixed expansions we have been using before in scalar equations, here we will use expansions of opposed parities in $y$ for different variables. As emphasized before, this remedy becomes clear after remapping the equation to $\theta$-space ($u^\dagger(\theta) = u(L\cot \theta)$, \emph{etc.}):
\begin{subequations}
\label{equato_theta}
\begin{gather}
\lambda\mathscr{M} \begin{pmatrix} 
u^\dagger  \\ h^\dagger \\ v^\dagger
\end{pmatrix} = \mathscr{L} \begin{pmatrix}
u^\dagger \\ h^\dagger \\ v^\dagger
\end{pmatrix}\,, \\
 \mathrm{where}\quad \mathscr{L} =\left(
\begin{array}{|cc|c@{}c|} 
\cline{1-2}
-a\sin\theta & -\mri k\sin\theta & & \multicolumn{1}{c}{L\cos\theta} \\
-\mri k & \multicolumn{1}{c|}{-a} & & \multicolumn{1}{c}{L^{-1}\sin^2\theta\partial_\theta} \\
\cline{1-2}\cline{4-4}
\multicolumn{1}{c}{-L\cos\theta} & \multicolumn{1}{c}{L^{-1}\sin^3\partial_\theta} & & \multicolumn{1}{|c|}{-a\sin\theta} \\
\cline{4-4}
\end{array}
\right) 
\, 
, 
\quad 
\mathscr{M} =\left(
\begin{array}{|cc|c@{}c|} 
\cline{1-2}
\sin\theta & 0 & & \multicolumn{1}{c}{0} \\
0 & \multicolumn{1}{c|}{1} & & \multicolumn{1}{c}{0} \\
\cline{1-2}\cline{4-4}
\multicolumn{1}{c}{0} & \multicolumn{1}{c}{0} & & \multicolumn{1}{|c|}{\sin\theta} \\
\cline{4-4}
\end{array}
\right)\label{Lthetablocks}\, .
\end{gather}
\end{subequations}
It becomes immediately clear that the sparsity of the discretization of the system is preserved if expansions of opposite parities are used for $(u^\dagger,h^\dagger) $ and $v^\dagger$. In the following, we present results obtained with a cosine expansion for $u^\dagger$ and $h^\dagger$ and a sine expansion for $v^\dagger$ (which yield to the structure depicted in figure~\ref{equato_spy} for matrices $\mathbfsf{M}$ and $\mathbfsf{L}$):
\begin{equation}
\label{equato_dagger}
	u^\dagger(\theta) = \sum_{m}\widetilde{u}_m \cos(m\theta)\, ,\quad
	h^\dagger(\theta) = \sum_{m}\widetilde{h}_m \cos(m\theta)\, ,\quad
	v^\dagger(\theta) = \sum_{m}\widetilde{v}_m \sin(m\theta)\, .
\end{equation} 
We compare in figure~\ref{equatorial_waves_eigenproblem} the frequencies obtained by solving the discretized eigenproblem to Rossby and inertio-gravity waves~(\ref{rossby_freq}), and Kelvin waves~(\ref{kelvin_freq}). The accuracy of a spectral method based on Chebyshev functions yields an excellent agreement. We also display on figure~\ref{equatorial_waves_eigenproblem} some eigenmodes. We illustrate how for a given $\Omega_n$, Rossby and inertio-gravity waves with different frequencies correspond to the roots of equation~(\ref{OMEGA}). These modes have a common azimuthal velocity $v$ but distinct meridional velocity $u$ and height of fluid $h$. We show plots of the amplitude of the profiles $|u(y)|$, $|v(y)|$, and $|h(y)|$ using a semi-logarithmic scale to illustrate the exponential decay of both variables as $|y|\rightarrow \infty$. Finally, the exponential convergence of the three spectra $|\widetilde{u}|$, $|\widetilde{v}|$, and $|\widetilde{h}|$ is shown on the last panel of figure~\ref{equatorial_waves_eigenproblem}.

We readily obtain a time-stepping scheme for the dynamics of equatorial waves by utilizing the matrices $\mathbfsf{M}$ and $\mathbfsf{L}$ with a implicit third-order Runge-Kutta scheme~\cite{ascherANM97}. We show temporal dynamics in figure~\ref{equatorial_waves_dynamics}. A initial perturbation localized around a latitude $y_0$ travels across the equator and is reflected when reaching the opposite latitude $-y_0$. In absence of dissipation, the wavepacket then continues bouncing back and forth between the bounding latitudes $-y_0$ and $y_0$, illustrating a well known feature (see for instance~\cite{gillBOOK}) of the equatorial region: this region behaves as a trapping wave guide. 
%
%
%
%
%

\section{Multidimensional domains}
\label{sec_2d}
\subsection{The spirit of the method}
The machinery of the Chebyshev functions on the infinite line is readily generalized for separable operators to bi- and tridimensional domains with at least one unbounded direction. For the sake of brevity, this paper describes only bidimensional domains $\left(x,y\right)\in\Omega$, the generalization to 3D being a mere bookkeeping exercise. As described in~\cite{heinrichsMoC89}, the philosophy consists in decomposing variables as a truncated expansion of a well-suited family of basis functions $\alpha_n(x)$ and $\beta_m(y)$ along each direction in the following fashion: 
\begin{equation}
f(x,y) \approx \sum_{n=1}^{N_x} \sum_{m=1}^{N_y} \widetilde{f}_{nm} \alpha_n(x) \beta_m(y)\, .
\end{equation}
 The procedure is as follows. First, select the bases $\alpha_n$ and $\beta_m$: Fourier modes should be used along periodic directions; Chebyshev polynomials are almost always the best choice along bounded directions; unbounded directions can be remapped to $[0,\pi]$, where cosines, sines, or the hybrid bases $\breve\beta$ or $\check{\beta}$ can be used, depending on the parity properties of the operators involved as discussed above.

Second, regularize the equations so that they contain only operators with a sparse representation. We have indicated a procedure for unbounded directions in section~\ref{general_case}; for bounded directions where Chebyshev polynomials are used, follow the quasi-inverse method of ~\cite{julienJCP06} equivalent to repeated integrations to remove all derivatives along these directions. 

Third, decompose the separable operators into a sum of the form $\mathscr{L} = \sum_i \mathscr{P}_i\mathscr{Q}_i $, where the $\mathscr{P}_i$ (resp. $\mathscr{Q}_i$) operators act on the variable $x$ (resp. $y$) only. Compute the matrices that represent the action of these operators on the selected bases $\alpha_n(x)$ and $\beta_m(y)$:
\begin{equation}
\mathbfsf{P}_i =  \frac{\inner{ \alpha_m}{\mathscr{P}_i \alpha_n}}{\norm{\alpha_m}^2},\qquad \mathbfsf{Q}_i =  \frac{\inner{ \beta_m}{\mathscr{Q}_i \beta_n}}{\norm{\beta_m}^2}\, .
\end{equation} 
These sparse matrices form the building blocks for the discretization of $\mathscr{L}$ which may now be assembled 
using the Kronecker product to obtain the matrix representing $\mathscr{L}$:
\begin{equation}
\mathbfsf{L} = \sum_i \mathbfsf{P}_i\otimes \mathbfsf{Q}_i\, .
\end{equation}
A similar treatment should be applied to the operator $\mathscr{M}$. The Kronecker product offers a simple way to compute the representation of operators $\mathscr{M}$ and $\mathscr{L}$ on the basis $\left(\alpha_0\beta_0,\alpha_0\beta_1,\dots,\alpha_0\beta_{N_y}, \alpha_1\beta_0, \dots \alpha_{N_x}\beta_{N_y} \right)$. The matrix of spectral coefficients $\widetilde{f}_{nm}$ can be vectorized into a long column vector of state 
\begin{equation}\boldsymbol{f} = \mathrm{vec}(f_{nm})\, ,\end{equation} which represents a stack of the columns of $\widetilde{f}_{nm}$. Finally, we obtain the familiar form $E\, \mathbfsf{M}\,\boldsymbol{f} = \mathbfsf{M}\, \boldsymbol{f}$.
 
 We illustrate this method on the three cases that occur with at least one unbounded direction. In this section, we consider a 2D quantum harmonic oscillator on the infinite plane $\Omega = \mathbb{R}^2$ and on the infinite strip with finite width $\Omega = \left[x_1, x_2 \right]\times \mathbb{R}$. In the next section, the case of a domain with an infinite direction and a periodic direction (infinite cylinder) is considered through the example of the Kelvin-Helmholtz instability. This is investigated through a time-stepping method in contrast with the eigenvalue problems considered in the present section.
 
\subsection{Bidimensional benchmark: the 2D QHO}
We consider a particle in the two-dimensional potential
\begin{equation}
V(x,y) = \frac{1}{2}x^2 - x + \frac{5}{6}y^2 - \frac{10}{3}y + \frac{\sqrt{3}}{3}  \left( xy - 2x -y\right) + \frac{13}{3}\, , \label{2d_potential}
\end{equation}
and obeying the 2D Schr\"odinger equation
\begin{equation}
E\, \psi = \left(- \partial_{xx} - \partial_{yy} + V(x,y)\right) \psi \,. \label{schrodinger2d}
\end{equation}
The motivation behind the choice of potential $V(x,y)$ is that this potential is nothing more than the simple potential $\bar{V}(X,Y) = X^2 + Y^2 /3$ rotated by an angle $\pi/3$ and translated by the quantity $(1,2)$. The resulting potential thus possesses no parity in $x$ nor $y$ and features a coupling term $xy$. Hence, equation (\ref{schrodinger2d}) is a test problem of reasonable complexity for our method. Despite this apparent complexity, the results are easily checked against analytic solutions. Indeed, since translations and rotations are unitary transforms, the eigenvalues of $V$ are equal to the eigenvalues of $\bar{V}$, namely:
\begin{equation}
E_{nm} = \left(2n+1\right) +\frac{1}{\sqrt{3}}  \left(2m+1\right)\,,\quad \left(n,m\right)  \in \mathbb{N}^2 \,.
\end{equation} Further, the eigenfunctions of $V$ are obtained by applying the same rotation and translation to those of $\bar{V}$:
\begin{equation}
\psi^{(nm)}(X,Y) = \mathrm{H}_n(X) \exp \left(-X^2 / 2\right) \times \mathrm{H}_m(Y/3^{1/4})\left(-Y^2 / \left[2\sqrt{3}\right]\right) \, .
\end{equation}
\subsection{The infinite plane $\Omega = \mathbb{R}^2$}
On the infinite plane, since we have demonstrated the efficiency and accuracy of the Chebyshev approximation, we simply decompose our variable on two sets of Chebyshev functions along $x$ and $y$. As in the 1D case, more intuition is generally available when dealing with trigonometric function instead of the full-fledged Chebyshev functions: hence we remap both the coordinates $x$ and $y$ using:
\begin{subequations}
\begin{gather}
x= L_x \cot(\chi)\,, \\y= L_y \cot(\theta)\, .
\end{gather}
\end{subequations}
where $L_x$ and $L_y$ are two real mapping parameters that can be optimised for accuracy, depending on the problem. Our Schr\"odinger equation (\ref{schrodinger2d}) for $\psi(x,y)$ is mapped into the following equation for $f(\chi,\theta) = \psi(x(\chi),y(\theta))$:
\begin{subequations}
\label{schrodinger_system}
\begin{equation}
E \mathscr{M} f(\chi,\theta) = \mathscr{L} f(\chi,\theta) \label{schroedinger2d_remapped} \,,
\end{equation}
where
\begin{align}
\mathscr{M} =& \sin ^2 \chi \sin^2 \theta\,, \\ 
\mathscr{L} =&  -\frac{1}{L_x^2}\left(  \sin^6 \chi \sin^2 \theta \partial_{\chi\chi} + 2\cos\chi \sin^5 \chi \sin^2\theta \partial_\chi\right) -\frac{1}{L_y^2}\left(  \sin^6 \theta \sin^2 \chi \partial_{\theta\theta} + 2\cos\theta\sin^5 \theta \sin^2\chi \partial_\theta \right) \nonumber\\ 
&+\frac{L_x^2}{2} \cos^2 \chi \sin^2 \theta -\left(1+\frac{2\sqrt{3}}{3}\right) L_x \cos\chi \sin\chi\sin^2\theta \nonumber\\
&+ \frac{5L_y^2}{6} \sin^2\chi \cos^2 \theta -\left(\frac{10}{3} +\frac{\sqrt{3}}{3}\right)L_y \cos\theta\sin\theta\sin^2\chi \nonumber \\
 &+\frac{\sqrt{3}L_xL_y}{3} \cos\chi\sin\chi\cos\theta\sin\theta + \frac{13}{3} \sin^2 \chi \sin^2\theta\, ,
\end{align}
\end{subequations}
after multiplication (regularization) by $\sin^2\chi \sin^2\theta$ to preserve sparsity. Due to the presence of parity mixing in both directions $\theta$ and $\chi$, we discretize this equation using mixed expansion in $\chi$ and $\theta$: for instance $\breve{\beta}(\chi)$ and $\breve{\beta}(\theta)$:
\begin{equation}
f(\chi,\theta) \approx \sum_{n,m=0}^{N} \widetilde{f}_{nm}\breve{\beta}_n(\chi)\breve{\beta}_m(\theta)\,.
\end{equation} Note that nothing would prevent us from using  different expansions in each direction, for instance, $\breve{\beta}(\chi)$ and $\mathring{\beta}(\theta)$. Doing so, however, would unnecessarily demand to compute the action of the operators on two different bases, yielding twice as much matrix building compared to using a common basis for both $\chi$ and $\theta$.
 The discretized version of system (\ref{schrodinger_system}), with $\boldsymbol{f}= \mathrm{vec}(\widetilde{f}_{nm})$:
\begin{subequations}
\label{shrodinger_matrices}\begin{equation}
E\,\mathbfsf{M}\, \boldsymbol{f} = \mathbfsf{L}\, \boldsymbol{f} 
\end{equation}
 can be expressed with a set of only five matrices:
\begin{align}
\mathbfsf{M} = \,& \mathbfsf{D} \otimes \mathbfsf{D}\,, \\
\mathbfsf{L} =\,& -\frac{1}{L_x^2}\left( \mathbfsf{A}\otimes \mathbfsf{D} + 2 \,\mathbfsf{B}\otimes \mathbfsf{D}\right) -\frac{1}{L_y^2}\left( \mathbfsf{D}\otimes \mathbfsf{A} +2\, \mathbfsf{D}\otimes \mathbfsf{B}\right) +\frac{L_x^2}{2} \mathbfsf{C}\otimes \mathbfsf{D} -\left(1+\frac{2\sqrt{3}}{3}\right) L_x \mathbfsf{E}\otimes \mathbfsf{D} \nonumber\\
&+ \frac{5L_y^2}{6} \mathbfsf{D}\otimes \mathbfsf{C} -\left(\frac{10}{3} +\frac{\sqrt{3}}{3}\right)L_y \mathbfsf{D}\otimes \mathbfsf{E} +\frac{\sqrt{3}L_xL_y}{3} \mathbfsf{E}\otimes \mathbfsf{E} + \frac{13}{3} \mathbfsf{D}\otimes \mathbfsf{D}\,.
\end{align}
\end{subequations}
where matrices $\mathbfsf{A}$, $\mathbfsf{B}$, $\mathbfsf{C}$, $\mathbfsf{D}$, and $\mathbfsf{E}$ are defined in equations (\ref{AB}) and (\ref{CDE}) and computed in appendix~\ref{app:chebfunc}.

\subsection{The infinite strip }
\begin{figure}
\includegraphics[width=\textwidth]{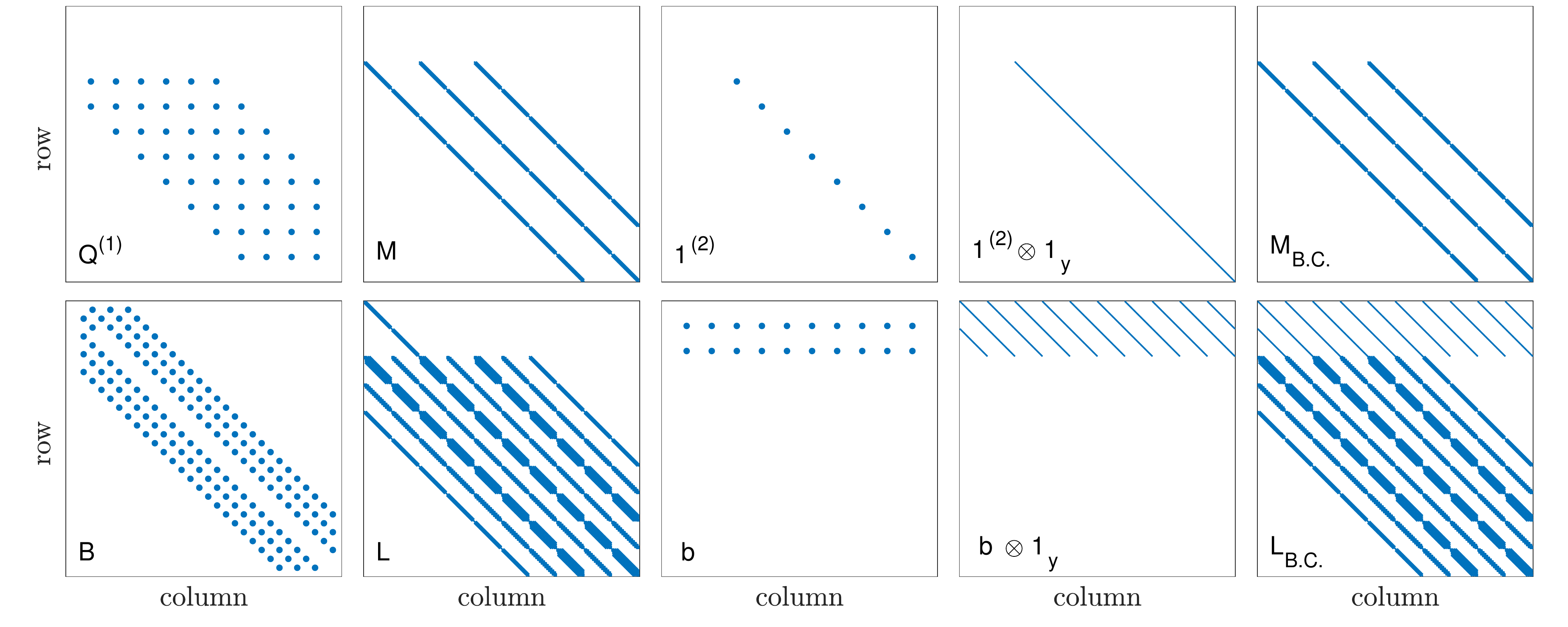}
\caption{Discretization of equation (\ref{schroedinger2d_remapped}) for the infinite strip geometry. Matrices $\mathbfsf{M}$ and $\mathbfsf{L}$ are obtained as Kronecker products of sparse matrices acting on the unbounded direction (for instance $\mathbfsf{B}$) or on the bounded direction (for instance $\mathbfsf{Q}$), and therefore inherit their sparsity. Along the bounded direction, boundary conditions are expressed with matrices $\mathbfsf{b}$ and $\mathbfsf{1}^{(2)}$. Finally, the discretized system including boundary conditions is represented by matrices $\mathbfsf{M}_{\mathrm{B.C.}}$ and $\mathbfsf{L}_{\mathrm{B.C.}}$ following equations~(\ref{2d_BC}). \label{fig_bc}}
\end{figure}
\begin{figure}
	\includegraphics[height = 0.29\textwidth]{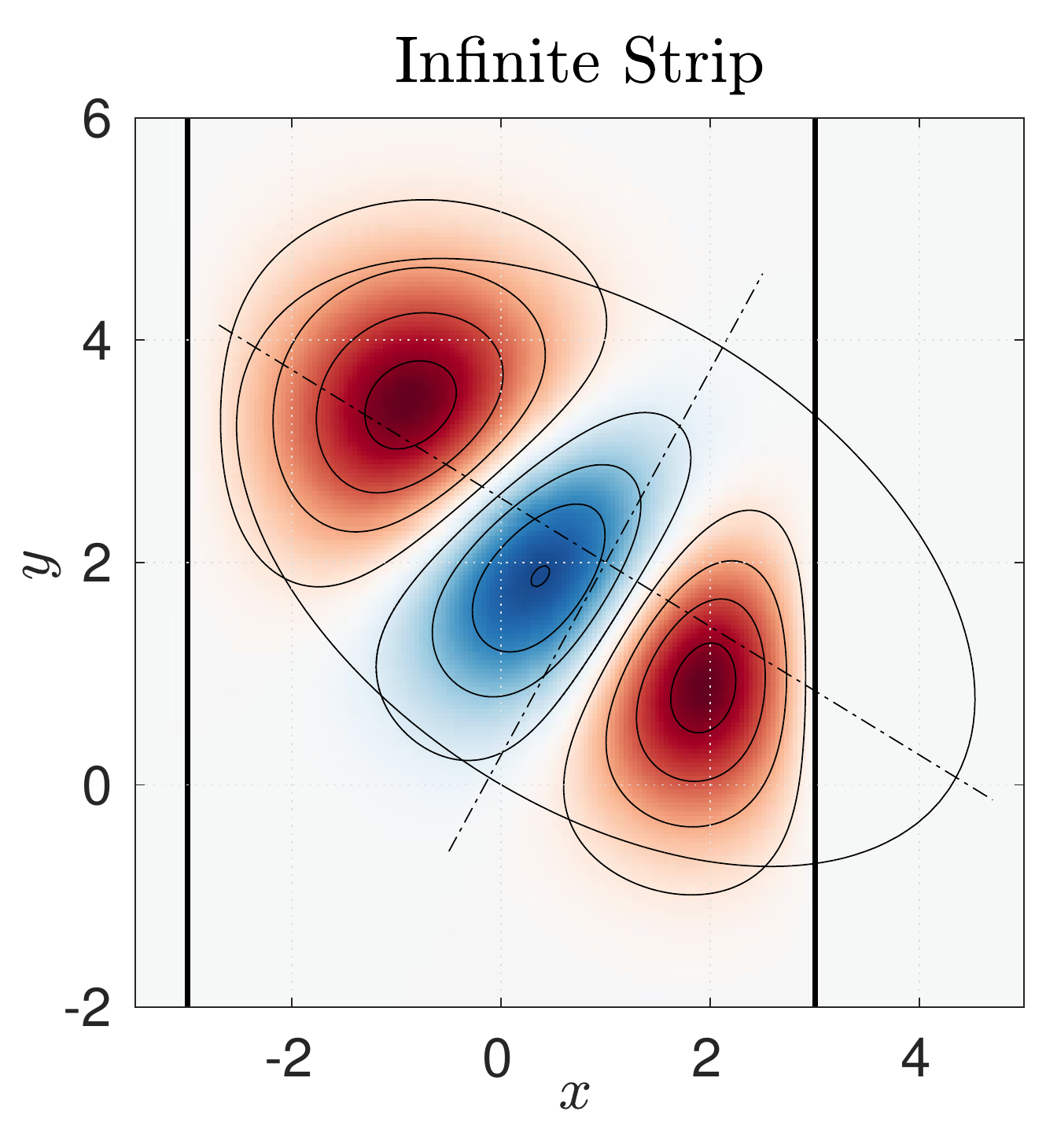}
	\includegraphics[height = 0.29\textwidth]{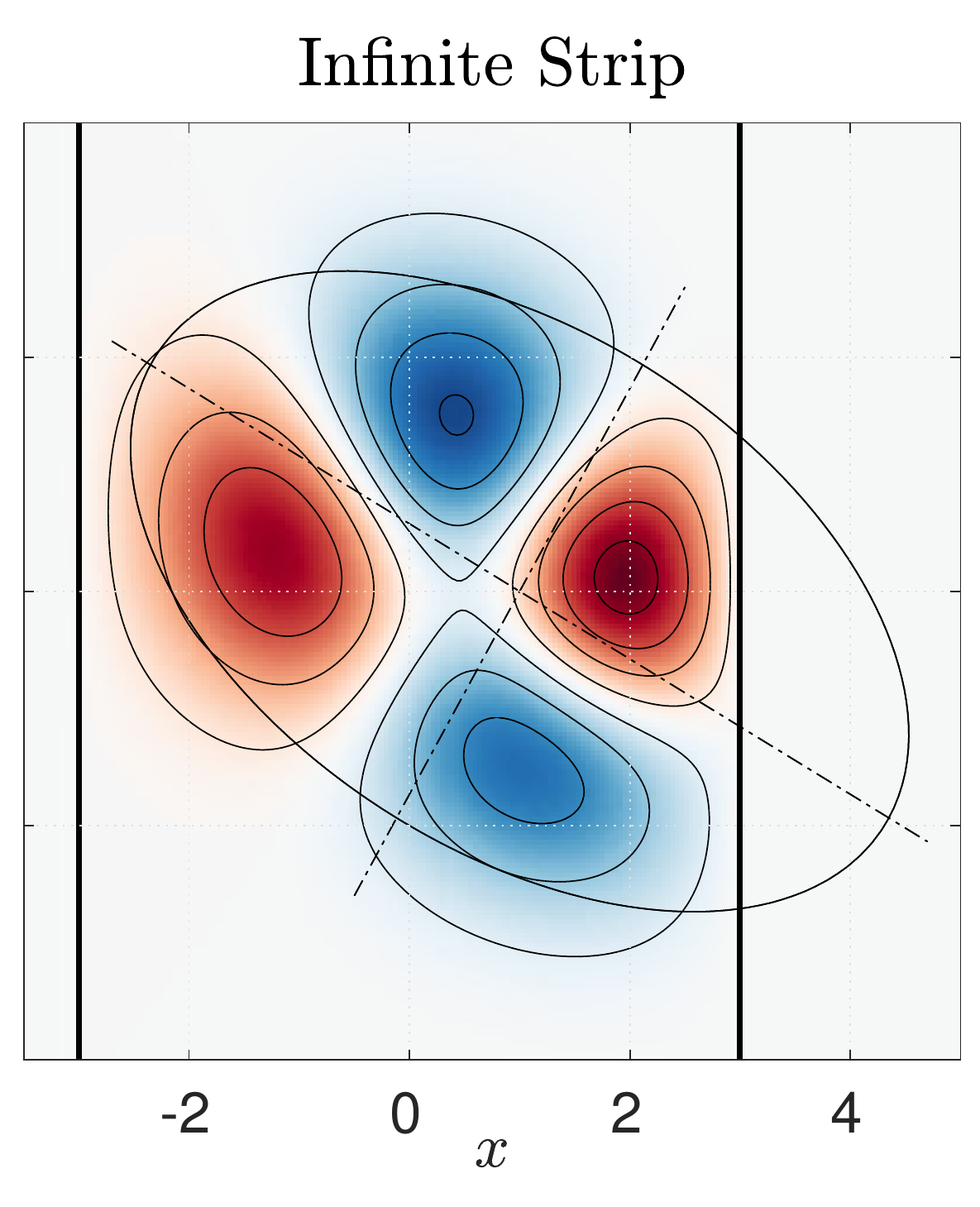}
	\includegraphics[height = 0.29\textwidth]{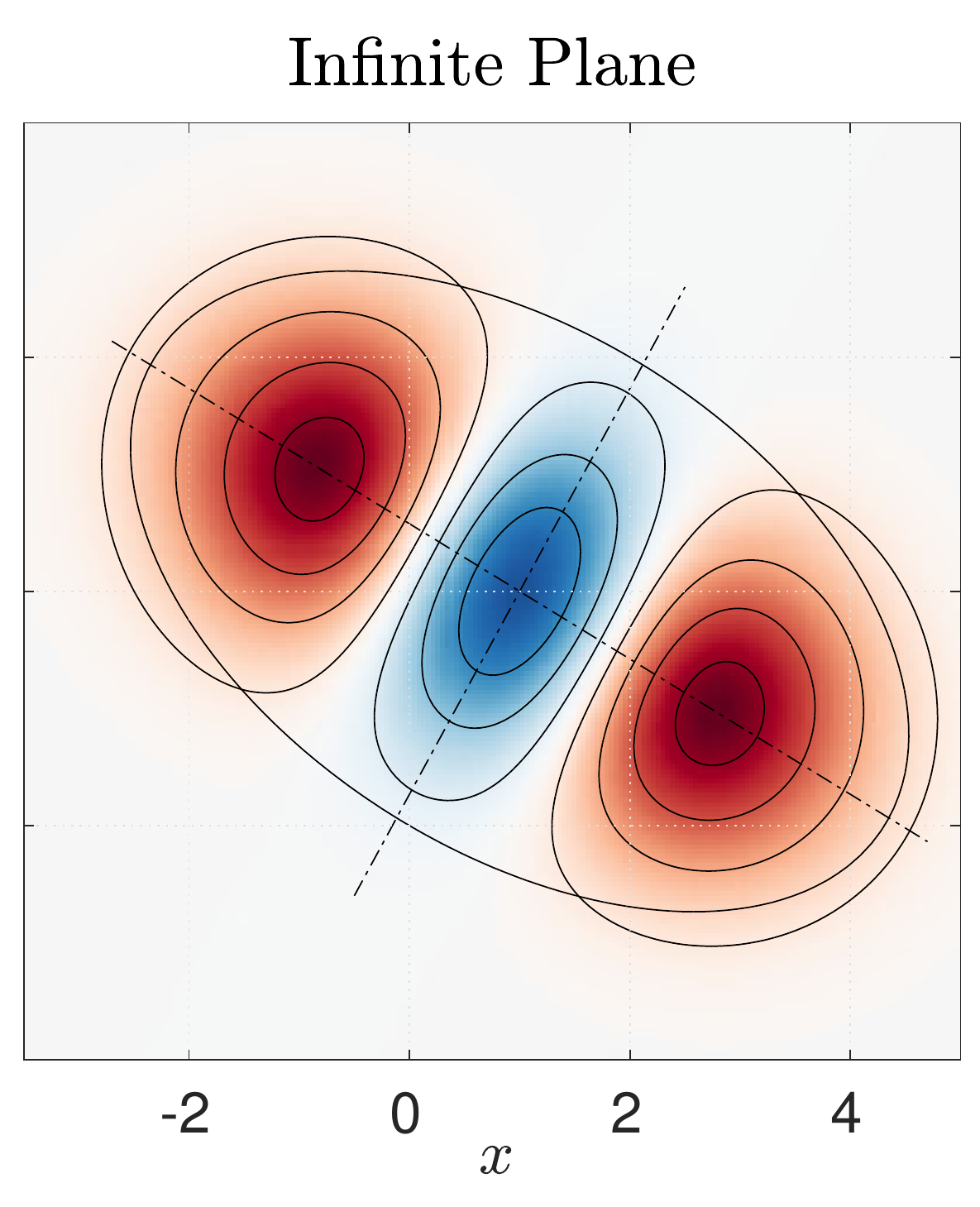}
	\includegraphics[height = 0.29\textwidth]{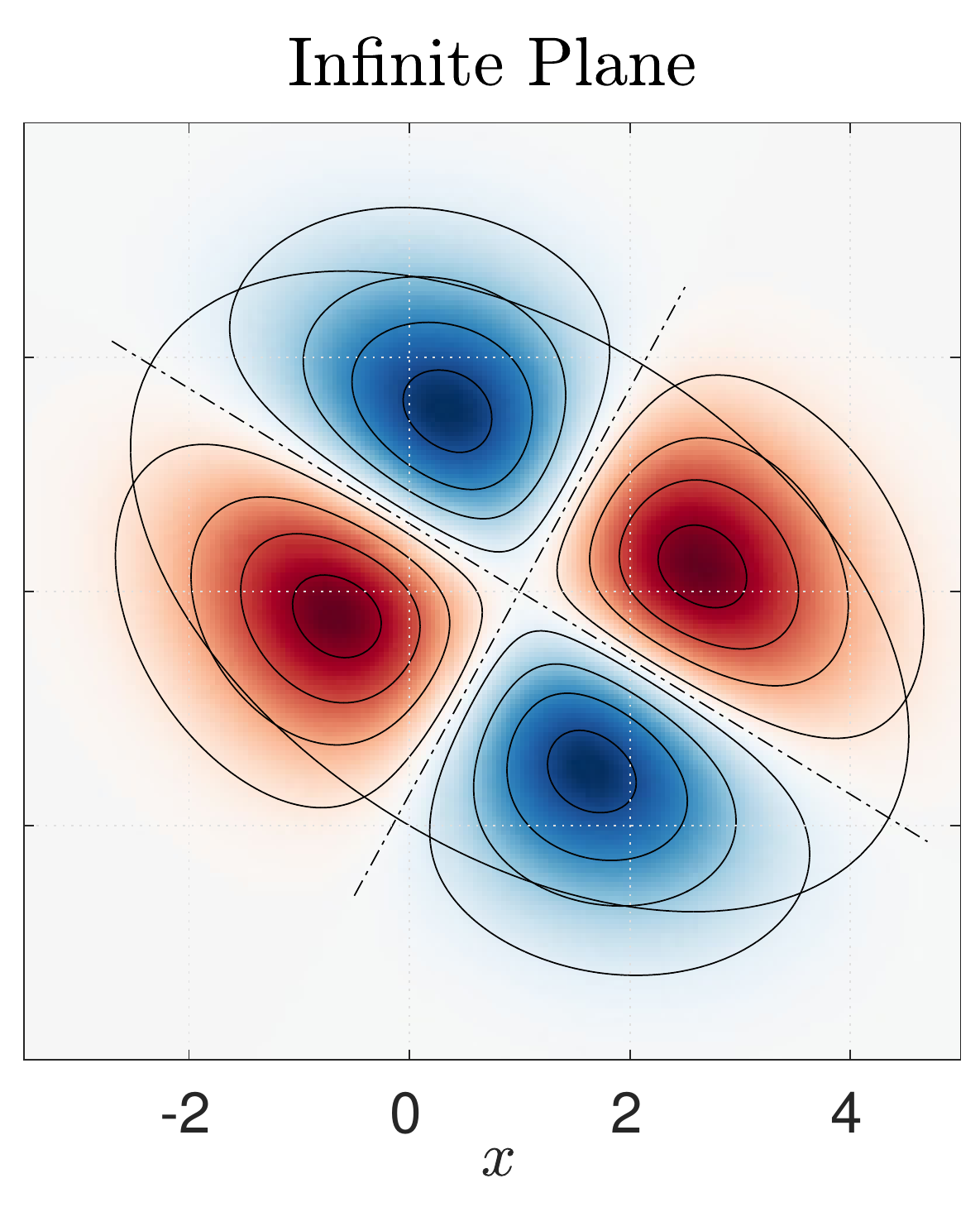}
	\caption{Solution $\psi(x,y)$. Modes $(1,3)$ and $(2,2)$ for a particle in the 2d quadratic potential given in equation~(\ref{2d_potential}), in presence of walls at $x=\pm 3$ (thick vertical lines, two left panels), or on an infinite plane (two right panels). To guide the eye, the dashed ellipses represent an equipotential line; the dashed lines indicate the two main directions of the potential. \label{particle_2d}}
\end{figure}

We now turn to solving equation~(\ref{schrodinger2d}) on an infinite strip $\Omega = [x_1,x_2] \times \mathbb{R}$ with Dirichlet boundary conditions $\psi(x_1)=\psi(x_2)=0$. As usual, the infinite $y-$direction is remapped to $\theta\in[0,\pi]$ and the bounded $x-$direction is remapped to $\chi \in [-1,1]$ via:
\begin{subequations}\begin{align}
x& = x_0 +  \chi\, \Delta x\,,\quad \mathrm{with} \quad x_0 = \frac{x_1+x_2}{2}\quad\mathrm{and}\quad \Delta x = \frac{x_2 - x_1}{2} \,, \\ 
\theta& = L \cot (\theta) \, .  
\end{align}
\end{subequations}
Since the spirit of this paper is to provide recipes for memory efficient discretization methods along the infinite line, it is desirable to couple them with memory efficient methods along the other directions too. When a truncated Chebyshev polynomial expansion is considered along a bounded direction, a sparse representation is obtained for operators containing only polynomials and derivatives. The quasi-inverse method~\cite{orszag} is employed that consists of integrating the governing equation or set of equations as many times as the highest derivative present. In our example, we will integrate twice to cancel out the second order $x$ derivative, so that equation~(\ref{schrodinger2d}) becomes:
\begin{multline}
	\label{constant_int}
E\sin^2{\theta}\iint d^2x f(x,\theta) =   - \sin^2 \theta f(x,\theta)- \frac{1}{L^2}\iint d^2x\left[ \sin^6\theta\partial_{\theta\theta} + 2 \cos\theta \sin^5 \theta \partial_\theta \right] f(x,\theta) \\ 
 +\iint d^2x  \left[ \frac{1}{2}x^2\sin^2\theta -\left(1+\frac{2\sqrt{{3}}}{3}\right) x\sin^2\theta + \frac{5}{6} \cos^2 \theta - \left(\frac{10}{3} +\frac{\sqrt{3}}{3} \right)\cos\theta\sin\theta + \frac{\sqrt{3}}{3} x \cos\theta\sin\theta \right] f(x,\theta) \\
 +ax + b\,.
\end{multline} Observe that by integrating twice we have introduced two arbitrary integration constants $a$ and $b$. The value of these yet floating integration constants will be determined by imposing boundary conditions as the last step of our discretization. For the time being, we define the discrete representation of three useful operators in Chebyshev space, denoted $\mathbfsf{Q}$:
\begin{subequations}
\begin{equation}
\mathbfsf{Q}^{(0)} =\frac{\inner{T_m}{\iint  T_n\mathrm{d}^2x}}{\norm{T_m}^2}\,,\qquad 
\mathbfsf{Q}^{(1)} =\frac{\inner{T_m}{\iint  x\,T_n\mathrm{d}^2x}}{\norm{T_m}^2}\,,\qquad \mathbfsf{Q}^{(2)} =\frac{\inner{T_m}{\iint x^2T_n\mathrm{d}^2x}}{\norm{T_m}^2}\,.\qquad  \tag{76a,b,c} \label{Q123}
\end{equation}
\end{subequations}
These matrices are banded with respectively two-, three- and four-bands. We give  their full expressions in the appendix~\ref{app:chebpol} and show that given the discretized matrix $\mathbfsf{x}$ for $x$ we arrive at $\mathbfsf{Q}^{(1)} = \mathbfsf{Q}^{(0)}\mathbfsf{x}$ and  $\mathbfsf{Q}^{(2)} =  \mathbfsf{Q}^{(0)}\mathbfsf{x}^2$ with simply matrix multiplications. Similarly to the previous section, we easily obtain our matrices $\mathbfsf{M}$ and $\mathbfsf{L}$ by combining the sparse representations of operators along different directions with the Kronecker product:
\begin{align}
\mathbfsf{M} =&\, \mathbfsf{Q}^{(0)}\otimes \mathbfsf{D}\, , \\
\mathbfsf{L} =&\, - \mathbfsf{1}_x \otimes \mathbfsf{D} - \frac{1}{L^2} \mathbfsf{Q}^{(0)}\otimes\left[ \mathbfsf{A} + 2 \mathbfsf{B}\right]  + \frac{1}{2}\mathbfsf{Q}^{(2)} \otimes \mathbfsf{D} 
\nonumber \\
& \,- \left( 1+\frac{2\sqrt{3}}{3}\right) \mathbfsf{Q}^{(1)}\otimes \mathbfsf{D} + \frac{5}{6} \mathbfsf{Q}^{(0)}\otimes \mathbfsf{C} - \left(\frac{10+\sqrt{3}}{3} \right) \mathbfsf{Q}^{(0)}\otimes \mathbfsf{E} 
+ \frac{\sqrt{3}}{3} \mathbfsf{Q}^{(1)}\otimes \mathbfsf{E}\, .
\end{align}

The last step consists in implementing boundary conditions: as mentioned above the first two coefficients of the expansion in $x$ are polluted by integration factors $a(\theta)$ and $b(\theta)$. As a consequence, the $2N_y$ lines in $\mathbfsf{M}$ and $\mathbfsf{L}$ that correspond to coefficients $f_{0m}$ and $f_{1m}$ (with $m\in[1,N_y]$) have to enforce Dirichlet boundary conditions, namely:
\begin{equation}
\forall m:\qquad\sum_n f_{nm} = 0\,,\quad \sum_n (-1)^{n}f_{nm} = 0 \,, 
\end{equation}
which we write
\begin{equation}
\forall m:\qquad\sum_n b_{0n}f_{nm} = 0\,,\quad \sum_n b_{1n}f_{nm} = 0 \,, 
\end{equation}
for a greater generality. Any boundary conditions can be implemented easily by using only matrix products in the following manner. We define a boundary-condition matrix $\mathbfsf{b}$ which is zero everywhere, except for the first two lines $b_{0n}$ and $b_{1n}$, and a diagonal matrix $\mathbfsf{1}^{(2)}$ equal to the identity everywhere except on the first two lines, which are identically zero. Define:
\begin{subequations}\label{2d_BC}
\begin{align}
\mathbfsf{M}_{\mathrm{B.C.}} &= \left( \mathbfsf{1}^{(2)}\otimes{\mathbfsf{1}_y}\right) \mathbfsf{M}\, , \\ 
\mathbfsf{L}_{\mathrm{B.C.}} &= \left( \mathbfsf{1}^{(2)}\otimes{\mathbfsf{1}_y}\right) \mathbfsf{L} + \left( \mathbfsf{b}\otimes \mathbfsf{1_y} \right)\, .
\end{align}
\end{subequations}
The mechanics of this simple algebra, illustrated on figure~\ref{fig_bc}, is the following: multiplying by $\mathbfsf{1}^{(2)}\otimes\mathbfsf{1}_y$ zeroes out the $2N_y$ lines of matrices $\mathbfsf{M}$ and $\mathbfsf{L}$ that correspond to projection onto the two lowest order Chebyshev modes that are polluted by arbitrary integration constants, as seen in equation~(\ref{constant_int}). In $\mathbfsf{L}_{\mathrm{B.C.}}$, these lines are replaced with the boundary condition by adding $\mathbfsf{b}\otimes \mathbfsf{1}_y$. Eigenmodes for both the infinite strip and the infinite plane are shown on figure~\ref{particle_2d}. The accuracy of this bidimensional example is omitted but has been checked to be as good as in the 1D case.

%
%
%
%
%

\section{Nonlinear time-stepping: the Kelvin-Helmholtz instability}
\label{sec_KH}
\begin{figure}
	\begin{center}
		\includegraphics[height = 0.35\textwidth]{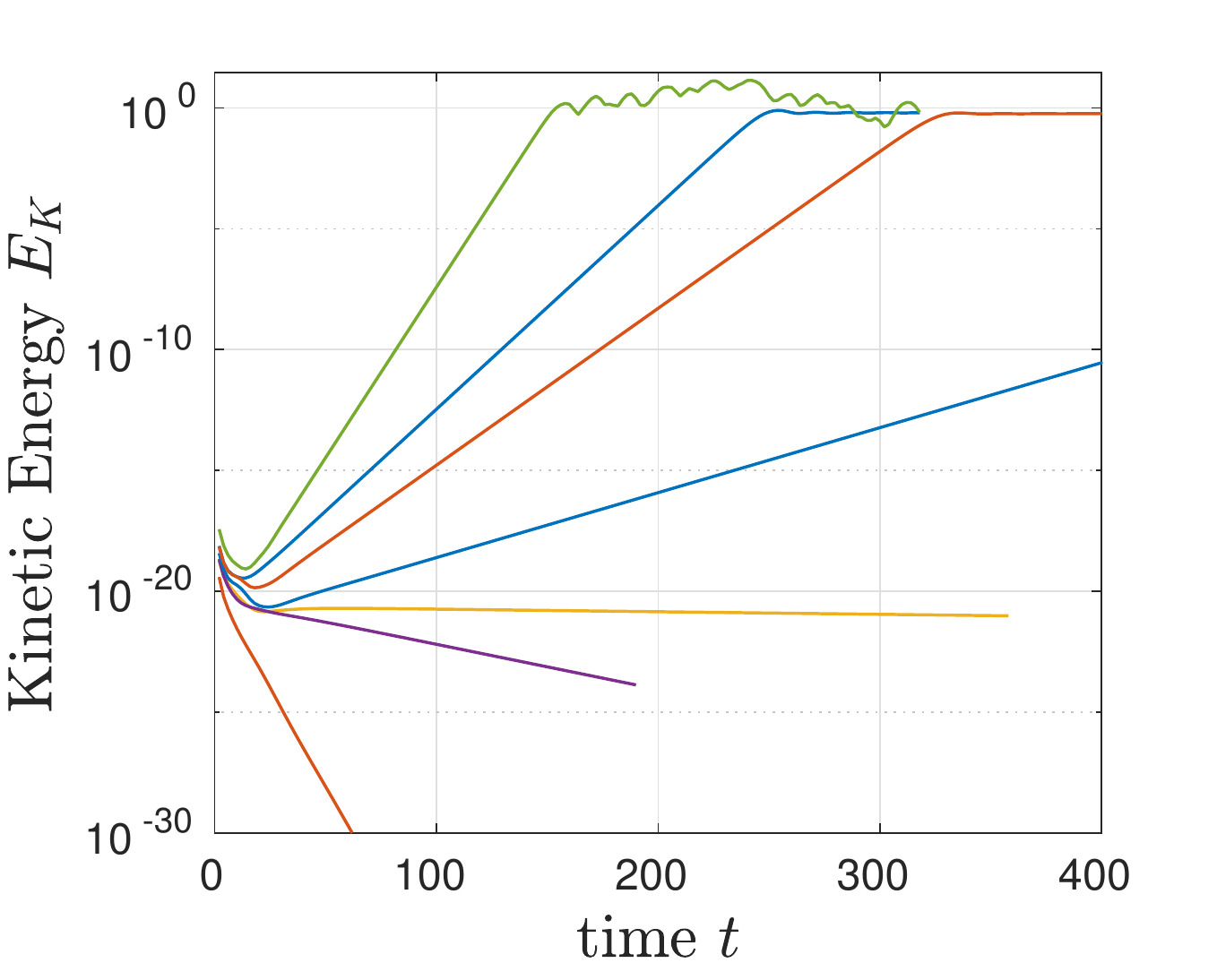}
		\includegraphics[height = 0.35\textwidth]{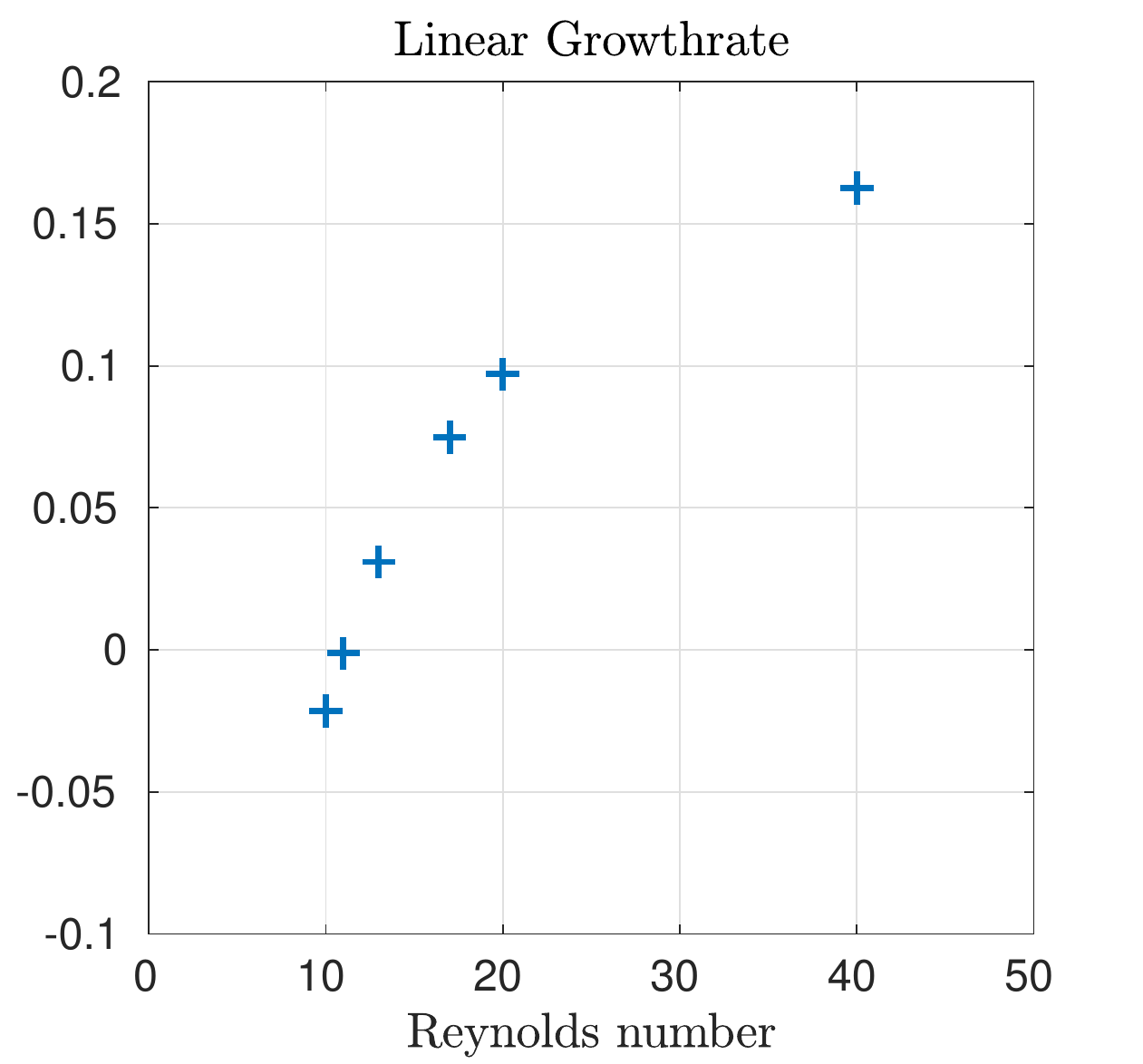}
\caption{\label{KH_linear} Stability of the base flow $U_0(y)\,\boldsymbol{e}_x$ (see equation~\ref{flow_u0}). Left panel: kinetic energy time series for various Reynolds numbers. From bottom to top $\mathrm{Re} = 5,10,11,13,17,20,40$. Right panel: growthrate $\lambda$ during the exponential growth phase as a function of Re. }
	\end{center}
\end{figure}

\subsection{Background parallel flow and formulation}
We consider a streamwise invariant parallel flow $U_0(y) \boldsymbol{e}_x$ imposed by external forces in a periodic domain along the streamwise direction $x\in\left[0,2\pi\right[$ with an unbounded spanwise direction $y\in\mathbb{R}$. Perturbations $\boldsymbol{u}(x,y,t) = -\partial_z \psi\, \boldsymbol{e}_x + \partial_x \psi \, \boldsymbol{e}_z$ to the background flow obey the incompressible governing equation:
\begin{equation}
\partial_t \nabla^2 \psi + U_0\, \partial_x \nabla^2 \psi - \partial_{zz}U_0\, \partial_x \psi + \mathcal{J}\left( \psi,\nabla^2 \psi\right) = \frac{1}{\mathrm{Re}} \nabla^4 \psi \,. \label{governing_KH}
\end{equation}
Such flows are known to be potentially linearly unstable if they present an inflection point $\partial_{zz}U_0 \neq 0$ (see for example~\cite{drazin2004}).  The Reynolds number, measuring the strength of the viscous dissipation force, controls the stability of the shear flow, i.e., the flow stabilizes as $Re$ is lowered. In this section, the stability of the flow
\begin{equation}
U_0(y) = \frac{y}{1+y^4}\label{flow_u0}
\end{equation}
is analyzed by means of a time-stepping integration of the full nonlinear governing equation~(\ref{governing_KH}).
\subsection{Discretization}
Following the usual prescription, we remap the equations to  $\theta$-space and multiply out with the factor $\left(L^4 \cos^4 \theta +\sin^4 \theta  \right)^3$ to obtain:
\begin{equation}
\mathscr{M}\partial_tf  = \mathscr{L}f+ \mathscr{N}(f)\,,\end{equation}
where $\mathscr{M}$ and $\mathscr{L}$ are linear operators and $\mathscr{N}(f)$ is the nonlinear advection term:
\begin{subequations}
\begin{align}
\mathscr{M}& = \left(L^4\cos^4\theta + \sin^4\theta\right)^3 \nabla^2 \\ 
\mathscr{L}& =- L \cos\theta\sin^3\theta \left(L^4\cos^4\theta + \sin^4\theta\right)^2 \partial_x \nabla^2 + 4L^3\cos^3\theta \left(3\cos^4\theta L^4 - 5 \sin^4\theta\right) \sin^5\theta \partial_x\nonumber\\ & \qquad \qquad+\frac{1}{\mathrm{Re}}  \left(L^4\cos^4\theta + \sin^4\theta\right)^3 \nabla^4 \\ 
\nabla^2& = \left( \frac{\sin^4\theta}{L^2}\partial_{\theta\theta} + \frac{2\sin^3\theta\cos\theta}{L^2} \partial_\theta + \partial_{xx}\right) \\ 
\mathscr{N}(f)& =  \left(L^4\cos^4\theta + \sin^4\theta\right)^3\frac{\sin^2\theta}{L} \left(-\partial_x f \partial_\theta \nabla^2 f + \partial_\theta f \partial_x\nabla^2 f \right)
\end{align}
\end{subequations}
We expand our variable $f$ as a Fourier series along the periodic direction $x$. Observing that parity mixing  is present amongst the operators,
 we chose to represent $f$ as a hybrid cosine and sine series in $\theta$:
\begin{equation}
f(x,\theta)\approx  \sum_{k=0}^{N_x-1}\sum_{n=1}^{N_y} \left( \widetilde{f}_{k+1,n} \breve{\beta}_n(\theta) \mathrm{e}^{\mri kx} \right) + \mathrm{c.c.}
\end{equation}
As the operators $\mathscr{M}$ and $\mathscr{L}$ are separable, we easily obtain their discrete representation $\mathbfsf{M}$ and $\mathbfsf{L}$ by using Kronecker products. We compute the nonlinear term pseudospectrally by first evaluating derivatives $\partial_x f$, $\partial_{\theta}\nabla^2 f$, \emph{etc.}, in spectral space. We then  transform these variables to physical space where the products are computed, and transform back to spectral space. A $2/3$ dealiasing rule, appropriate for quadratic nonlinearities,  is used. Additional details on the computation of $\mathbfsf{N}(f)$, particularly concerning the implementation of the transform between spectral and physical space for the functions $\breve{\beta}_n$ by means of FFT algorithms, are relegated to appendix~\ref{app:transforms}.

\subsection{Semi-implicit time-stepping schemes}
IMEX schemes are a popular class of time-stepping schemes in computational fluid mechanics: these schemes combine an implicit treatment of the linear terms, whereas nonlinearities are treated explicitly. This popularity stems from the salient features of the Navier-Stokes equations which typically contains stiff linear terms, advocating for the use of stable and accurate implicit schemes. An implicit treatment of the nonlinear term would be costly and ultimately requires iterative methods compared to direct solves.   Fortunately an explicit treatment yields satisfactory results. The simplest IMEX scheme, used in the following, is the order one Backward Euler which obtains $f^{(i+1)}$ from $f^{(i)}$ after a timestep $\tau$ by solving the linear system:
\begin{equation}
\big(\mathbfsf{M} - \tau \mathbfsf{L}\big) f^{(i+1)} = \mathbfsf{M}f^{(i)} + \tau \mathbfsf{N}
\left(f^{(i)}\right)\, .\end{equation}
 
\subsection{Stability of the flow: numerical results}
Equation~(\ref{governing_KH}) is investigated by time-integration starting from an initial condition consisting of small amplitude noise uniformly
distributed over the resolved spatial scales. The time series of the kinetic energy $E_K$ is displayed on figure~\ref{KH_linear} for seven values of  the Reynolds number $Re$. Three phases are distinguished. 

\begin{figure}
	\begin{center}
		\includegraphics[width = 0.32\textwidth]{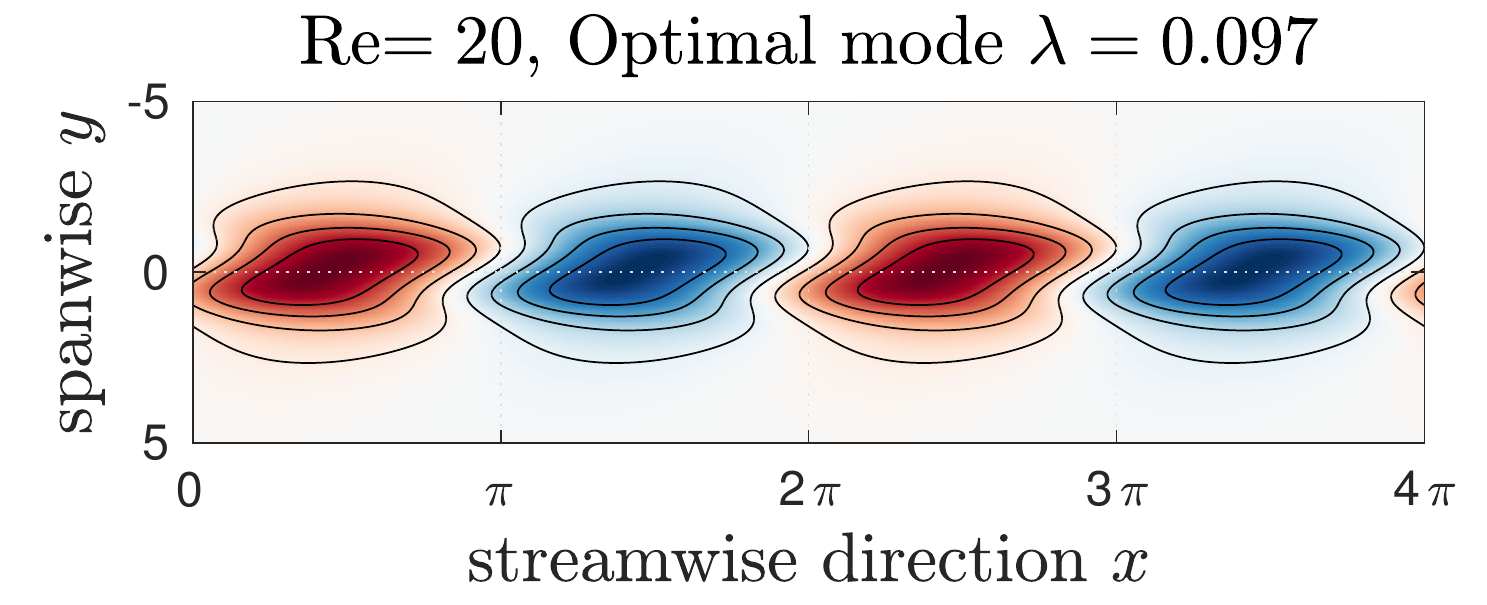}
		\includegraphics[width = 0.32\textwidth]{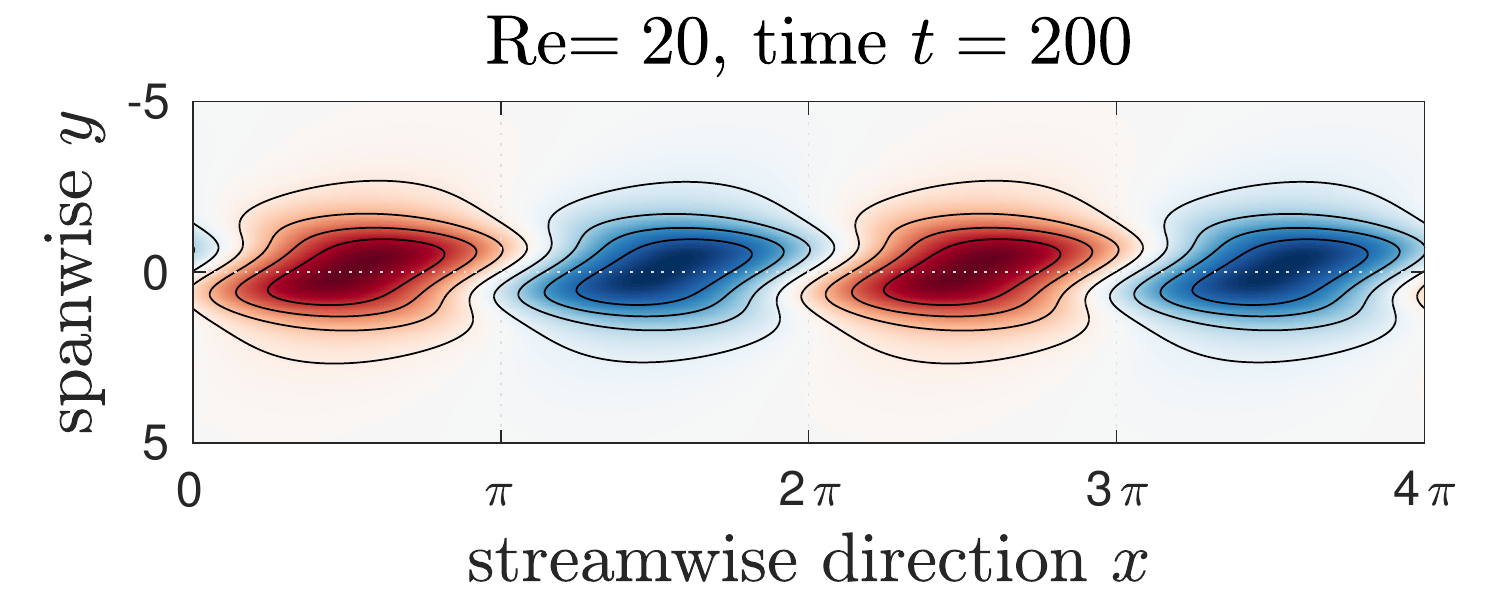}
		\includegraphics[width = 0.32\textwidth]{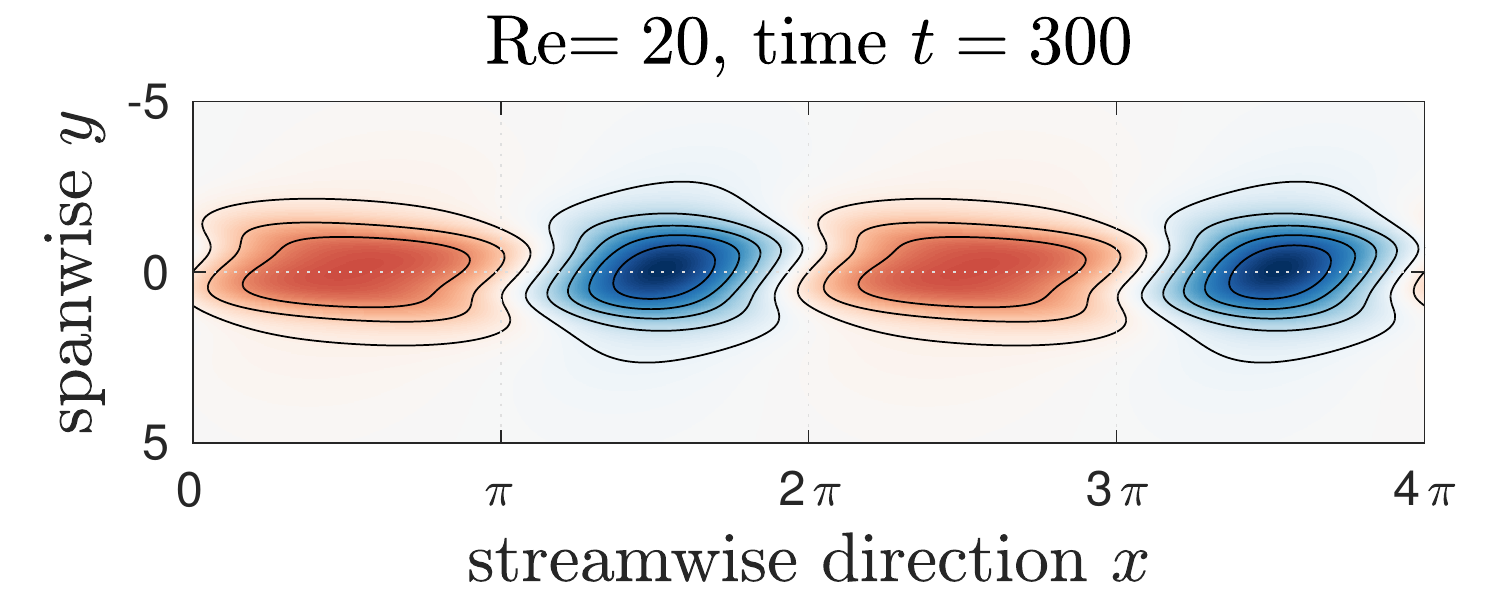}
		\includegraphics[width = 0.32\textwidth]{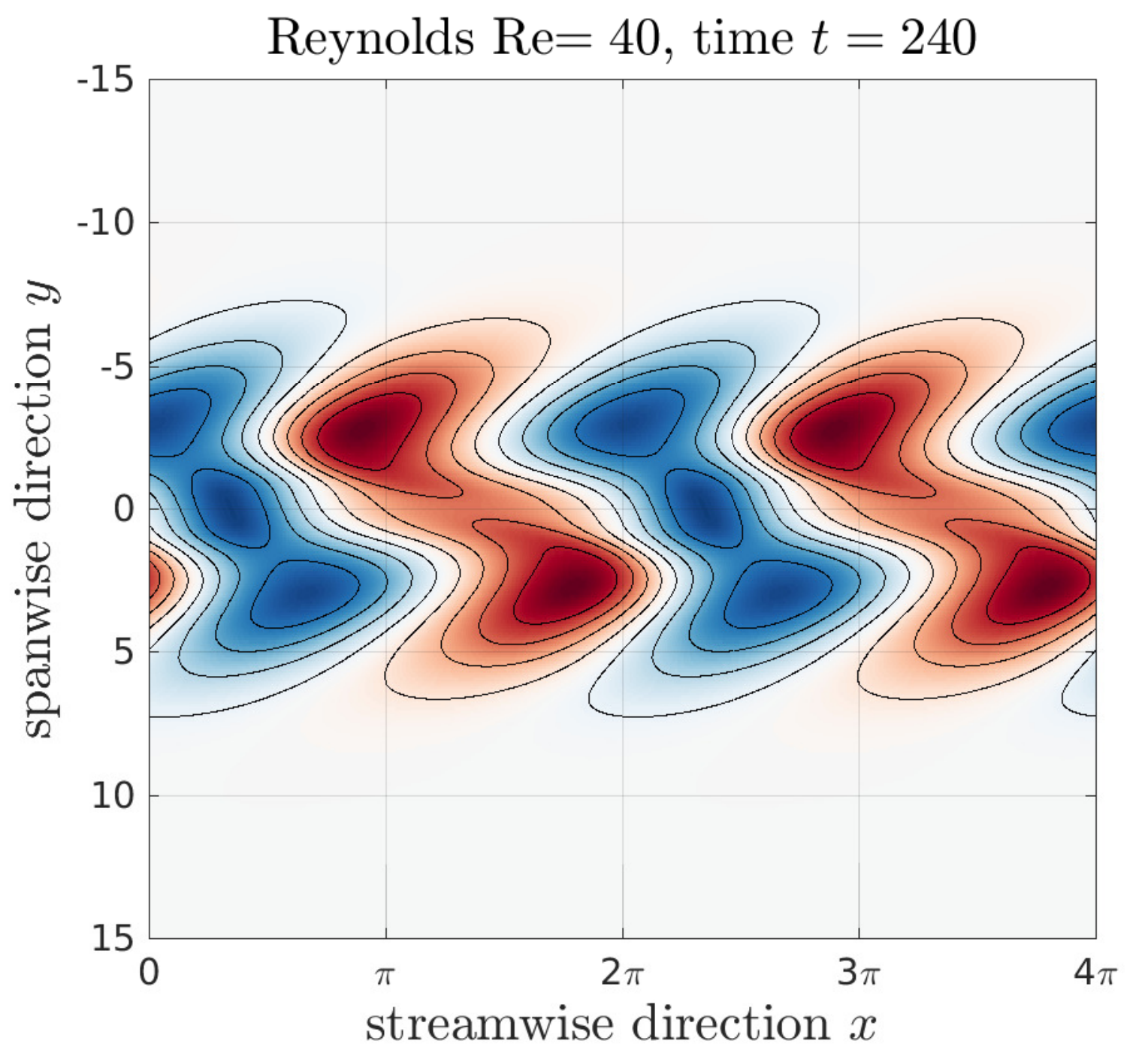}
		\includegraphics[width = 0.32\textwidth]{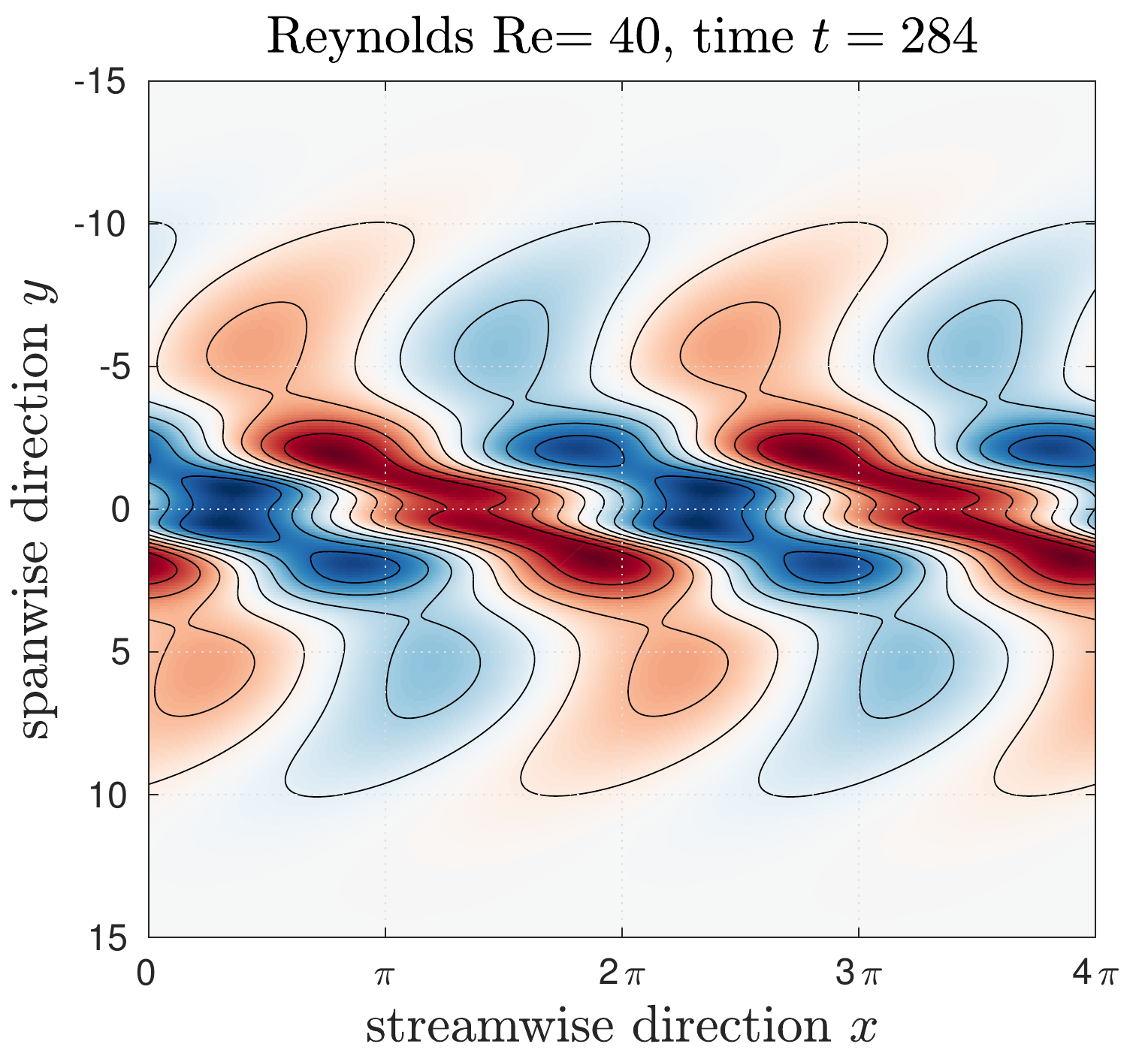}
		\includegraphics[width = 0.32\textwidth]{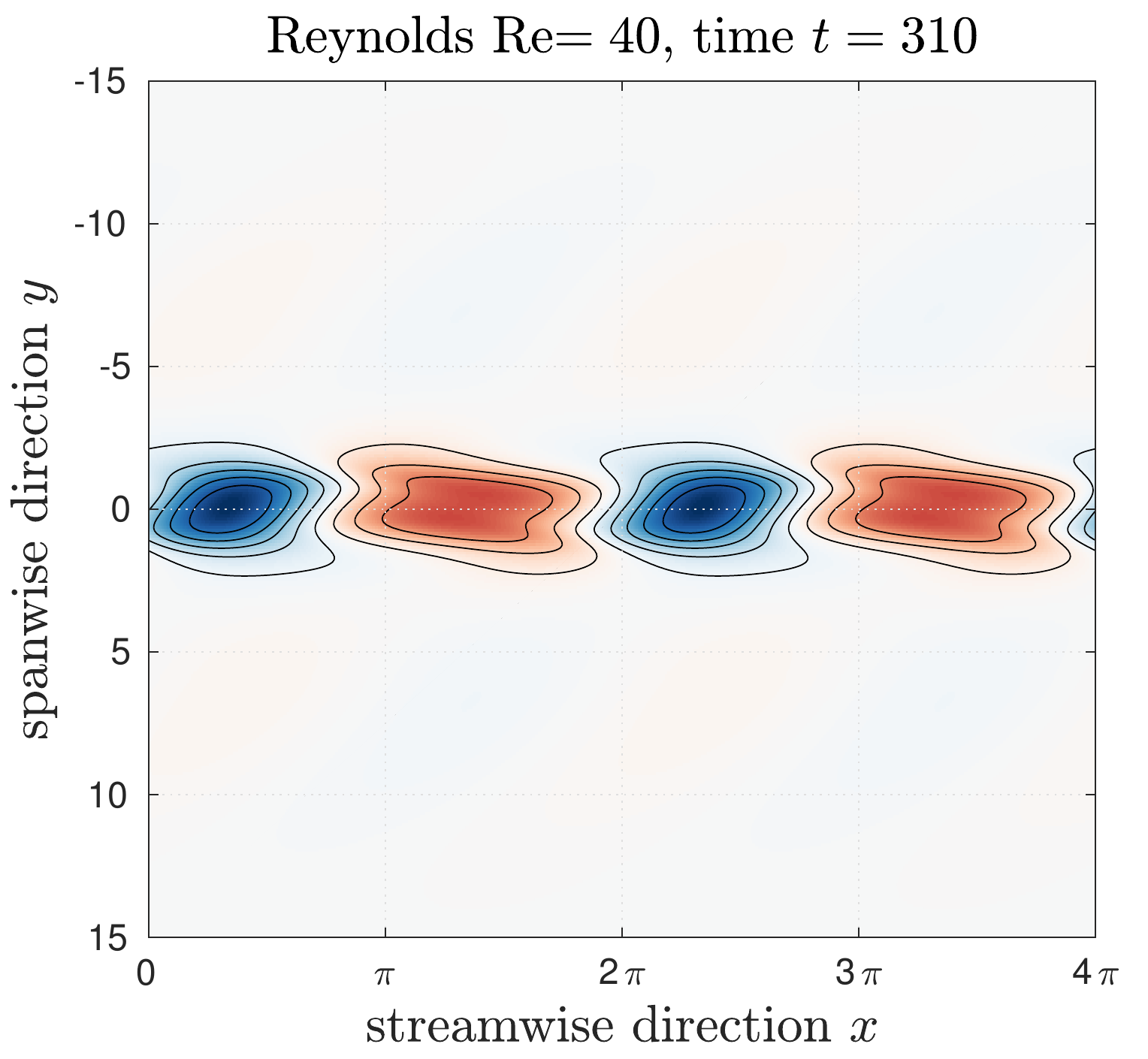}
		\caption{\label{KH_NL} Planform of the streamfunction of the perturbation $\psi(x,y)$. Top row, $\mathrm{Re}=20$: Optimal mode obtained by solving the eigenproblem (left); snapshot during the exponential growth phase (center); and snapshot after saturation (right). Bottom row, $\mathrm{Re}=40$: snapshots at three different instants after saturation, illustrating the intermittency of the flow.  }
	\end{center}
\end{figure}
During the initial phase, as most of the modes that compose the initial condition are stable, the total energy first plummets (for $t\lesssim 20$). During the second phase, an optimal mode emerges from the sea of decaying modes: this mode is either the most unstable (fastest growing) mode or the least damped mode, depending on the value of Re. The planform of the mode that dominates the temporal dynamics is depicted in figure~\ref{KH_NL} for $\mathrm{Re}=20$ and compared with the optimal mode obtained by means of the corresponding eigenproblem. As expected, the agreement is excellent. As long as the dynamics remains in the linear regime ($|\psi|\ll 1$). The total kinetic energy is dominated by the energy of the optimal mode alone which follows an exponential growth trend in time $E_K \propto \mathrm{exp}(2\lambda t)$. The optimal growth rate $\lambda(Re)$ can be fitted for each values of Re. Figure~\ref{KH_linear} shows how the optimal growth rate $\lambda(Re)$ varies as a function of $Re$. We obtain the critical Reynolds number $\mathrm{Re}_c \approx 11.066$ corresponding to the threshold of the instability by fitting the growthrate $\lambda$ with a polynomial in $\mathrm{Re}$ around the onset.

Finally, the instability saturates as $\psi$ becomes of order $O(1)$ (figure~\ref{KH_linear}). As can be observed from the timeseries at modest supercriticality $\mathrm{Re}\le 20$, the saturated state becomes steady after a short overshoot. The steady planform of the saturated state is shown in figure~\ref{KH_NL} for $\mathrm{Re}=20$ and displays a noticeable but vertically contained distortion compared to the linear optimal mode. By contrast, when the supercriticality is increase (e.g. for $\mathrm{Re} = 40$), the saturated state becomes unsteady, showing intermittent bursts of activity, as reported in the bottom row of figure~\ref{KH_NL}. The unstable layer of fluid around $y=0$ then sheds vortices to the quiescent surrounding regions. Here, we notice the limitations of our method. As no proper physical trapping mechanism is present, viscous damping remains the only source of control, thus the outwardly propagating vortices can violate the CFL criterion as they propagate fast enough to reach regions where the collocation grid becomes coarser and coarser. If viscosity acts on timescales longer than the time needed for these vortices to reach large values of $\vert y\vert$, these vortices will become underresolved before they are damped out. The accuracy and stability of the methods is then jeopardized. If no additional physics is introduced to resolve this issue, the solution is to increase the number of basis function along $y$ and to use a larger $L$ to increase the size of the region where the dynamics is safely resolved. Note that a similar increase in resolution would be necessary if one were to chose to introduce boundary conditions to tackle the problem of an unstable shear flow in an unbounded $y-$direction: these boundary conditions would have to be placed ``far enough'', so that  they do not affect the dynamics. Increasing the distance between boundaries would equally lead to increasing the number of grid points or Chebyshev polynomials in the $y-$direction.

%
%
%
%
%

\section{Conclusion}
\label{sec_conclusion}
In this paper, we have illustrated on fundamental examples originated from topics as varied as quantum mechanics, geophysical fluid dynamics, or fluid mechanics, the implementation of sparse spectral methods for solving PDE systems in unbounded domains of arbitrary dimensions.

 Central to our technique are hybrid expansions $\{TS_n\}$ and $\{ST_n\}$ (equation~\ref{mixed_cheb_series}), composed of Chebyshev functions $\{TB_n\}$ and $\{SB_n\}$ which preserve the sparsity of the discretization of differential operators on the infinite line even in presence of parity flipping differential operators in the general form of polynomial or rational fractions. Chebyshev functions correspond to remapped $\cos$ and $\sin$ functions to the infinite line, so that our hybrid expansions are remapped bases of interleaved sines and cosines, $\{\breve{\beta}_n\}$ and $\{\mathring{\beta}_n\}$ defined in equations~\ref{def:betas}. For exponentially decaying functions, we have shown that our hybrid Chebyshev expansions inherit the exponential convergence properties of standard Chebyshev expansions.
 
Numerical analyses of systems of differential equations are greatly facilitated, or made possible at all, by the sparsity of the discretization of rational differential operators that results from this choice of hybrid bases. We have shown examples of investigation of the linear dynamics  by solving sparse eigenproblems or by means of strongly accurate and stable implicit time-marching schemes, along with examples of analyses of non linear dynamics, tackled by means of popular IMEX schemes. The method is easily generalized to domains of higher dimensionality by means of Kronecker products.

The work in this paper was supported in part by the National Science Foundation under Grant DMS-1317666. The authors would like to acknowledge useful comments from Dr. Ian Grooms and two anonymous referees.

\section{Bibliography}
\bibliographystyle{unsrt}
\bibliography{EC_ii_a}

 %
 %
 %
 %
%
 %
 %
 %
 
 %
 %
 %
 %
 %
 %
 %
 %
 %
 
 %
 %
 %
%
 %
 %
 %
 %
 
 %
 %
 %
 %
 %
 %
 %
 %
 %
 
 %
 %
 %
%
 %
 %
 %
 %
 
 %
 %
 %
 %
 %
 %
 %
 %
 %
 
 %
 %
 %
 %
 %
 %
 %
 %
 
 %
 %
 %
 %
 %
 %
 %
 %
 %
 
 %
 %
 %
 %
 %
 %
 %
 %
 
 %
 %
 %
 %
 %
 %
 %
 %
 %
 
 %
 %
 %
 %
 %
%
 %
 %

\newpage

\section{Appendix: Convergence of the expansions.}
\label{app:convergence}
We discuss in this appendix the speed of convergence of the expansion of functions $\psi(y)$ on the $TS(y)$, and $ST(y)$ bases (eq.~\ref{def:TSST}). We first present tools to assess the speed of convergence of functions on the $[0,\pi]$ interval using the $\mathring\beta$ and $\breve{\beta}$ bases (eq.~\ref{def:beta_explicit}). We then discuss a few cases of functions on the infinite line, distinguished by their asymptotic behaviour at $|y|\rightarrow \infty$.
\subsection{Convergence of Fourier series for $2\pi$-periodic functions.}
It is well-known (see e.g.~\cite{appel}) that a $2\pi$ periodic function which is of $\mathscr{C}^{k-1}$ class (i.e. with a continuous $k-1$ derivative) on $\mathbb{R}$ and piecewise of $\mathscr{C}^{k}$ class on $\mathbb{R}$ possesses Fourier coefficients $\tilde f_n = o(1/n^k)$. This behaviour is referred to as algebraic convergence. By extension, functions of $\mathscr{C}^\infty$ class on $\mathbb{R}$ possesses Fourier coefficients that decay asymptotically faster than any power of $1/n$. Among the class of $\mathscr{C}^\infty$ functions on $\mathbb{R}$, the most coveted are analytic functions, the Fourier coefficients of which decay exponentially with $\tilde{f}_n \sim \exp(-an)$, with $a>0$. Such a convergence is dubbed geometric convergence. Remapping infinite intervals onto the $[0, \pi]$ segment shall typically bring us to consider an intermediate class of functions: non analytic $\mathscr{C}^{\infty}$ functions with essential singularities. For instance, $\exp(-\cot(\theta)^2)$ is a $2\pi$-periodic function of class $\mathscr{C}^{\infty}$ on $\mathbb{R}$ with essential singularities at $\theta= n\pi$ (with $n\in\mathbb{Z}$). It has been shown in~\cite{boydJCP82} that  Fourier coefficients of $\mathscr{C}^{\infty}$ functions which possess a singularity of the form $\exp(-1/\left|\theta\right|^k)$ decay like $\tilde f_n \sim \exp (-an^{2k/[2k+1]})$, with $a>0$.
\subsection{Convergence of cosine, sine, or mixed trigonometric series on the $[0,\pi]$ segment.}
Let $f(\theta)$ be a function of $\mathscr{C}^{\infty}$ class on $[0,\pi]$, that satisfies $f(0)= f(\pi)=0$. The behaviour of the trigonometric series of $f$ depends on symmetry considerations. To arm ourselves for manipulating such functions, we define the projectors on the symmetric and antisymmetric components of $f$ with respect to the center of the interval $\pi/2$ (see figure~\ref{continuation_and_probs}, panels a and c):
\begin{subequations}
\label{eq_decomposition}
\begin{align}
\mathscr{S}_{\pi/2} [f(\theta)] &= \frac{f(\theta) + f(\pi - \theta)}{2}\, ,\\
\mathscr{A}_{\pi/2} [f(\theta)] &= \frac{f(\theta) - f(\pi - \theta)}{2} = f(\theta) - \mathscr{S}_{\pi/2} [f(\theta)] \,.
\end{align}
\end{subequations}
Using these projectors, four distinct continuations of $f$ to the interval $[0,2\pi]$, with different symmetries, can be built. The four continuations (illustrated on figure~\ref{continuation_and_probs}, panels b and d) naturally coincide with $f$ on the $[0,\pi]$ segment and are as defined as follows for $\theta \in [-\pi,0]$:
\begin{subequations}
\label{g_continuations}
\begin{align}
g_1(\theta) &=+\mathscr{S}_{\pi/2}[f(-\theta)] + \mathscr{A}_{\pi/2}[f(-\theta)] \,, \\
g_2(\theta) &=+\mathscr{S}_{\pi/2}[f(-\theta)] - \mathscr{A}_{\pi/2}[f(-\theta)] \,, \\
g_3(\theta) &=-\mathscr{S}_{\pi/2}[f(-\theta)] + \mathscr{A}_{\pi/2}[f(-\theta)] \,, \\
g_4(\theta) &=-\mathscr{S}_{\pi/2}[f(-\theta)] - \mathscr{A}_{\pi/2}[f(-\theta)] \,. 
\end{align}
\end{subequations}
Each of these continuations corresponds to one of our four choices of basis: $\left\{\cos (2p\theta)\right\}$ or $\left\{\sin ([2p+1]\theta)\right\}$ for the symmetric component $\mathscr{S}_{\pi/2} [f(\theta)]$, and $\left\{\cos ([2p+1]\theta)\right\}$ or $\left\{\sin (2p\theta)\right\}$ for the antisymmetric component $\mathscr{A}_{\pi/2} [f(\theta)]$. Inspecting the parity around $\theta=0$, one easily identifies that $g_1$ corresponds to the pure cosine expansion, $g_4$ to the pure sine expansion, $g_2$ to the mixed $\breve{\beta}$ expansion, and $g_3$ to the $\mathring{\beta}$ expansion:
\begin{subequations}
\begin{align}
g_1(\theta) = \sum_{n=0}^{N-1} \tilde{f}_n\cos(n\theta)\,,& \quad
g_2(\theta) = \sum_{n=1}^{N} \breve{f}_n\breve{\beta_n}(\theta)\,, \\
g_3(\theta) = \sum_{n=1}^{N} \mathring{f}_n\mathring{\beta_n}(\theta)\,,& \quad
g_4(\theta) = \sum_{n=1}^{N}   \widehat{f}_n\sin(n\theta)\,. 
\end{align}
\end{subequations}
The behaviour of these four expansions, and notably their speed of convergence, is bound with the class of the continued fonction $g_i$ they correspond to. We discuss several cases below.
\begin{center}
\begin{figure}
\includegraphics[height = 0.31 \textwidth]{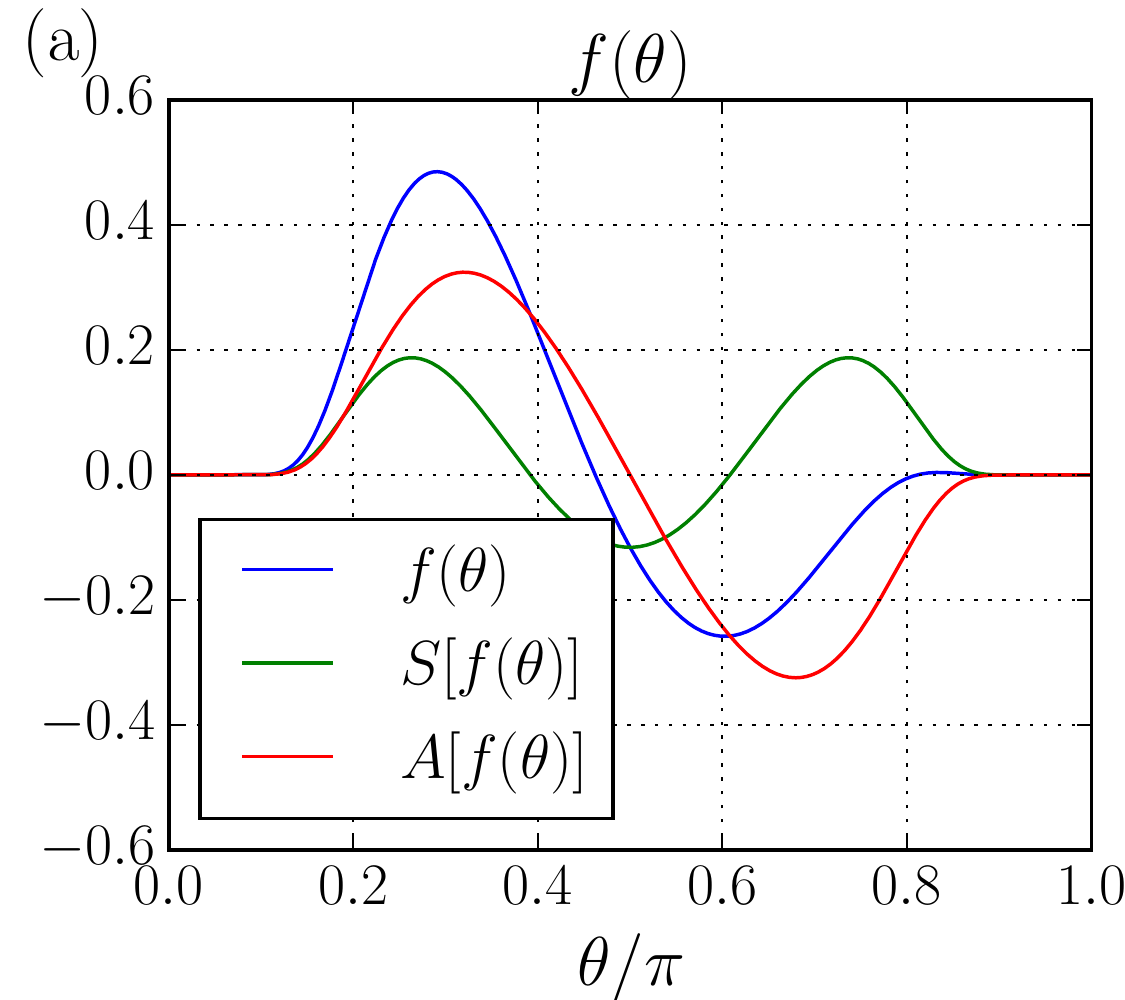}
\includegraphics[height = 0.31 \textwidth]{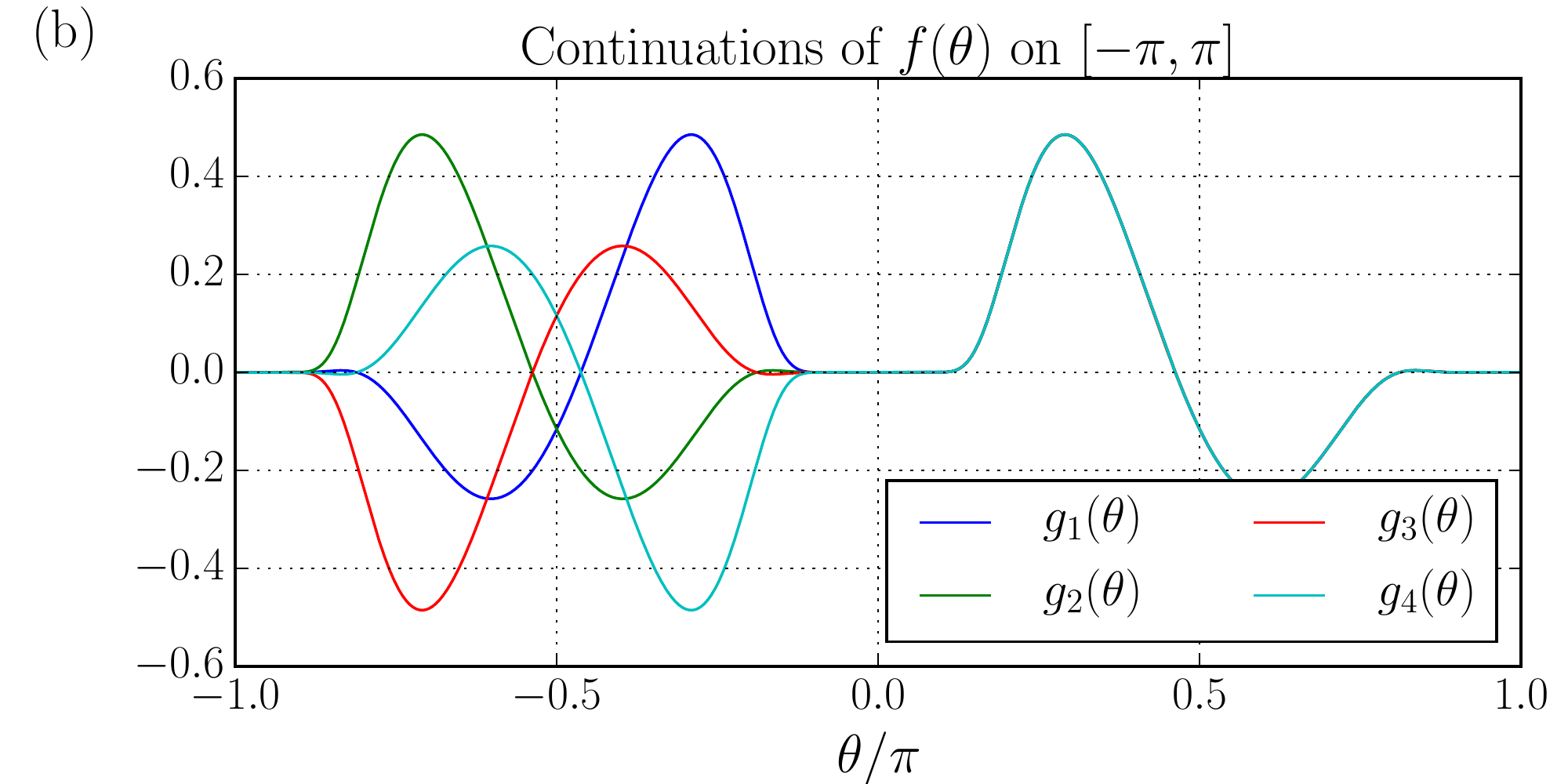}

\includegraphics[height = 0.31 \textwidth]{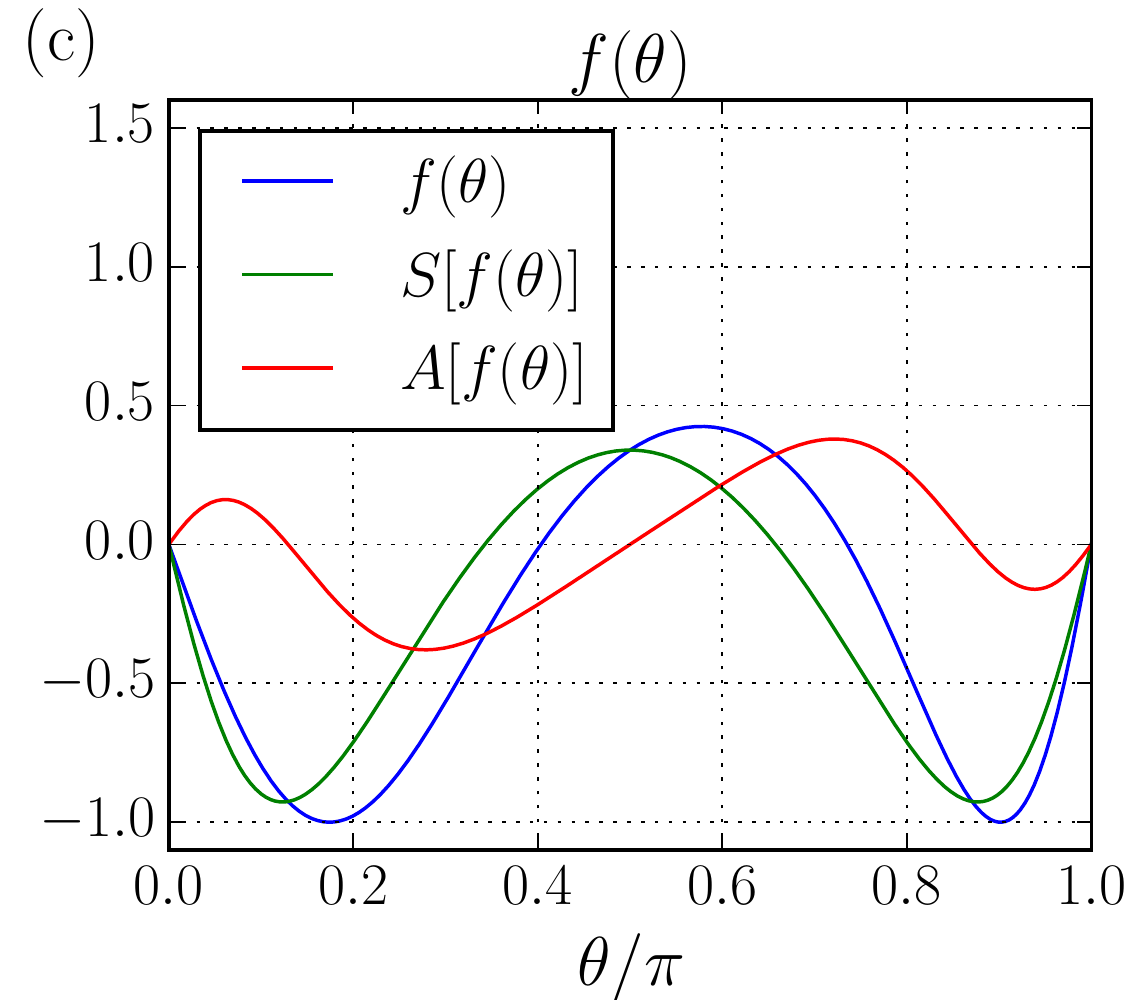}
\includegraphics[height = 0.31 \textwidth]{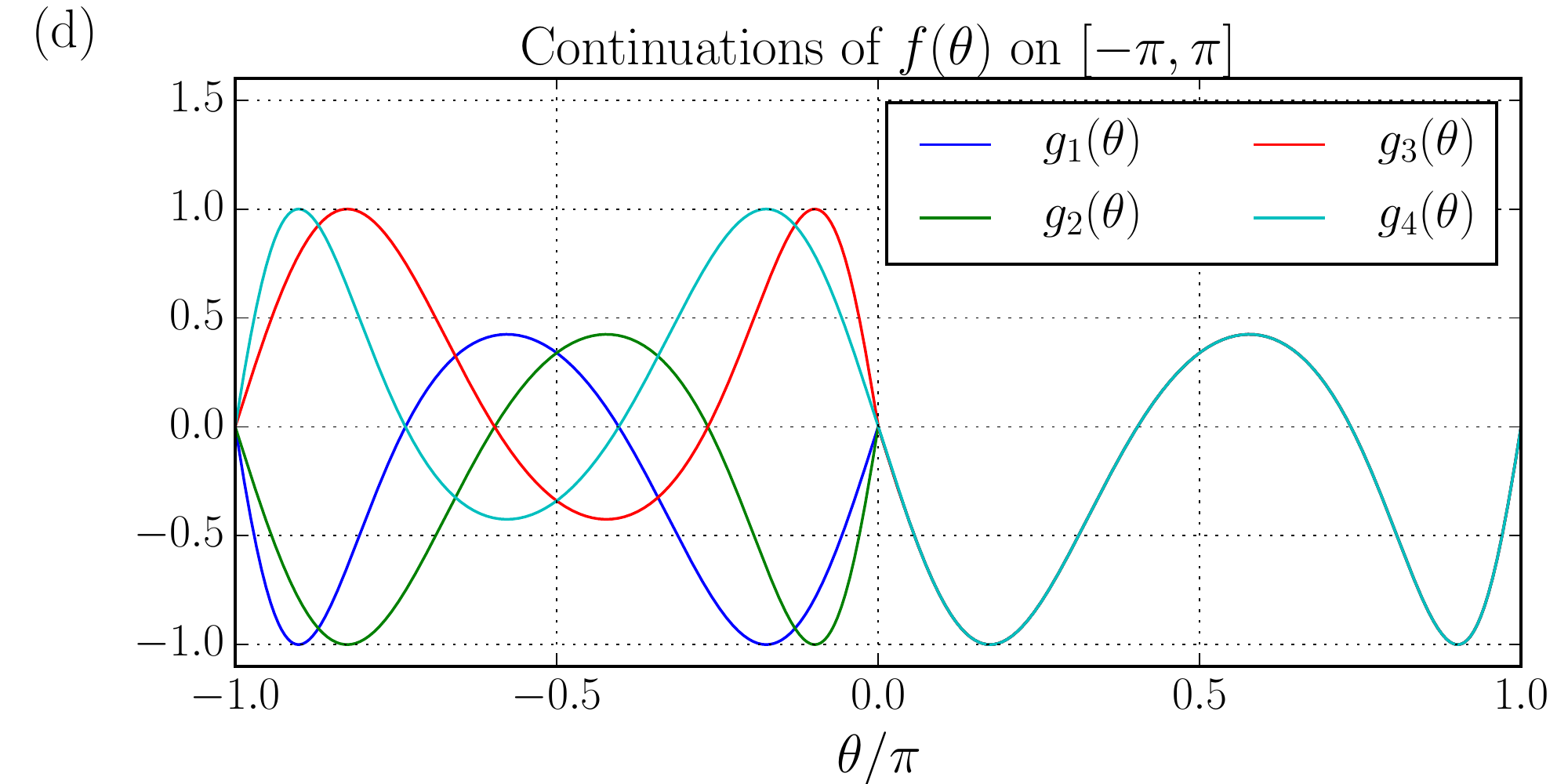}
\caption{
\label{continuation_and_probs} 
Example of continuations for a function with essential singularities at the end of the interval (top row, panels (a) and (b)) and for an analytic function (bottom row, panels (c) and (d)). Left panels (a) and (c) show the decomposition $f(\theta) = \mathscr{S}[f(\theta)] + \mathscr{A}[f(\theta)]$ (see equation~\ref{eq_decomposition}). Right panels (b) and (d) show the four possible continuations (see equation~\ref{g_continuations}). Note that all the continuations are of $\mathscr{C}^\infty$ class in (b) whereas it is so for $g_2$ solely in panel (d): other continuations are piecewise of class $\mathscr{C}^1$, with a discontinuous slope at $\theta = 0,\, \pm pi,\, \pm 2 \pi,\, \dots$.
}
\end{figure}
\end{center}
\paragraph{Functions with essential singularities at $0$ and $\pi$--} Functions with essential singularities at the end of the $[0, \pi]$ segment (common in the context of the remapped infinite line) are advantageous and give complete flexibility concerning the choice of the basis. Indeed, consider $f$ such that:
\begin{equation}
f(\varepsilon) \sim \exp\left(-\frac{1}{\left| \varepsilon\right|^k}\right) ,\quad \mathrm{and}\quad f(\pi-\varepsilon) \sim \exp\left(-\frac{1}{\left| \varepsilon\right|^q}\right) \,.
\end{equation}
Each of the for continuations $g_1,\dots,g_4$ will be of class $\mathscr{C}^{\infty}$ on $\mathbb{R}$, with essential singularities. Therefore, the coefficients of each expansion (pure or mixed trigonometric) will converge at a subgeometric rate. In this case, the user has a total freedom of choice: mixed expansions will inherit the desirable properties of pure expansions.
\paragraph{Non singular functions--} Non singular functions are ironically the least flexible as far as the choice of the basis is concerned. Let the Taylor expansions of $f$ at $\theta=0$ and $\theta = \pi$ be: 
\begin{equation}
f(\varepsilon) = \sum_{n=0}^{\infty} a_n\varepsilon^n ,\quad \mathrm{and}\quad f(\pi-\varepsilon) = \sum_{n=0}^{\infty} b_n\varepsilon^n \,.
\end{equation}
Our decomposition based on the symmetry around $\pi/2$ can be taken one step further, to distinguish between even and odd functions around $0$. The function $f$ can be decomposed into:
\begin{equation}
f(\theta) = \mathscr{SE}(\theta) + \mathscr{SO}(\theta) + \mathscr{AE}(\theta) + \mathscr{AO}(\theta)
\end{equation}
with
\begin{subequations}
\begin{align}
\mathscr{SE}(\theta) = \sum_{p=0}^\infty \left(\frac{a_{2p} + b_{2p}}{2}\right) \theta^{2p},\,\quad
\mathscr{SO}(\theta) = \sum_{p=0}^\infty \left(\frac{a_{2p+1} + b_{2p+1}}{2}\right) \theta^{2p+1}\,,
\\
\mathscr{AE}(\theta) = \sum_{p=0}^\infty \left(\frac{a_{2p} - b_{2p}}{2}\right) \theta^{2p},\,\quad
\mathscr{AO}(\theta) = \sum_{p=0}^\infty \left(\frac{a_{2p+1} - b_{2p+1}}{2}\right) \theta^{2p+1}\,.
\end{align}
\end{subequations}
As one attempts to continue $f$ over $[-\pi,0]$ by building $g$ of the form:
\begin{equation}
g(-\theta)= \sigma_{\mathscr{S}} \left(\mathscr{SE}(\theta) + \mathscr{SO}(\theta) \right) + \sigma_{\mathscr{A}} \left( \mathscr{AE}(\theta) + \mathscr{AO}(\theta) \right)\,,
\end{equation} 
it becomes obvious that none of the four choices for $\sigma_\mathscr{S}, \sigma_{\mathscr{A}} = \pm 1$ will yield an analytic function unless $\mathscr{S}[f(\theta)]$ and $\mathscr{A}[f(\theta)]$ are purely odd or purely even at $\theta=0$: some derivatives will be discontinuous at $0$. Such functions $f(\theta)$ will invariably have algebraically decaying Fourier coefficients, regardless of the expansion chosen. \\
Functions such that $\mathscr{S}[f(\theta)]$ and $\mathscr{A}[f(\theta)]$ have a definite parity around zero have potential for being continued into a $2\pi$-periodic analytic function on $\mathbb{R}$, but will offer little flexibility concerning the choice of basis that yields geometric convergence. As an illustration, consider $f$ such that $f(\theta) = \mathscr{SE}(\theta) + \mathscr{AO}(\theta)$. Building $g_2(\theta) = +\mathscr{S}_{\pi/2}[f(-\theta)] - \mathscr{A}_{\pi/2}[f(-\theta)]$ is the sole way to continue $f$ into an analytic function. As a consequence, a $\{\breve{\beta}\}$ expansion alone yields geometrically decaying coefficients.
\subsection{Convergence of $SB$, $TB$, or mixed $TS$ or $ST$ bases on $(-\infty, \infty)$.}
By virtue of the discussion above, the speed of convergence of pure $\{TB\}$ and $\{SB\}$ or hybrid $\{TS\}$ and $\{ST\}$ Chebyshev expansions for a function $\psi(y)$ is controlled by the asymptotic behaviour of $\psi$ as $|y|\rightarrow \infty$. Functions $\psi(y)$ dominated at $|y|\rightarrow \infty$ by an exponentially decaying function $\exp(-b\left|y\right|^a)$ with $a,b>0$, are mapped onto functions $f(\theta)$ with essential singularities at  both ends of the $[0,\pi]$ segment. Indeed, upon remapping $y = L \cot \theta$ and choosing $L=1$ for simplicity and without loss of generality, we find that $f(\theta) = \psi(y(\theta))$ is dominated around $\theta \downarrow 0$ by $\exp(-b/\left|\theta\right|^a)$ (by virtue of $\cot \theta = 1/\theta + O(\theta)$). In the light of our reasoning on the $[0,\pi]$ segment, the four mixed and hybrid expansions will converge subgeometrically. \\
Functions that decay algebraically $\psi(y) \sim 1/y^k$ for $|y|\rightarrow \infty$ will map to analytic functions on $[0,\pi]$. As stressed above, such a situation is unfortunate as it might not allow to choose freely the basis of expansion (unless one is to pay the price of algebraic convergence).

\section{Appendix: Orthogonality of the hybrid bases $\{\breve{\beta}_n\}$ and $\{\mathring{\beta}_n\}$.}
\label{app:ortho}
Recalling from equation~(\ref{def:beta_breve_condensed}) that:
\begin{equation}
\breve{\beta}_{2p} = \sin(2p\theta),\quad \mathrm{and} \quad \breve{\beta}_{2p+1} = \cos(2p\theta)\,,
\end{equation}
the orthogonality of the basis is immediate. We prove that $\inner{\breve{\beta}_m}{\breve{\beta}_n} \propto \delta^n_m$, where $\delta^n_m$ is the kronecker delta symbol, by remarking the following: we have $\inner{\breve{\beta}_1}{\breve{\beta}_1} = 1$, and for $p$ or $q$ not both equal to zero:
\begin{equation}
\inner{\breve{\beta}_{2p}}{\breve{\beta}_{2q}} = \frac{1}{2} \delta^{p}_q\,,\quad
\inner{\breve{\beta}_{2p+1}}{\breve{\beta}_{2q+1}}= \frac{1}{2} \delta^{p}_q\,, \quad
\inner{\breve{\beta}_{2p}}{\breve{\beta}_{2q+1}} = 0\,, \quad
\inner{\breve{\beta}_{2p+1}}{\breve{\beta}_{2q}} = 0\,. 
\end{equation}
Similarly, using equation~(\ref{def:beta_mathring_condensed}):
\begin{equation}
\mathring{\beta}_{2p-1} = \cos([2p-1]\theta),\quad \mathrm{and} \quad \mathring{\beta}_{2p} = \sin([2p-1]\theta)\,,
\end{equation}
one immediately gets the orthogonality $\inner{\mathring{\beta}_m}{\mathring{\beta}_n} =\frac{1}{2} \delta^n_m$:
\begin{equation}
\inner{\mathring{\beta}_{2p}}{\mathring{\beta}_{2q}} = \frac{1}{2} \delta^{p}_q\,, \quad
\inner{\mathring{\beta}_{2p+1}}{\mathring{\beta}_{2q+1}} = \frac{1}{2} \delta^{p}_q\,, \quad
\inner{\mathring{\beta}_{2p}}{\mathring{\beta}_{2q+1}} = 0\,, \quad
\inner{\mathring{\beta}_{2p+1}}{\mathring{\beta}_{2q}} = 0\,. 
\end{equation}
\section{Appendix: Matrix elements}
\subsection{Discretization along an unbounded direction: Chebyshev functions}
\label{app:chebfunc}
This appendix documents a method to obtain discretized matrices for operators composed of superpositions of powers of cosines, sines, and derivatives. Such operators are typically obtained after the procedure described in~\ref{general_case} and exemplified in~\ref{sec_arbitrary_well}. We will consider the four matrices $\elemssc$, $\elemcss$, $\elemccc$ and $\elemscs$ as our elementary building blocks. These matrices represent the action of $\sin\theta$ on cosines and sines, and the action of $\cos\theta$ on cosines and sines, respectively. Their matrix elements are given by:
\begin{subequations}
\begin{align}
\elemssc_{mn} &= \frac{\inner{\sin(m\theta)}{\sin\theta\cos([n-1]\theta)}}{\norm{\sin(m\theta)}^2} = \left\{\begin{array}{ccl} \delta^1_m&\mathrm{for}&n=1\\
\frac{1}{2}\left(\delta^{n}_{m} - \delta^{n-2}_m \right)&\mathrm{for}&n\ge 2
\end{array}\right. 
\\
\elemcss_{mn}& = \frac{\inner{\cos([m-1]\theta)}{\sin\theta\sin(n\theta)}}{\norm{\cos([m-1]\theta)}^2}
 =\frac{1}{2}\left(-\delta^{n+1}_{m-1} + \delta^{n-1}_{m-1} \right)
\\
\elemccc_{mn} & = 
\frac{\inner{\cos([m-1]\theta)}{\cos\theta\cos([n-1]\theta)}}{\norm{\cos([m-1]\theta)}^2}
 =
\left\{\begin{array}{ccl} \delta^2_{m}&\mathrm{for}&n=1\\ \frac{1}{2}\left(\delta^{n}_{m-1} + \delta^{n-2}_{m-1} \right) &\mathrm{for}&n\ge 2 \end{array}\right.
\\
\elemscs_{mn} & = 
\frac{\inner{\sin(m\theta)}{\cos\theta\sin(n\theta)}}{\norm{\sin(m\theta)}^2} 
= \frac{1}{2}\left(\delta^{n+1}_{m} + \delta^{n-1}_m \right) 
\end{align}
We also define differentiation matrices:
\begin{align}
\mathcal{D}^{sc} &= \frac{
\inner{\sin(m\theta)}{\partial_\theta \cos([n-1]\theta)}}
{\norm{\sin(m\theta)}^2} =- \left(n-1\right)\delta_m^{n-1} \,, \\
\mathcal{D}^{cs} &= \frac{
\inner{\cos([m-1]\theta)}{\partial_\theta \sin(n\theta)}}
{\norm{\sin([m-1]\theta)}^2} = n\, \delta^n_{m-1} \,. 
\end{align}
\end{subequations}
From these building blocks, more complex operators are readily constructed using matrix products, 
 with the only reservation that the lower right corner of the matrix, which corresponds to modes neighbouring the truncation, deserves specific attention. Indeed, should no care be taken, the lower right corner of the matrix product would be inexact, due to the proximity of truncation. A rule of thumb is that, when two matrices $\mathbfsf{L}_1$ and $\mathbfsf{L}_2$ represent the action of operators $\mathscr{L}_1$ and $\mathscr{L}_2$ and have a bandwidth $(1+2p)$ (with $p$ subdiagonals and $p$ superdiagonals), the action of the operator $\mathscr{L}_3=\mathscr{L}_1 \mathscr{L}_2$ is \emph{not} represented by the product $\mathbfsf{L}_1\mathbfsf{L}_2$, for the last $p$ columns of this product are spurious. This problem can be understood as some kind of aliasing related to linear operators and is easily circumvented by using a larger truncation $N' > N$ to discretize $\mathscr{L}_1$ and $\mathscr{L}_2$, so that to obtain large square matrices $\mathbfsf{L}_1'$ and $\mathbfsf{L}_2'$ of size $N'$. Then, one computes the product $\mathbfsf{L}_3'=\mathbfsf{L}_1'\mathbfsf{L}_2'$ of size $N'$, and finally one extracts a smaller square matrix $\mathbfsf{L}_3$ of size $N$ that rigorously discretize $\mathscr{L}_3$. Below we denote with an overbar $\overline{\mathbfsf{L}}^N$ the extraction from a given square matrix $\mathbfsf{L}$ of a square submatrix of size $N$. Using our prescription, we can safely use a matrix product to obtain the matrices 
$\mathbfsf{A}$, $\mathbfsf{B}$, $\mathbfsf{C}$, $\mathbfsf{D}$ and $\mathbfsf{E}$, defined in equation (43), as products of elementary matrices $\mathcal{E}$ of suitably enlarged size $N'> N$. We distinguish below the case of parity-conserving and parity-flipping operators.
\paragraph{The case of parity-conserving operators} Our prescription is first illustrated with the parity conserving operator $\mathscr{D} = \sin ^2 \theta $, in view of obtaining its discretization $\mathbfsf{D}$ on the basis $\{\breve{\beta}_n\}$, as defined in  equation~(42d). Recalling that this basis contains interleaved even harmonics of sines and cosines [see equations~(\ref{def:beta_breve_condensed}) and~(\ref{def:beta_breve_explicit})]:
\begin{equation}
\left\{\breve{\beta}_n\right\}_{1\le n \le N} = \left\{1,\,\sin2\theta,\,\cos 2\theta,\,\sin 4\theta,\, \cos 4\theta,\dots \right\}\,,
\end{equation}
we easily deduce that even elements $\breve\beta_{2p}=\sin (2p\theta)$ are antisymmetric functions whereas odd elements $\breve\beta_{2p-1} = \cos ([2p-2]\theta)$ are symmetric functions. The symmetry of the operator $\mathscr{D}$ ensures us that this operator cannot couple symmetric and antisymmetric functions. Hence we only compute the coupling among (symmetric) cosines $\mathbfsf{D}^{c}$ and among (antisymmetric) sines $\mathbfsf{D}^{s}$ as products of elementary matrices $\mathcal{E}$ of size $N'\ge N+1$:
\begin{subequations}
\begin{align}
\mathbfsf{D}^{c} &= \frac{\inner{\cos([m-1]\theta)}{\sin^2\theta\cos([n-1]\theta)}}{\norm{\cos([m-1]\theta)}^2} = \overline{\elemcss\elemssc}^N\,, \\ 
\mathbfsf{D}^{s} &=
\frac{\inner{\sin(m\theta)}{\sin^2\theta\sin(n\theta)}}{\norm{\sin(m\theta)}^2}= \overline{\elemssc\elemcss}^N\,.
\end{align}
\end{subequations}
Finally, we obtain the expression of $\mathbfsf{D}$, the discretization of $\mathscr{D}=\sin^2\theta$ on the basis $\{ \breve{\beta}_n\}$, by interweaving adequately the matrices $\mathbfsf{D}^{c}$ and $\mathbfsf{D}^{s}$:
\begin{subequations}
\begin{align}
\mathbfsf{D}_{2p-1,2q-1} & = \mathbfsf{D}^c_{2p-1,2q-1}\,, \\
\mathbfsf{D}_{2p,2q} & = \mathbfsf{D}^s_{2p,2q}\,. 
\end{align}\label{interweaving}
\end{subequations}
We indicate below the construction by matrix products for matrices of operators $\mathscr{A} = \sin^6 \partial^{2}_{\theta\theta}$, $\mathscr{B}= \cos \theta \sin^5 \theta \partial_\theta$, and $\mathscr{C}=\cos^2\theta$. These parity conserving operators are all treated in a similar fashion as $\mathscr{D}$: as justified above, we only compute coupling among cosines and sines:
\begin{subequations}
\begin{align}
\mathbfsf{A}^{c} &= \frac{\inner{\cos([m-1]\theta)}{\mathscr{A}\cos([n-1]\theta)}}{\norm{\cos([m-1]\theta)}^2} = \overline{\elemcss\elemssc\elemcss\elemssc\elemcss\elemssc\mathcal{D}^{cs}\mathcal{D}^{sc}}^N\,, \\ 
\mathbfsf{A}^{s} &=
\frac{\inner{\sin(m\theta)}{\mathscr{A}\sin(n\theta)}}{\norm{\sin(m\theta)}^2}= \overline{\elemssc\elemcss\elemssc\elemcss\elemssc\elemcss\mathcal{D}^{sc}\mathcal{D}^{cs}}^N\,,\\
\mathbfsf{B}^{c} &= \frac{\inner{\cos([m-1]\theta)}{\mathscr{B}\cos([n-1]\theta)}}{\norm{\cos([m-1]\theta)}^2} = \overline{\elemccc\elemcss\elemssc\elemcss\elemssc\elemcss\mathcal{D}^{sc}}^N\,, \\ 
\mathbfsf{B}^{s} &=
\frac{\inner{\sin(m\theta)}{\mathscr{B}\sin(n\theta)}}{\norm{\sin(m\theta)}^2}= \overline{\elemscs\elemssc\elemcss\elemssc\elemcss\elemssc\mathcal{D}^{cs}}^N\,.
\end{align}
\end{subequations}
The final interweaving to obtain $\mathbfsf{A}$ and $\mathbfsf{B}$ is identical as in the case of $\mathbfsf{D}$, given in equations~(\ref{interweaving}). The matrix $\mathbfsf{C}$ is directly obtained from the identity $\cos^2\theta = 1 - \sin^2\theta$. Thus, denoting $\mathbfsf{1}$ the identity matrix :
\begin{equation}
\mathbfsf{C}=\mathbfsf{1} - \mathbfsf{D}\,.
\end{equation}
\paragraph{The case of parity-flipping operators}
Finally, we illustrate the treatment of a parity breaking term by discretizing the operator $\mathscr{E}=\cos\theta\sin\theta$. The parity-flipping nature of $\mathscr{E}$ guarantees that, in contrast with parity-conserving operators, this operator does not couple elements of the basis with the same parity.
Instead, the coupling occurs between sines and cosines, leading us to compute $\mathbfsf{E}^{sc}$ and $\mathbfsf{E}^{cs}$:
\begin{subequations}
\begin{gather}
\mathbfsf{E}^{sc} = 
\frac{
\inner{\sin(m\theta)}{\mathscr{E} \cos([n-1]\theta)}
}
{\norm{\sin(m\theta)}^2}
= \overline{\elemscs \elemssc}^N\,,\\ 
\mathbfsf{E}^{sc} = \frac{
\inner{\cos([m-1]\theta)}{\mathscr{E} \sin(n\theta)}
}
{\norm{\cos([m-1]\theta)}^2}
= \overline{\elemccc \elemcss}^N
 \,.
\end{gather}
\end{subequations}
Finally, the matrix $\mathbfsf{E}$ defined in equation (42e) is obtained by the following interweaving, that differs from the parity-conserving case:
\begin{subequations}
\begin{align}
\mathbfsf{E}_{2p,2q-1} = \mathbfsf{E}^{sc}_{2p,2q-1}\,, \\
\mathbfsf{E}_{2p-1,2q} = \mathbfsf{E}^{sc}_{2p-1,2q}\,.
\end{align}
\end{subequations}
\subsection{Discretization along a bounded direction: Chebyshev polynomials and the quasi-inverse technique}\label{app:chebpol}
Central to the quasi-inverse technique is the three-term relation for the antiderivative of Chebyshev polynomials, first emphasized in the context of numerical resolution of ODEs by Clenshaw~\cite{clenshawPCPS57}:
\begin{subequations}
\begin{gather}
\int T_0(x)\mathrm{d}x  = T_1(x)+k\,, \\ 
\int T_1(x)\mathrm{d}x  =\frac{1}{4}T_2(x)+k\,, \\
\int T_n(x)\mathrm{d}x  =\frac{1}{2}\left( \frac{T_{n+1}(x)}{n+1} + \frac{T_{n-1}(x)}{n-1}\right)+k,\:\mathrm{for}\: n\ge 2\,.
\end{gather}
\end{subequations}
Upon integration, an arbitrary constant $k$ naturally appears. This constant will be set by enforcing boundary conditions: the consequence is that, for now, the top row of the matrix representing integration can be left equal to zero.
\begin{equation}
\mathbfsf{I}_{mn} = \frac{\inner{T_{m-1}}{\int T_{n-1}}}{\norm{T_{m-1}}^2} = \left\{\begin{gathered}
\delta_m^2\:\mathrm{for}\: n=1 \,, 
\\
\frac{1}{4}\delta_m^3\:\mathrm{for}\: n=2\,, 
\\
\frac{1}{2m}\delta^{n+1}_m - \frac{1}{2m}\delta^{n-1}_m\,.
\end{gathered} \right.
\end{equation}
Another handy three-term relationship corresponds to the multiplication by $x$:
\begin{subequations}
\begin{gather}
xT_0(x) = T_1(x)\,, \\
xT_n(x) = \frac{1}{2}\big( T_{n+1}(x) + T_{n-1}(x) \big),\:\mathrm{for}\:n\ge 1\,,
\end{gather}
\end{subequations}
which has the discretized matrix:
\begin{equation}
\mathbfsf{x}_{mn} = \frac{\inner{T_{m-1}}{x T_{n-1}}}{\norm{T_{m-1}}^2} = \left\{\begin{gathered}
\delta_m^2\:\mathrm{for}\: n=1\,,\\
\frac{1}{2}\delta_m^{n+1} +\frac{1}{2}\delta_m^{n-1} \,.\\ 
\end{gathered} \right.
\end{equation}
Hence the matrices $\mathbfsf{Q}$ defined in equations~(\ref{Q123}) are again obtained by truncated matrix products of large matrices:
\begin{subequations}
\begin{gather}
\mathbfsf{Q}^{(0)} = \overline{\mathbfsf{I}^2}^N \,, \\ 
\mathbfsf{Q}^{(1)} = \overline{\mathbfsf{I}^2\mathbfsf{x}}^N \,, \\ 
\mathbfsf{Q}^{(2)} = \overline{\mathbfsf{I}^2\mathbfsf{x}^2}^N \,.
\end{gather}
\end{subequations}

\end{document}